\documentclass[aps,prd,twocolumn,preprintnumbers,nofootinbib,longbibliography,floatfix,superscriptaddress]{revtex4-1}

\usepackage{multirow,csvsimple,verbatim,graphicx,bm,color,float,amsmath,amssymb,siunitx,hyperref,bold-extra,afterpage}
\usepackage[caption=false]{subfig}
\usepackage[utf8]{inputenc}

\definecolor{darkblue}{rgb}{0,0,0.7}
\definecolor{darkred}{rgb}{0.7,0,0}
\definecolor{darkgreen}{rgb}{0,0.7,0}
\definecolor{orange}{rgb}{0.9, 0.25, 0}
\definecolor{purple}{rgb}{0.4, 0.04, 0.2}

\begin{document}
\DeclareRobustCommand{\Sec}[1]{Sec.~\ref{#1}}
\DeclareRobustCommand{\Secs}[2]{Secs.~\ref{#1} and \ref{#2}}
\DeclareRobustCommand{\App}[1]{App.~\ref{#1}}
\DeclareRobustCommand{\Tab}[1]{Table~\ref{#1}}
\DeclareRobustCommand{\Tabs}[2]{Tables~\ref{#1} and \ref{#2}}
\DeclareRobustCommand{\Fig}[1]{Fig.~\ref{#1}}
\DeclareRobustCommand{\Figs}[2]{Figs.~\ref{#1} and \ref{#2}}
\DeclareRobustCommand{\Figss}[3]{Figs.~\ref{#1}, \ref{#2} and \ref{#3}}
\DeclareRobustCommand{\Eq}[1]{Eq.~(\ref{#1})}
\DeclareRobustCommand{\Eqs}[2]{Eqs.~(\ref{#1}) and (\ref{#2})}
\DeclareRobustCommand{\Ref}[1]{Ref.~\cite{#1}}
\DeclareRobustCommand{\Refs}[1]{Refs.~\cite{#1}}

\title{Exploring the Space of Jets with CMS Open Data}

\author{Patrick T. Komiske}
\email{pkomiske@mit.edu}
\affiliation{Center for Theoretical Physics, Massachusetts Institute of Technology, Cambridge, MA 02139, USA}
\affiliation{Department of Physics, Harvard University, Cambridge, MA 02138, USA}

\author{Radha Mastandrea}
\email{rmastand@mit.edu}
\affiliation{Center for Theoretical Physics, Massachusetts Institute of Technology, Cambridge, MA 02139, USA}

\author{Eric M. Metodiev}
\email{metodiev@mit.edu}
\affiliation{Center for Theoretical Physics, Massachusetts Institute of Technology, Cambridge, MA 02139, USA}
\affiliation{Department of Physics, Harvard University, Cambridge, MA 02138, USA}

\author{Preksha Naik}
\email{prekshan@mit.edu}
\affiliation{Center for Theoretical Physics, Massachusetts Institute of Technology, Cambridge, MA 02139, USA}

\author{Jesse Thaler}
\email{jthaler@mit.edu}
\affiliation{Center for Theoretical Physics, Massachusetts Institute of Technology, Cambridge, MA 02139, USA}
\affiliation{Department of Physics, Harvard University, Cambridge, MA 02138, USA}

\begin{abstract}
We explore the metric space of jets using public collider data from the CMS experiment.
Starting from \SI{2.3}{{fb}^{-1}} of proton-proton collisions at  $\sqrt{s}=7$ TeV collected at the Large Hadron Collider in 2011, we isolate a sample of 1,690,984 central jets with transverse momentum above \SI{375}{GeV}.
To validate the performance of the CMS detector in reconstructing the energy flow of jets, we compare the CMS Open Data to corresponding simulated data samples for a variety of jet kinematic and substructure observables.
Even without detector unfolding, we find very good agreement for track-based observables after using charged hadron subtraction to mitigate the impact of pileup.
We perform a range of novel analyses, using the ``energy mover's distance'' (EMD) to measure the pairwise difference between jet energy flows.
The EMD allows us to quantify the impact of detector effects, visualize the metric space of jets, extract correlation dimensions, and identify the most and least typical jet configurations.
To facilitate future jet studies with CMS Open Data, we make our datasets and analysis code available, amounting to around two gigabytes of distilled data and one hundred gigabytes of simulation files.
\end{abstract}

\preprint{MIT-CTP 5129}
\maketitle
{\small\tableofcontents}

\section{Introduction}

Ever since the first evidence for jet structure~\cite{Hanson:1975fe}, the fragmentation of short-distance quarks and gluons into long-distance hadrons has been a rich area for experimental and theoretical investigations into quantum chromodynamics (QCD).
A variety of observables have been proposed over the decades to probe the jet formation process~\cite{Bjorken:1969wi,Ellis:1976uc,Georgi:1977sf,Farhi:1977sg,Parisi:1978eg,Donoghue:1979vi,Rakow:1981qn}, especially with recent advances in the field of jet substructure~\cite{Seymour:1991cb,Seymour:1993mx,Butterworth:2002tt,Butterworth:2007ke,Butterworth:2008iy,Abdesselam:2010pt,Altheimer:2012mn,Altheimer:2013yza,Adams:2015hiv,Larkoski:2017jix,Asquith:2018igt,Marzani:2019hun}.
The stress-energy flow~\cite{Tkachov:1995kk,Sveshnikov:1995vi,Cherzor:1997ak} is a particularly powerful probe of jets, since it in principle contains all the information about a jet that is infrared and collinear (IRC) safe~\cite{Kinoshita:1962ur,Lee:1964is,Sterman:1977wj}.
A variety of observables have been built around the energy flow concept~\cite{Berger:2002jt,Berger:2003iw,Larkoski:2013eya,Moult:2016cvt,Komiske:2017aww}, including recent work on machine learning for jet substructure~\cite{Komiske:2018cqr,Lim:2018toa,Chakraborty:2019imr}.

The unprecedented release of public collider data by the CMS experiment~\cite{Chatrchyan:2008aa} starting in November 2014~\cite{CERNOpenDataPortal} has enabled new exploratory studies of jets.
The first such jet analyses~\cite{Larkoski:2017bvj,Tripathee:2017ybi} were performed using the CMS 2010 Open Data~\cite{CMS:JetPrimary2010B}, corresponding to \SI{31.8}{pb^{-1}} of \SI{7}{TeV} data from Run 2010B at the Large Hadron Collider (LHC).
Among other aspects of jets, these studies explored the groomed momentum fraction $z_g$~\cite{Larkoski:2015lea}, which has subsequently been measured in proton-proton and heavy-ion collisions by CMS~\cite{Sirunyan:2017bsd}, ALICE~\cite{Acharya:2019djg}, and STAR~\cite{Kauder:2017mhg}.
The CMS Open Data release from LHC Run 2011A includes detector-simulated Monte Carlo (MC) samples, facilitating machine learning studies~\cite{Madrazo:2017qgh,Andrews:2018nwy,Andrews:2019faz}, an underlying event study~\cite{PaktinatMehdiabadi:2019ujl}, as well as a novel search for dimuon resonances~\cite{Cesarotti:2019nax}.
CMS has also released data from Runs 2012B and 2012C, which have been used to search for non-standard sources of parity violation in jets~\cite{Lester:2019bso} and extract standard model cross sections~\cite{Apyan:2019ybx}.
Beyond CMS, archival ALEPH data~\cite{ALEPHPreservationPolicy} have been used by \Ref{Heister:2016stz} to search for new physics and by \Refs{Kile:2017ryy,Kile:2017ccn,Kile:2017psu} to perform QCD studies.
While analyses using public collider data cannot match the sophistication or scope of official measurements by the experimental collaborations, they can enable proof-of-concept collider investigations and help stress-test archival data strategies.

In this paper, we perform the first exploratory study of the ``space'' of jets using the CMS 2011 Open Data.
This data and MC release corresponds to \SI{2.3}{{fb}^{-1}} of proton-proton collisions collected at a center-of-mass energy of $\sqrt{s}=7$ TeV.
The key idea, as proposed in \Ref{Komiske:2019fks}, is to compute the pairwise distance between jet energy flows, and then use this information to construct a metric space.
This enables a variety of distance-based jet analyses, including quantitative characterizations and qualitative visualizations.
Because this is an exploratory study, we do not unfold for detector effects nor estimate systematic uncertainties, but the general agreement between the CMS Open Data and simulated MC samples provides evidence for the experimental robustness of these methods.

The metric we use is the ``energy mover's distance'' (EMD)~\cite{Komiske:2019fks}, inspired by the famous earth mover's distance~\cite{DBLP:journals/pami/PelegWR89,Rubner:1998:MDA:938978.939133,Rubner:2000:EMD:365875.365881,DBLP:conf/eccv/PeleW08,DBLP:conf/gsi/PeleT13} sharing the same acronym.
The EMD has units of energy (i.e.~GeV) and quantifies the amount of ``work'' in energy times angle to make one jet radiation pattern look like another, including the cost of creating energy for jets with different $p_T$.
While we focus on the EMD between pairs of jets in this study, the same concept could be applied to pairs of events as a whole.
Crucially, the CMS Open Data contains full information about reconstructed particle flow candidates (PFCs)~\cite{CMS-PAS-PFT-09-001,CMS-PAS-PFT-10-001,Sirunyan:2017ulk}, which provide a robust proxy for the energy flow of a jet.
It also contains information about primary vertices, allowing us to mitigate pileup (multiple proton-proton collisions per beam crossing) through charged hadron subtraction (CHS)~\cite{CMS:2014ata}.
Because of the improved resolution and pileup insensitivity of charged particles (i.e.~tracks), we use a track-based variant of EMD for these exploratory studies.

We base our study on the CMS 2011 \texttt{Jet} Primary Dataset~\cite{CMS:JetPrimary2011A} and focus on the \texttt{HLT\_Jet300} single-jet trigger, which we show is fully efficient to reconstruct jets with transverse momentum ($p_T$) above \SI{375}{GeV}.
We also use dijet MC samples~\cite{CMS:QCDsim0-5,CMS:QCDsim5-15,CMS:QCDsim15-30,CMS:QCDsim30-50,CMS:QCDsim50-80,CMS:QCDsim80-120,CMS:QCDsim120-170,CMS:QCDsim170-300,CMS:QCDsim300-470,CMS:QCDsim470-600,CMS:QCDsim600-800,CMS:QCDsim800-1000,CMS:QCDsim1000-1400,CMS:QCDsim1400-1800,CMS:QCDsim1800}, generated with \textsc{Pythia 6}~\cite{Sjostrand:2006za} and simulated using \textsc{Geant 4}~\cite{Agostinelli:2002hh}, to understand the performance of the CMS detector in reconstructing the jet energy flow.
In order to facilitate future jet studies on the CMS Open Data, we make our MIT Open Data (MOD) software framework available~\cite{EnergyFlow,MODRepo}, along with the distilled data~\cite{MOD:ZenodoCMS} and MC~\cite{MOD:ZenodoMC170,MOD:ZenodoMC300,MOD:ZenodoMC470,MOD:ZenodoMC600,MOD:ZenodoMC800,MOD:ZenodoMC1000,MOD:ZenodoMC1400,MOD:ZenodoMC1800}  files needed to recreate the majority of our studies.

The remainder of this paper is organized as follows.
We begin in \Sec{sec:cmsopendata} by describing the CMS Open Data and the MOD software framework used for our analysis.
In \Sec{sec:jetsubstructure}, we validate the \texttt{Jet} Primary Dataset by comparing the basic kinematic and substructure properties of jets between the CMS data and MC samples.
The core of our analysis is in \Sec{sec:emd}, where we perform a variety of exploratory studies using the EMD.
We conclude in \Sec{sec:conclusion} with a discussion of future jet studies on public collider data.

\newpage

\section{Processing the CMS Open Data}
\label{sec:cmsopendata}

In this section, we describe the main steps for processing the CMS Open Data.
Our eventual analyses will be based on a single unprescaled trigger above its turn-on threshold, but we include additional details here about the analysis pipeline in order to demonstrate the general capabilities of our framework.
The reader already familiar with how CMS data is processed can safely skip to \Sec{subsec:selection}, where we describe the baseline jet selection criteria used for our substructure and EMD studies.

\subsection{Jet Primary Dataset}
\label{subsec:jetprimary}

The CMS Open Data is available on the CERN Open Data Portal~\cite{CERNOpenDataPortal}, which currently hosts data collected by CMS in 2010~\cite{CMS2010Release}, 2011~\cite{CMS2011Release}, and 2012~\cite{CMS2012Release}, as well as specialized samples for machine learning studies~\cite{CMSMLRelease}.
It also contains limited datasets from ALICE~\cite{ALICE2014Release}, ATLAS~\cite{ATLAS2016Release}, and LHCb~\cite{LHCbMasterclass}, as well as data from the OPERA neutrino experiment~\cite{OPERA2018Release}.
Accompanying the CMS 2011 Open Data is a virtual machine which runs version 5.3.32 of the CMS software (CMSSW) framework.
This open data initiative complements efforts like \textsc{HEPData}~\cite{HEPData}, \textsc{Rivet}~\cite{RIVET}, and \textsc{Reana}~\cite{REANA} to preserve the results and workflows of official collider analyses (see further discussion in \Ref{OpenNotEnough}).

The CMS Open Data is grouped into primary datasets that contain a subset of the triggers used for event selection~\cite{Khachatryan:2016bia}.
There are 19 primary datasets included in the 2011 release, along with corresponding MC samples (see \Sec{subsec:simulation} below).
All of the primary datasets are provided by CMS in their analysis object data (AOD) format, which provides high-level reconstructed objects used for the bulk of official CMS analyses in Run 1.
A subsample of some primary datasets (e.g.\ \texttt{Jet}~\cite{CMS:JetRAW2011A} and \texttt{MinimumBias}~\cite{CMS:MinimumBiasRAW2011A}) are also provided in the RAW format, containing the full readout of the CMS detector.

Our analysis is based on the \texttt{Jet} Primary Dataset~\cite{CMS:JetPrimary2011A}, which includes a variety of single jet and dijet triggers.
This primary dataset contains 30,726,331 events spread across 1,223 AOD files, totaling \SI{4.7}{TB}.
The 2011A data-taking period is subdivided into 318 runs, and the runs are subdivided into 109,428 luminosity blocks (LBs)~\cite{CMS:luminosity2011}.
A luminosity block is the smallest unit of data-taking for which there there is calibrated luminosity information, and during one block, the triggers are guaranteed to have consistent requirements and prescale factors (see \Sec{subsec:trigger_and_lumi} below).
Of the events in the \texttt{Jet} Primary Dataset, 26,275,768 are contained in ``valid" LBs which are certified by CMS for use in physics analyses~\cite{CMS:ValidatedRuns2011A}.

Each event in the AOD format has a complete list of PFCs, which are particle-like objects containing the reconstructed four-momentum and a probable particle identification (PID) code.
In addition, the AOD format has AK5 jets, which are clusters of PFCs identified by the anti-$k_t$ jet algorithm~\cite{Cacciari:2008gp} with radius parameter $R = 0.5$.
Jet energy correction (JEC) factors are obtained for the AK5 jets, including a correction for pileup using the area-median subtraction procedure~\cite{Cacciari:2008gn}.
The jets also have the information needed to impose jet quality criteria (JQC).

\subsection{MIT Open Data Framework}
\label{subsec:modformat}

Because of the technical challenges involved in using CMSSW, we only use it to extract information from the AOD files, performing the actual physics analyses outside of the virtual machine.
Building on the MOD software framework introduced in \Ref{Tripathee:2017ybi}, we use a custom \texttt{MODProducer} module in CMSSW to translate each AOD file into a plain text MOD file.
We then use a custom framework called \texttt{MODAnalyzer} to read in each MOD file and perform various jet analysis tasks using \textsc{FastJet} 3.3.1~\cite{Cacciari:2011ma}.
Finally, we convert the MOD files into HDF5~\cite{hdf5} files for universal usability.

As described in more detail in \Sec{subsec:selection}, we consider the hardest and second-hardest jets for our analysis, after correcting the jet $p_T$ using the JEC factors and imposing the ``medium'' JQC~\cite{CMS:2010xta,2011JInst...611002C}.
To access the constituents of jets, we first recluster the complete set of PFCs into AK5 jets and then compare against the CMS-provided preclustered AK5 objects.
Due to rare numerical rounding issues, there are cases where the AK5 objects disagree, and we discard jets whose transverse momenta differ from the CMS-provided jets by more than one part in $10^{6}$ or whose four-vectors are more than $10^{-6}$ apart in the rapidity-azimuth plane.
When the AK5 objects agree, we associate them in the HDF5 files.

A number of substantial improvements have been made to \texttt{MODProducer} compared to \Ref{Tripathee:2017ybi}.
We have added additional physics information in the MOD format, including metadata about files, LBs, and triggers.
We have added primary vertex information to implement CHS for pileup mitigation (see \Sec{subsec:jetconstituents} below), made possible because the AOD files have a \texttt{VertexCollection} handle that can assign a charged-particle track to the closest collision vertex.
We also added the ability to process MC files provided by CMS in the AODSIM format, including both generation-level particles and reconstructed PFCs.

After the jet selection stage in \texttt{MODAnalyzer}, the rest of our workflow is in \textsc{Python} 3.
We used \textsc{NumPy}~\cite{NumPy} for data manipulation, \textsc{Matplotlib}~\cite{Matplotlib} to produce figures, \textsc{Python Optimal Transport}~\cite{flamary2017pot} to calculate the EMD, and \textsc{EnergyFlow} 0.13~\cite{EnergyFlow} for a variety of jet analysis tasks.
In addition, we embedded our code in \textsc{Jupyter} notebooks~\cite{Jupyter} for enhanced transparency and portability.
To assist future jet studies on the CMS Open Data, our complete set of \textsc{Jupyter} notebooks is available~\cite{MODRepo}, and the corresponding reduced jet datasets are on the \textsc{Zenodo} platform~\cite{MOD:ZenodoCMS,MOD:ZenodoMC170,MOD:ZenodoMC300,MOD:ZenodoMC470,MOD:ZenodoMC600,MOD:ZenodoMC800,MOD:ZenodoMC1000,MOD:ZenodoMC1400,MOD:ZenodoMC1800}.

\subsection{Triggers, Prescales, and Luminosities}
\label{subsec:trigger_and_lumi}

\begin{table*}[t]
\begin{center}
\begin{tabular}{ l @{$\quad$}  r @{$\quad$} r @{$\quad$} r @{$\quad$} r @{$\quad$} r @{$\quad$} r @{$\quad$} }
\hline
\hline
Trigger Name & LBs & Events & Fired & $\mathcal{L}^\text{trig}_{\rm eff}$ [$\text{nb}^{-1}$] & $\langle p^\text{trig} \rangle$ & $\sigma^\text{trig}_{\rm eff}$ [$\text{nb}$] \\
\hline
\hline
\texttt{HLT\_Jet30} & 109,196 & 26,254,892 & 1,884,768 & 12.567 & 185,672.632 & 149,981.925 \\
\texttt{HLT\_Jet60} & 109,196 & 26,254,892 & 1,829,490 & 293.986 & 7,936.716 & 6,223.060 \\
\texttt{HLT\_Jet80} & 102,304 & 24,742,482 & 1,512,638 & 901.352 & 2,293.846 & 1,678.188 \\
\texttt{HLT\_Jet110} & 109,196 & 26,254,892 & 2,212,878 & 6,172.430 & 378.016 & 358.510 \\
\texttt{HLT\_Jet150} & 102,304 & 24,742,482 & 2,616,716 & 33,521.114 & 61.679 & 78.062 \\
\texttt{HLT\_Jet190} & 109,196 & 26,254,892 & 2,715,282 & 114,843.687 & 20.317 & 23.643 \\
\texttt{HLT\_Jet240} & 109,196 & 26,254,892 & 2,806,220 & 392,659.479 & 5.942 & 7.147 \\
\texttt{\textbf{HLT\_Jet300}} & \bf 98,462 & \bf 22,788,815 & \bf 4,616,184 & \bf 2,284,792.618 & \bf 1.000 & \bf 2.020 \\
\texttt{HLT\_Jet370} & 109,196 & 26,254,892 & 1,514,305 & 2,333,280.071 & 1.000 & 0.649 \\
\texttt{HLT\_Jet800} & 47,156 & 10,578,173 & 23,332 & 1,414,462.687 & 1.000 & 0.016 \\
\hline
\texttt{HLT\_DiJetAve30} & 98,462 & 22,788,815 & 1,394,369 & 20.585 & 110,990.490 & 67,735.556 \\
\texttt{HLT\_DiJetAve60} & 98,462 & 22,788,815 & 1,440,740 & 539.491 & 4,235.090 & 2,670.555 \\
\texttt{HLT\_DiJetAve80} & 91,570 & 21,276,405 & 1,059,885 & 1,474.722 & 1,369.123 & 718.702 \\
\texttt{HLT\_DiJetAve110} & 98,462 & 22,788,815 & 1,714,381 & 10,583.561 & 215.881 & 161.985 \\
\texttt{HLT\_DiJetAve150} & 91,570 & 21,276,405 & 2,162,760 & 59,292.115 & 34.053 & 36.476 \\
\texttt{HLT\_DiJetAve190} & 98,462 & 22,788,815 & 2,343,401 & 208,109.103 & 10.979 & 11.260 \\
\texttt{HLT\_DiJetAve240} & 98,462 & 22,788,815 & 2,697,899 & 800,844.351 & 2.853 & 3.369 \\
\texttt{HLT\_DiJetAve300} & 98,462 & 22,788,815 & 2,356,128 & 2,284,792.618 & 1.000 & 1.031 \\
\texttt{HLT\_DiJetAve370} & 98,462 & 22,788,815 & 741,410 & 2,284,792.618 & 1.000 & 0.324 \\
\hline
\texttt{HLT\_DiJetAve15U} & 10,734 & 3,466,077 & 225,367 & 1.841 & 26,335.253 & 122,404.801 \\
\texttt{HLT\_DiJetAve30U} & 10,734 & 3,466,077 & 353,409 & 45.628 & 1,062.680 & 7,745.523 \\
\texttt{HLT\_DiJetAve50U} & 10,734 & 3,466,077 & 339,051 & 298.084 & 162.664 & 1,137.434 \\
\texttt{HLT\_DiJetAve70U} & 10,734 & 3,466,077 & 624,758 & 2,061.075 & 23.525 & 303.122 \\
\texttt{HLT\_DiJetAve100U} & 10,734 & 3,466,077 & 301,727 & 4,314.114 & 11.239 & 69.940 \\
\texttt{HLT\_DiJetAve140U} & 10,734 & 3,466,077 & 415,806 & 25,144.074 & 1.928 & 16.537 \\
\texttt{HLT\_DiJetAve180U} & 10,734 & 3,466,077 & 255,163 & 48,487.453 & 1.000 & 5.262 \\
\texttt{HLT\_DiJetAve300U} & 10,734 & 3,466,077 & 21,347 & 48,487.453 & 1.000 & 0.440 \\
\hline
\texttt{HLT\_Jet240\_CentralJet30\_BTagIP} & 47,156 & 10,578,173 & 2,216,488 & 1,414,462.687 & 1.000 & 1.567 \\
\texttt{HLT\_Jet270\_CentralJet30\_BTagIP} & 47,156 & 10,578,173 & 1,280,355 & 1,414,462.687 & 1.000 & 0.905 \\
\texttt{HLT\_Jet370\_NoJetID} & 109,196 & 26,254,892 & 1,711,067 & 2,333,280.071 & 1.000 & 0.733 \\
\hline
\hline
Missing & 89 & &  & 6.066 &  &  \\
Zeroed & 143 & 20,876 &  & &  &  \\
Total & 109,428 & 26,275,768 & 26,275,768 & 2,333,286.137 &  &  \\
\hline
\hline
\end{tabular}
\caption{Triggers in the CMS 2011A \texttt{Jet} Primary Dataset~\cite{CMS:JetPrimary2011A}, restricted to LBs that have been identified as valid for physics analyses by CMS~\cite{CMS:ValidatedRuns2011A} and that have non-zero recorded luminosity~\cite{CMS:luminosity2011}.
Shown are the number of valid LBs and events for which the trigger is present and the number of valid events for which the trigger fired.
Also provided are the effective luminosity $\mathcal{L}^\text{trig}_{\rm eff}$ defined in \Eq{eq:efflumidef}, and the average prescale value $\langle p^\text{trig} \rangle$ and effective cross section $\sigma^\text{trig}_{\rm eff}$ defined in \Eq{eq:averageprescale_sigmaeff}.
As discussed in \App{app:missinglumi}, there are 89 ``missing'' LBs in the CMS 2011A luminosity table~\cite{CMS:luminosity2011} that are not represented in the \texttt{Jet} Primary Dataset, but they have a negligible impact on our analysis.
We also omit 143 ``zeroed'' LBs during which events were detected but zero luminosity was recorded.
The \texttt{HLT\_Jet300} trigger (bolded) is the one used for the jet studies in \Secs{sec:jetsubstructure}{sec:emd}.
}
\label{table:full_list_of_triggers}
\end{center}
\end{table*}

\begin{figure*}[t]
  \centering
  \subfloat[]{
  \includegraphics[height=0.485\textwidth]{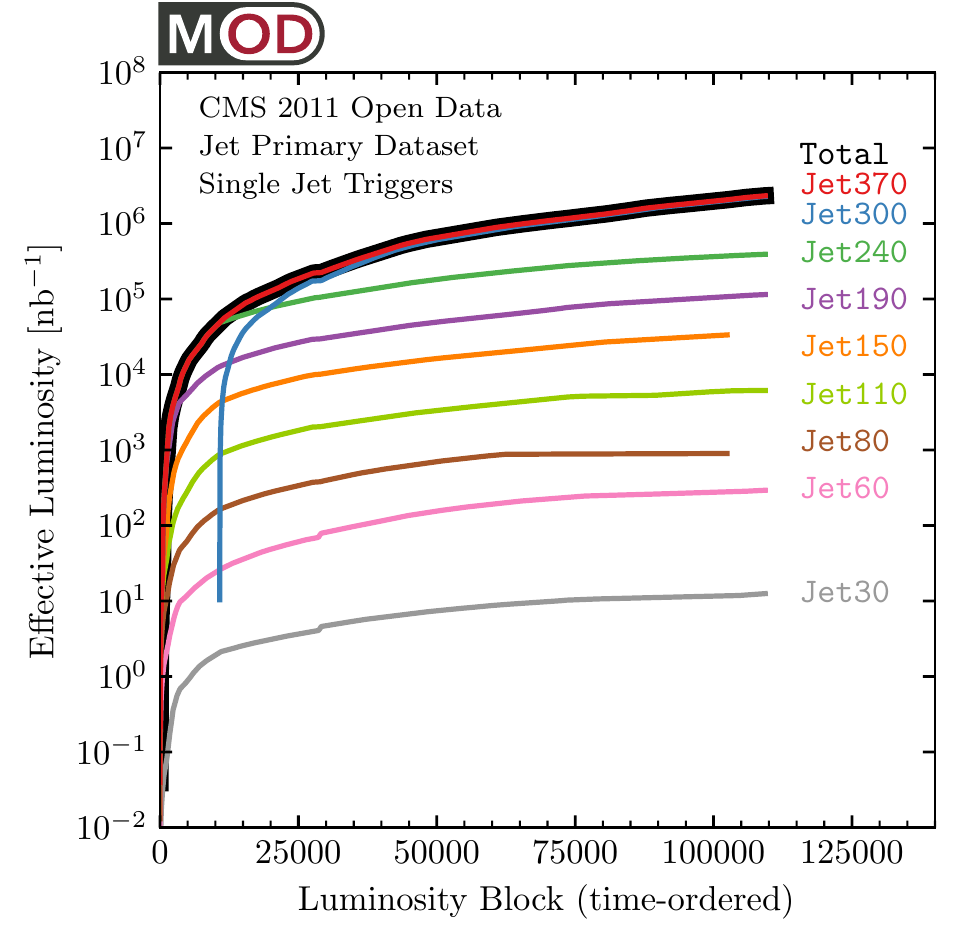}
  \label{figures:integrated_effective_luminosity}
  }
  \subfloat[]{
  \includegraphics[height=0.485\textwidth]{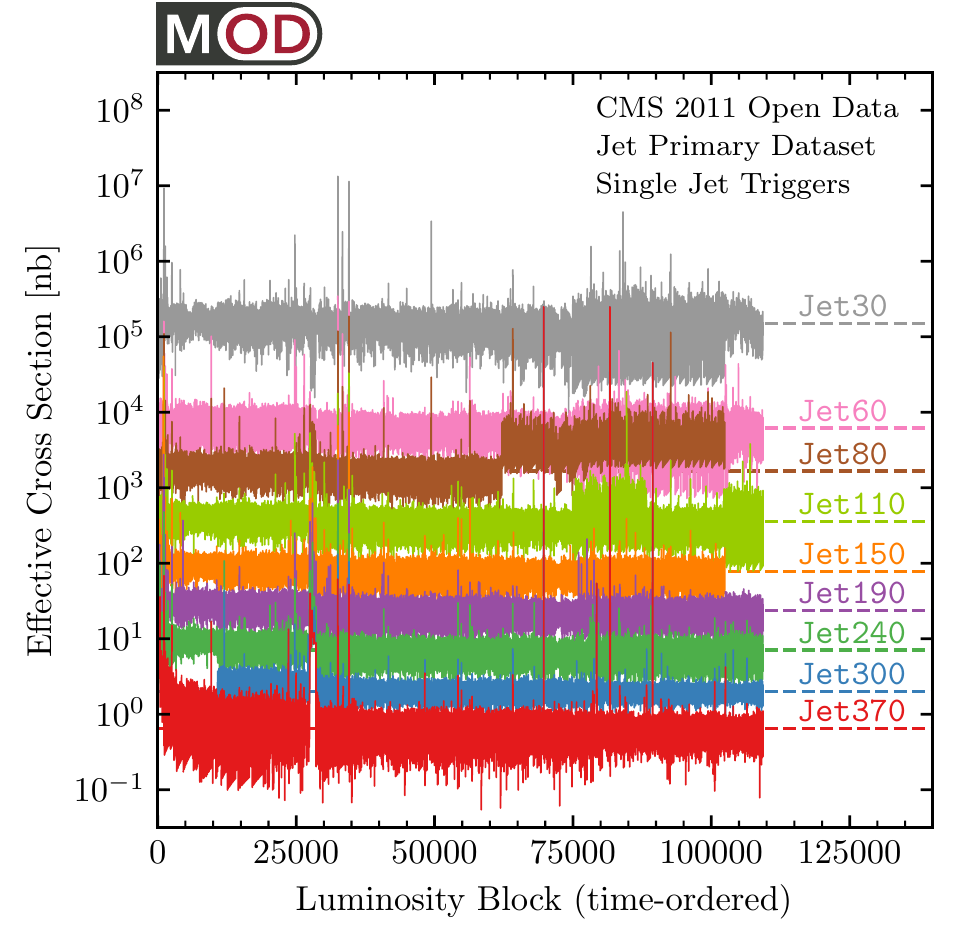}
  \label{figures:integrated_effective_xsec}
  }
  \caption{%
  (a) Effective luminosity for the single-jet triggers as a function of the cumulative number of LBs, ordered in time.
  Note that the \texttt{Jet300} trigger used for our jet studies turns on after around \SI{50}{pb^{-1}} has already been collected, but this is a relatively small fraction of the total \SI{2.3}{fb^{-1}} collected over the course of Run 2011A.
  The luminosity profile as a function of date is shown in \Fig{figure:del_rec_300} of \App{app:missinglumi}.
  (b) Effective cross section for the single-jet triggers in each LB where the trigger fired.
  The flatness of these curves indicates that the trigger behavior is roughly constant across the entire run, apart from moments where the trigger criteria or prescale factors changed.
   The horizontal dashed lines correspond to the total effective cross section for that trigger from \Tab{table:full_list_of_triggers}.
}
     \label{figures:integrated_effective_luminosity_and_xsec}
\end{figure*}

The \texttt{Jet} Primary Dataset contains 30 triggers~\cite{Khachatryan:2016bia}.
We summarize these triggers in \Tab{table:full_list_of_triggers}, indicating the number of valid LBs and events for which the trigger is present, as well as the number of valid events for which the trigger fired.
There are single jet and dijet triggers, where the trigger names include the nominal $p_T$ requirement for the jet(s).
For simplicity, we do not distinguish between trigger versions, denoted by suffixes like \texttt{\_v2}, in our analysis.
(The documentation for \Ref{CMS:JetPrimary2011A} lists 5 \texttt{L1FastJet} trigger variants in the \texttt{Jet} Primary Dataset, but as far as we can tell, these triggers were introduced after Run 2011A was complete.)

There are 7 triggers that were operational during the entire 2011A run, corresponding to 109,339 LBs.
This can be compared to the luminosity information in \Ref{CMS:luminosity2011}, which lists 109,428 valid LBs in this run, leaving 89 LBs unaccounted for in the \texttt{Jet} Primary Dataset.
These ``missing'' LBs only contribute \SI{6}{{nb}^{-1}} to the recorded integrated luminosity, so their absence has a negligible impact on our studies.
We investigate the missing LBs in more detail in \App{app:missinglumi}.
There are also 643 LBs that are on the list of validated runs~\cite{CMS:ValidatedRuns2011A} but absent from the luminosity table~\cite{CMS:luminosity2011}; we omit these from our analysis under the assumption that they are not in fact valid runs.
Finally, we omit 143 valid LBs that contain events but have zero recorded luminosity, and we investigate these ``zeroed'' LBs further in \App{app:missinglumi}.

Because the total data-taking rate is limited, the lower $p_T$ jet triggers are prescaled to only fire a fraction of the time they are active.
The prescale factors satisfy $p^\text{trig} \geq 1$, with $p^\text{trig} = 1$ indicating an unprescaled trigger.
(Strictly speaking, there are separate and independent prescale factors for the Level 1 (L1) trigger and the high-level trigger (HLT), but we always use $p^\text{trig}$ to refer to the product of these factors.)
The trigger prescale factors are fixed within a LB but can change between LBs.
The effective luminosity for a given trigger is:
\begin{equation}
\label{eq:efflumidef}
\mathcal{L}_{\rm eff}^\text{trig} = \sum_{b \in \text{LBs}} \frac{\mathcal{L}_b}{p_b^\text{trig}},
\end{equation}
where $b$ labels a LB, $\mathcal{L}_b$ is the recorded integrated luminosity in that block, and $p_b^\text{trig}$ is the associated prescale factor.
The effective luminosities for the \texttt{Jet} Primary Dataset triggers are reported in \Tab{table:full_list_of_triggers}, along with their average prescale factors and effective cross sections:
\begin{equation}
\label{eq:averageprescale_sigmaeff}
\langle p^\text{trig} \rangle = \frac{\mathcal{L}^\text{trig}_{\rm total}}{\mathcal{L}_{\rm eff}^\text{trig}}, \qquad \sigma^\text{trig}_{\rm eff} = \frac{N^\text{trig}}{\mathcal{L}_{\rm eff}^\text{trig}},
\end{equation}
where $\mathcal{L}^\text{trig}_{\rm total} = \sum_b \mathcal{L}_b$ is the total luminosity of the run while the trigger was present, and $N^\text{trig}$ is the total number of events for which the trigger fired.

Our analysis is based on the substructure of individual jets, so we focus our attention on the 9 single-jet triggers in \Tab{table:full_list_of_triggers}, omitting \texttt{HLT\_Jet800} since it contains relatively few events.
Their effective luminosities as a function of the number of cumulative time-ordered LBs are plotted in \Fig{figures:integrated_effective_luminosity}.
We see that as the integrated luminosity increases, some of jet triggers have to be prescaled.
We also see that the \texttt{HLT\_Jet300} trigger only starts acquiring data partway through the 2011A run, coinciding with the \texttt{HLT\_Jet240} trigger being prescaled.

In \Fig{figures:integrated_effective_xsec}, we plot the effective cross section in each time-ordered LB for the 9 single-jet triggers.
The trigger behaviors are relatively stable over the course of the 2011A run, though there is a noticeable shift in the \texttt{HLT\_Jet80} trigger when its selection criteria changed.
One can also see when the \texttt{HLT\_Jet300} trigger turned on and when the \texttt{HLT\_Jet80} and \texttt{HLT\_Jet150} triggers were turned off.

Since \texttt{HLT\_Jet300} is the lowest $p_T$ single-jet trigger that is unprescaled, it will be the sole trigger used in our substructure and EMD studies (see further discussion in \Sec{subsec:selection}).
For reference, the recorded luminosity for \texttt{HLT\_Jet300} as a function of time is plotted in \Fig{figure:del_rec_300} of \App{app:missinglumi}.

\subsection{Monte Carlo Event Samples}
\label{subsec:simulation}

\begin{table}
\begin{center}
\begin{tabular}{ r @{ -- } l c @{$\quad$} r @{$\quad$} l @{$\quad$}  c}
\hline
\hline
$\hat{p}_T^{\rm min}$ & $\hat{p}_T^{\rm max}$  [GeV] & Files &  Events  & $\sigma_{\rm eff}^\text{MC}$ [$\text{nb}$] & DOI \\
\hline\hline
0 & 5 & 55 & 1,000,025 & $4.84\times 10^7$& \cite{CMS:QCDsim0-5}\\
5 & 15 & 83 & 1,495,884 & $3.68\times 10^7$& \cite{CMS:QCDsim5-15}\\
15 & 30 & 5,519 & 9,978,850 & $8.16\times 10^5$& \cite{CMS:QCDsim15-30}\\
30 & 50 & 277 & 5,837,856 & $5.31\times 10^4$& \cite{CMS:QCDsim30-50}\\
50 & 80 & 299 & 5,766,430 & $6.36\times 10^3$& \cite{CMS:QCDsim50-80}\\
80 & 120 & 317 & 5,867,864 & $7.84\times 10^2$& \cite{CMS:QCDsim80-120}\\
120 & 170 & 334 & 5,963,264 & $1.15\times 10^2$  & \cite{CMS:QCDsim120-170}\\
\hline
170 & 300 &  387 & 5,975,592 & $2.43\times 10^{1}$ & \cite{CMS:QCDsim170-300}\\
300 & 470 & 382 & 5,975,016 & $1.17\times 10^{0}$ & \cite{CMS:QCDsim300-470}\\
470 & 600 & 274 & 3,967,154 & $7.02\times 10^{-2}$ & \cite{CMS:QCDsim470-600}\\
600 & 800 & 271 & 3,988,701 & $1.56\times 10^{-2}$ & \cite{CMS:QCDsim600-800}\\
800 & 1000 &  295 & 3,945,269 & $1.84\times 10^{-3}$ & \cite{CMS:QCDsim800-1000}\\
1000 & 1400 & 131 & 1,956,893 & $3.32\times 10^{-4}$ & \cite{CMS:QCDsim1000-1400}\\
1400 & 1800 &  182 & 1,991,792 & $1.09\times 10^{-5}$ & \cite{CMS:QCDsim1400-1800}\\
1800 & $\infty$ & 75 & 996,500 & $3.58\times 10^{-7}$ & \cite{CMS:QCDsim1800}\\
\hline
\hline
\end{tabular}
\caption{
Information about the MC event samples provided by CMS~\cite{CMS:QCDsim0-5,CMS:QCDsim5-15,CMS:QCDsim15-30,CMS:QCDsim30-50,CMS:QCDsim50-80,CMS:QCDsim80-120,CMS:QCDsim120-170,CMS:QCDsim170-300,CMS:QCDsim300-470,CMS:QCDsim470-600,CMS:QCDsim600-800,CMS:QCDsim800-1000,CMS:QCDsim1000-1400,CMS:QCDsim1400-1800,CMS:QCDsim1800} from the \textsc{Pythia} 6 hard QCD scattering process.
Shown are the generator-level $\hat{p}_T$ ranges, the number of files and events in each sample, and the effective cross section $\sigma^\text{MC}_{\rm eff}$.
Only the 8 samples with $\hat{p}_T > \SI{170}{GeV}$ are required for the jet studies in \Secs{sec:jetsubstructure}{sec:emd}.
}
\label{table:qcd_datasets_cross_section_info}
\end{center}
\end{table}

A key feature of the CMS 2011 data release compared to the initial one from 2010 is the inclusion of MC event samples.
(Some MC samples corresponding to the 2010 dataset have been subsequently released.)
For our analysis, we use samples of hard QCD scattering generated by \textsc{Pythia} 6.4.25~\cite{Sjostrand:2006za} with tune Z2~\cite{Field:2011iq}.
As summarized in \Tab{table:qcd_datasets_cross_section_info}, there are 15 samples with non-overlapping hard-scattering parton $\hat{p}_T$ ranges~\cite{CMS:QCDsim0-5,CMS:QCDsim5-15,CMS:QCDsim15-30,CMS:QCDsim30-50,CMS:QCDsim50-80,CMS:QCDsim80-120,CMS:QCDsim120-170,CMS:QCDsim170-300,CMS:QCDsim300-470,CMS:QCDsim470-600,CMS:QCDsim600-800,CMS:QCDsim800-1000,CMS:QCDsim1000-1400,CMS:QCDsim1400-1800,CMS:QCDsim1800}, totaling \SI{13.4}{TB}.
They are labeled by CMS as \texttt{QCD\_Pt-MINtoMAX\_TuneZ2\_7TeV\_pythia6}, where $\hat{p}_T \in [\text{MIN},\text{MAX}]\,\text{GeV}$.
These events are then simulated and reconstructed using the CMS detector simulation based on \textsc{Geant 4}~\cite{Agostinelli:2002hh}.
Throughout this paper, we use ``generation'' to refer to the output of the parton shower generator, and ``simulation'' to refer to the output of the detector simulation.

Both the generation-level and simulation-level objects are stored in AODSIM format by CMS, and we convert them to our MOD format using \texttt{MODProducer}.
Apart from the generation-level event record from \textsc{Pythia}, the AODSIM format is very similar to AOD.
In particular, AODSIM includes reconstructed AK5 jets, simulated trigger information, as well as the addition of pileup.
We store the simulated PFCs, the final-state particles in the \textsc{Pythia} event record, and the $2 \to 2$ hard-scattering process for anticipated future studies related to parton flavor.
If an association between simulation-level and generation-level jets is needed, jets are matched if their jet axes are within $\Delta R=0.5$ of each other.
To enable future jet flavor studies, generation-level jets are also matched to hard-process partons if they are less than $\Delta R=1.0$ apart.

Because of the steep dependence of the QCD dijet cross section on $\hat{p}_T$, the MC events have different weights, though the weights for all events in a single MC sample are the same.
Therefore, when filling histograms, we have to weight each MC event by the generated cross section $\sigma_\text{eff}^\text{MC}$ divided by the number of events in the MC sample, as given in \Tab{table:qcd_datasets_cross_section_info}.
As discussed in \App{app:pileupreweighting}, we also weight the MC events according to the number of primary vertices in order to match the distribution of pileup seen in the data.

One subtlety in using the generation-level \textsc{Pythia} information is that there is a cutoff on the hadron lifetime above which they are considered stable.
This cutoff is set to $c \, \tau_{\rm stable} = \SI{10}{mm}$, which means that various hadrons with non-zero strangeness are considered stable, notably the $K_S^0$ meson.
Typically, these strange hadrons decay within the CMS detector volume and are often reconstructed as if the decay products came from the primary vertex.
For example, $K_S^0 \to \pi^+ \pi^-$ will typically be reconstructed as two pion-labeled PFCs.
This leads to a mismatch in observables like track multiplicity unless we manually decay these strange hadrons.
As a workaround, we load the generation-level event record into \textsc{Pythia} 8.235~\cite{Sjostrand:2014zea} and adjust the hadron lifetime threshold to $c \, \tau_{\rm stable} = \SI{1000}{mm}$.
Because the kinematics and flavors of the hadron decay will not be the same as in the CMS detector simulation, there is a slight mismatch when comparing a generation-level event to its simulation-level counterpart, though this issue does not arise when comparing histograms.

\subsection{Jet and Trigger Selection}
\label{subsec:selection}

\begin{figure*}[p]
  \centering
  \subfloat[]{
	\includegraphics[height=0.46\textwidth]{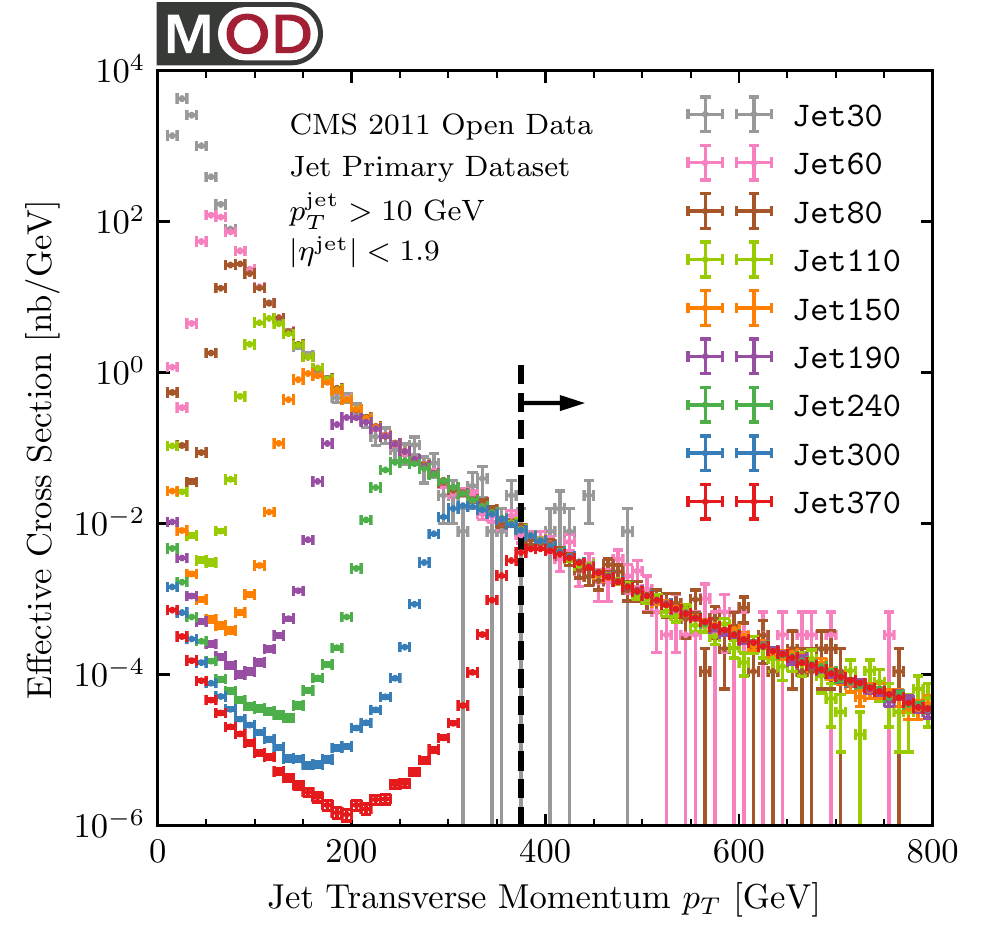}
	\label{fig:data_raw_pt_spectrum}
	}
  \subfloat[]{
  \includegraphics[height=0.46\textwidth]{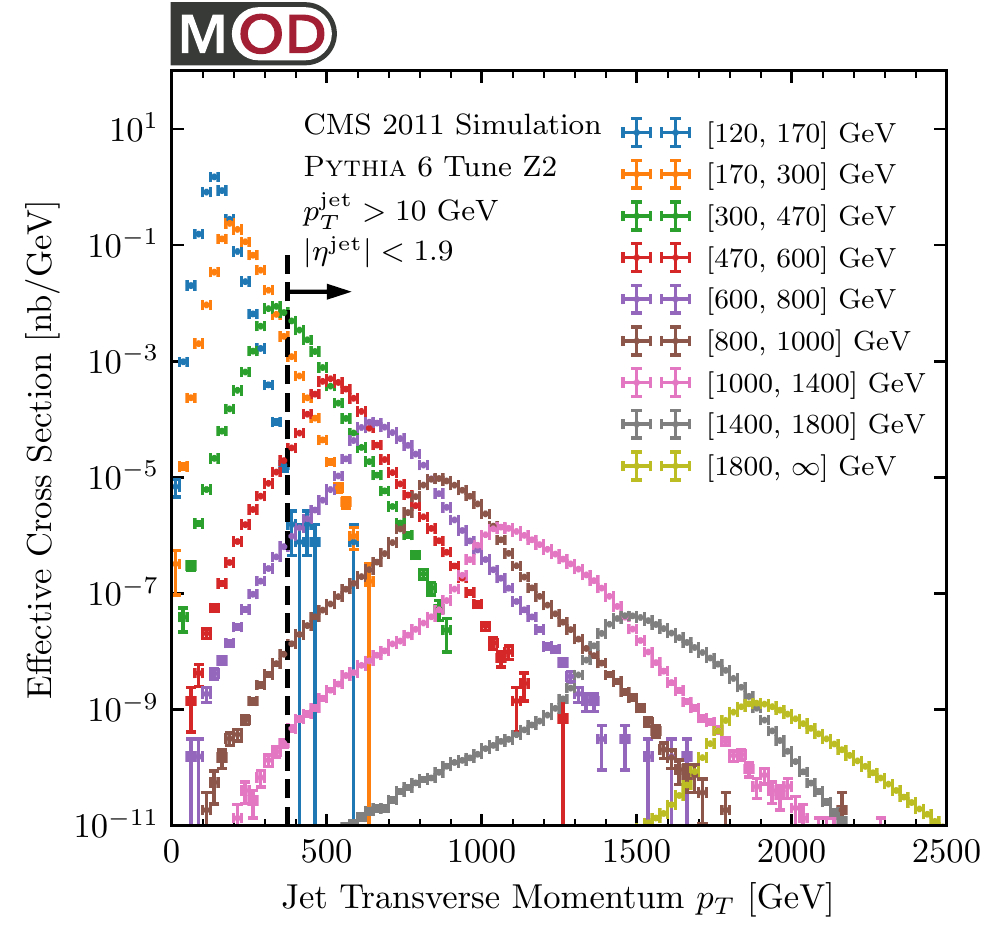}
  \label{fig:sim_raw_pt_spectrum}
  }
  \caption{
  The $p_T$ spectrum for the hardest jet in (a) the 9 single-jet triggers and (b) the 9 relevant simulated MC samples, restricted to $|\eta^{\rm jet}| < 1.9$.
  These jet spectra have JEC factors included and medium JQC imposed.
  The vertical dashed lines at \SI{375}{GeV} indicate the jet $p_T$ threshold used in this analysis.
 }
     \label{figures:raw_pt_spectrum}
\end{figure*}

\begin{figure*}[p]
\centering
  \subfloat[]{
	\includegraphics[width=0.49\textwidth]{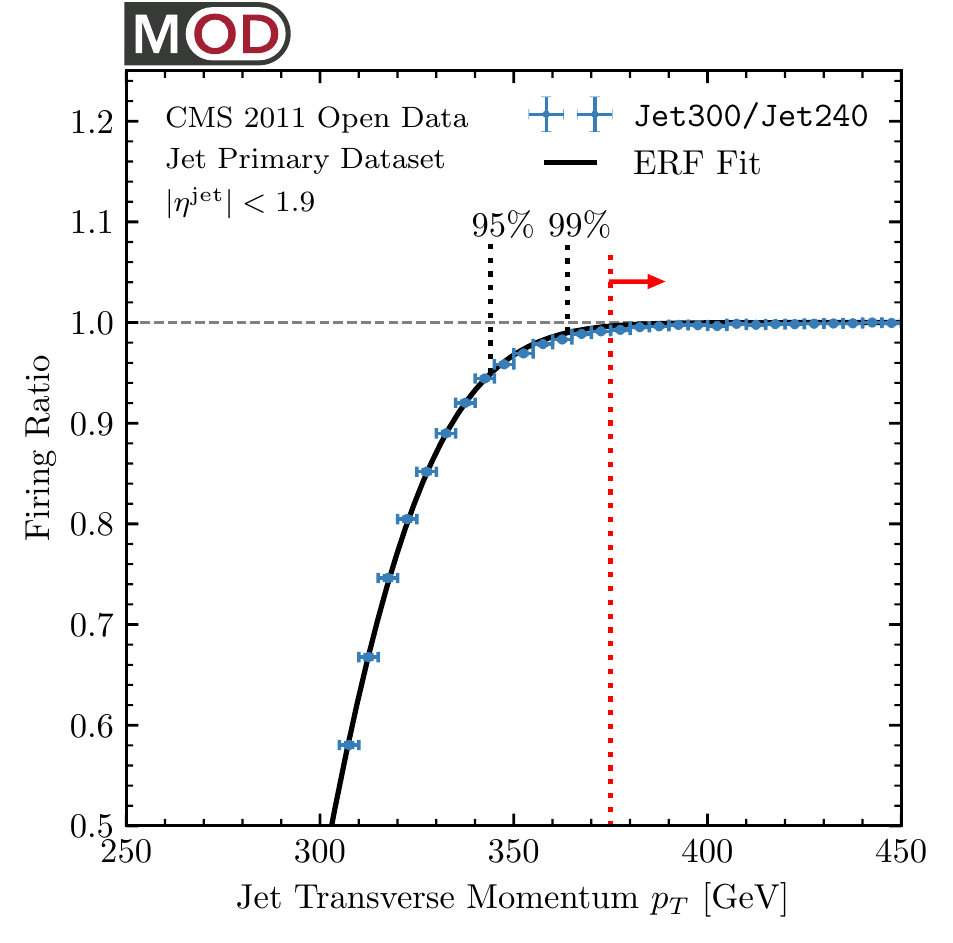}
	\label{fig:data_trigger_ratio}
	}
  \subfloat[]{
  \includegraphics[width=0.49\textwidth]{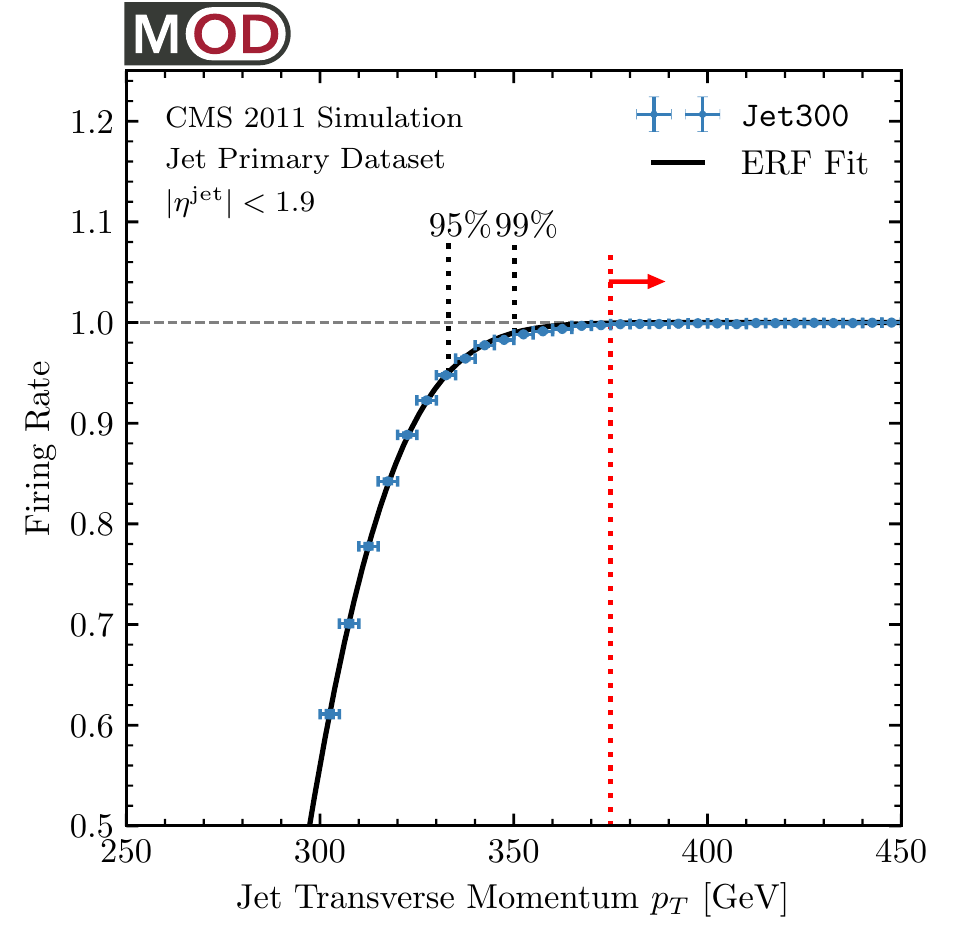}
  \label{fig:sim_trigger_efficiency}
  }
\caption{
Trigger turn on behavior as a function of reconstructed hardest jet $p_T$ for the \texttt{Jet300} trigger, including JEC factors.
Shown are (a) the relative efficiency of the \texttt{Jet300} trigger with respect to \texttt{Jet240} in the CMS Open Data, and (b) the absolute efficiency of the \texttt{Jet300} trigger in the MC simulation.
Both of these curves are fit to an error function (ERF) to estimate the efficiency boundaries.
From these, we conclude that the \texttt{Jet300} trigger is fully efficient above $p^{\rm jet}_T > \SI{375}{GeV}$.
This analysis is repeated for the other triggers in \Fig{fig:trigger_turn_on_all_triggers} of \App{app:additionalplots}.
}
\label{fig:trigger_turn_on_sim}
\end{figure*}

The jet studies in \Secs{sec:jetsubstructure}{sec:emd} are based on the two hardest $p_T$ jets in an event.
This is motivated by the fact that $2 \to 2$ QCD dijet production at leading order yields two jets of equal $p_T$.
Therefore, considering the substructure of just the hardest $p_T$ jet (as in the studies of \Refs{Larkoski:2017bvj,Tripathee:2017ybi}) is IRC unsafe, since an infinitesimally soft emission can change the relative jet ordering.
On the other hand, considering more than two jets requires information beyond leading order, so we only consider the two hardest $p_T$ jets in our analysis.
(See \Ref{Cacciari:2019qjx} for further discussions of single-jet inclusive cross section definitions.)

The CMS single-jet triggers are designed to fire any time an event has a jet whose $p_T$ is above a given threshold.
We independently analyze the two hardest jets in an event, correcting their $p_T$ values by the appropriate JEC factors.
When we perform our substructure analysis, we require that the jets satisfy $|\eta^{\rm jet}| < 1.9$ to make sure that the $R = 0.5$ jets are reconstructed fully within the tracking volume that covers $|\eta^{\rm tracker}| < 2.4$.
We impose ``medium'' JQC (see \Tab{table:jqc})~\cite{CMS:2010xta,2011JInst...611002C} throughout this study.

\begin{figure*}[t]
  \subfloat[]{
	\includegraphics[height=0.44\textwidth]{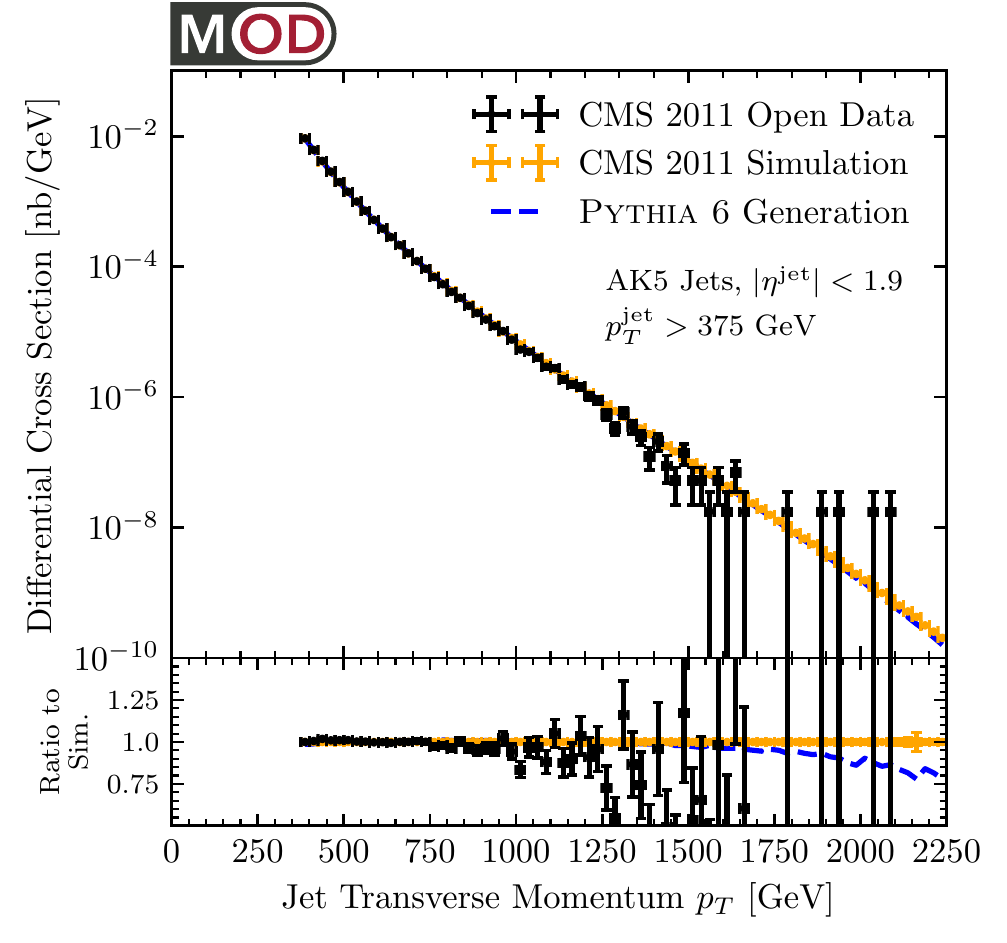}
	\label{fig:pTspectrum}
	}
\subfloat[]{
	\includegraphics[height=0.44\textwidth]{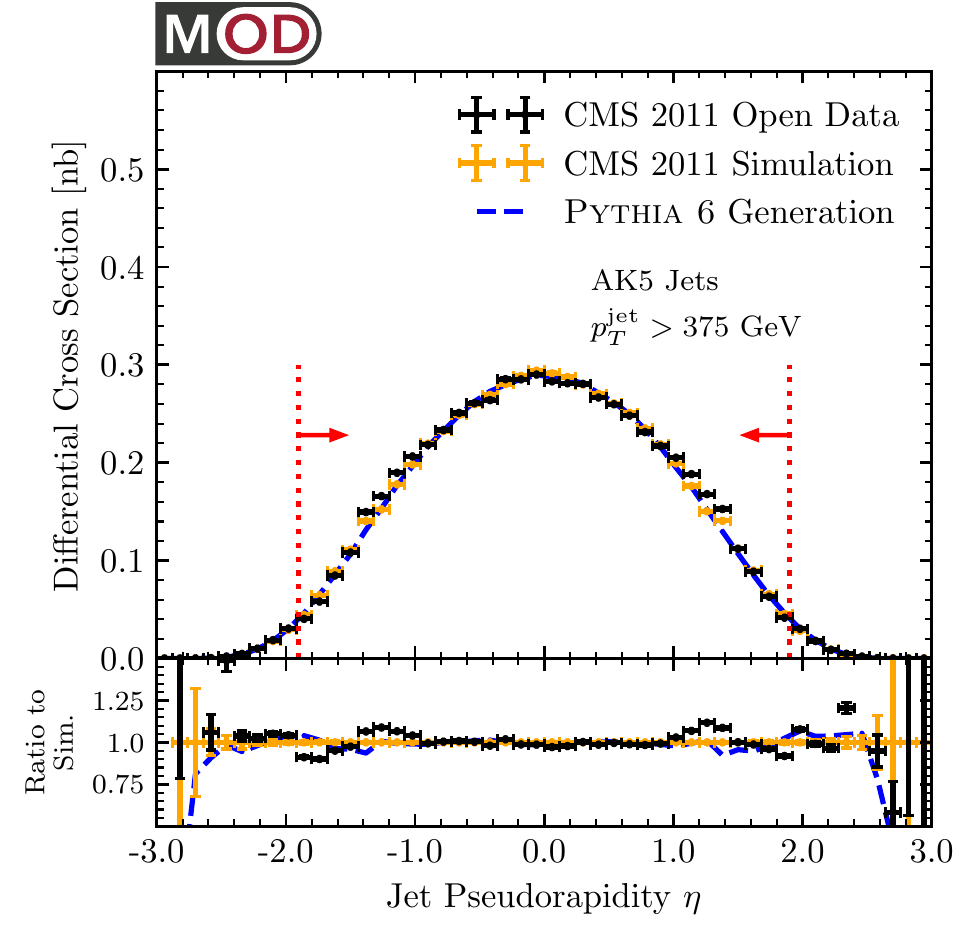}
	\label{fig:etaspectrum}
	}
  \caption{
  (a) Jet transverse momentum spectrum, comparing the CMS Open Data to MC event samples at the simulation level and generation level.
  We consider up to two of the hardest $p_T$ jets, restricted to $|\eta^\text{jet}| < 1.9$ and $p_T^\text{jet} > \SI{375}{GeV}$.
  In addition to having a $p_T$-dependent NLO $K$-factor, the MC events have been normalized to match the lowest $p_T$ bin.
  (b) Jet pseudorapidity spectrum, with the $|\eta^\text{jet}|$ requirement removed.
  For both jet spectra, we see very good agreement between data and simulation, indicating that we have properly processed the CMS Open Data, including appropriate JEC factors.
In these and all subsequent plots, the error bars indicate statistical uncertainties only, with no attempt at estimating systematic uncertainties.
The jet azimuth spectrum is shown in \Fig{fig:phispectrum} of \App{app:additionalplots}.
}
  \label{fig:overalljetkinematics}
\end{figure*}

In \Fig{fig:data_raw_pt_spectrum}, we show the $p_T$ spectrum of just the hardest jet in the CMS 2011 Open Data, separated into the 9 single-jet triggers.
(The spectrum for the two hardest jets will be shown in \Fig{fig:pTspectrum}.)
We see that the triggers start to collect an appreciable number of jets when the jet $p_T$ matches the trigger name, asymptoting to a common smooth $p_T$ spectrum. 
The small population of jets at low $p_T$ values below the turn on is due primarily to trigger misfirings, for example from fake jets that do not satisfy the jet quality criteria.
In \Fig{fig:sim_raw_pt_spectrum}, we show the same $p_T$ spectrum in the CMS simulation, separated into the 9 most relevant MC samples for our analysis (out of 15 total).
We see that the MC files have support mainly in their designated $\hat{p}_T$ ranges, albeit with a spread due to phenomena like initial state radiation (ISR) that change the overall event kinematics.

To simplify our physics studies, we use just one of the single-jet triggers.
As mentioned above, we select \texttt{HLT\_Jet300} since this has the lowest $p_T$ threshold among the unprescaled single-jet triggers.
Looking at \Fig{fig:data_raw_pt_spectrum}, we can estimate that \texttt{Jet300} is fully efficient above $p_T > \SI{375}{GeV}$.
Looking at \Fig{fig:sim_raw_pt_spectrum}, we see that all of the MC samples with $\hat{p}_T > \SI{170}{GeV}$ contribute appreciably to the $p_T > \SI{375}{GeV}$ region, corresponding to 8 required MC event samples.

To determine where the \texttt{Jet300} trigger is fully efficient, we compare its behavior to the \texttt{Jet240} trigger; see related trigger efficiency studies in \Refs{Khachatryan:2016bia,ATLAS:2016qun}.
In \Fig{fig:data_trigger_ratio}, we consider events where the \texttt{Jet300} trigger is present and the \texttt{Jet240} trigger fired.
We then plot the fraction of events where \texttt{Jet300} fired as a function of jet $p_T$.
Fitting the resulting fraction to an error function, we estimate that the \texttt{Jet300} trigger is 99\% efficient (relative to \texttt{Jet240}) at \SI{367}{GeV}, justifying our choice of $p_T > \SI{375}{GeV}$.
We can cross check our trigger efficiency study using the simulated MC samples.
In \Fig{fig:sim_trigger_efficiency}, we plot the fraction of events where the simulated  \texttt{Jet300} trigger fired as a function of jet $p_T$.
Doing the same error function fit, we find that the simulated  \texttt{Jet300} trigger is 99\% efficient (relative to an absolute scale) at \SI{350}{GeV}, which is again consistent with our $p_T > \SI{375}{GeV}$ choice.
For completeness, we provide efficiency plots for all of the triggers in \Fig{fig:trigger_turn_on_all_triggers} of \App{app:additionalplots}.
Since we are performing an exploratory jet study, we do not correct for this small trigger inefficiency in our analysis.

\begin{table}
\begin{center}
\begin{tabular}{ r c c c }
\hline
\hline
& Loose & Medium & Tight\\
\hline
\hline
Neutral Hadron Fraction & $<0.99$ & $<0.95$ & $<0.90$ \\
Neutral Electromagnetic Fraction & $<0.99$ & $<0.95$ & $<0.90$ \\
Number of Constituents & $>1$ & $>1$ & $>1$ \\ \hline
Charged Hadron Fraction & $>0.00$ & $>0.00$ & $>0.00$ \\
Charged Electromagnetic Fraction & $<0.99$ & $<0.99$ & $<0.99$ \\
Number of Charged Constituents & $>0$ & $>0$ & $>0$ \\
\hline
\hline
\end{tabular}
\caption{The jet quality criteria based on CMS recommendations for $|\eta^\text{jet}|<2.4$. For $|\eta^\text{jet}|>2.4$, where tracking information is not available, the charged particle criteria are not applied and all particles are treated as neutral.
For our analysis, we impose the ``medium'' criteria.
}
\label{table:jqc}
\end{center}
\end{table}

\begin{table*}[p]
\begin{center}
\begin{tabular}{r @{$\quad$}  r   @{$\quad$} r @{$\quad$} r}
\hline
\hline
& CMS 2011 Open Data & CMS 2011 Simulation& \textsc{Pythia} 6 Generation\\
\hline
\hline
Total Events & 30,726,331  & 28,796,917 & 21,802,470 \\
Valid & 26,254,892  &  &   \\
\texttt{Jet300} Trigger Present & 22,788,815 &  &  \\
\texttt{Jet300} Trigger Fired & 4,616,184 & 22,108,599 & \\
\hline
Two Hardest Jets, $p_T^\text{jet}>\SI{10}{GeV}$ &9,106,775 & 44,217,198 & 43,604,940 \\
 $p_T^\text{jet} > \SI{375}{GeV}$  & 1,785,625 & 35,155,818 & 35,267,080 \\
AK5 Match  & 1,785,625 &  35,155,790 & \\
\hline
Medium JQC  & 1,731,255 & 35,145,175 & \\
 $|\eta^\text{jet}| < 1.9$  & 1,690,984 & 34,969,900 & 35,089,120  \\
 $p_T^\text{jet} \in [375,425]\,\text{GeV}$  & 879,046 & 2,379,525& 2,203,305  \\
\hline
\hline
\end{tabular}
\caption{
Initial workflow and event selection for the jet studies in \Secs{sec:jetsubstructure}{sec:emd}.
The selections in the first block ensure that the \texttt{Jet300} trigger fired in a valid LB, the requirements in the second block ensure that the \texttt{Jet300} trigger is fully efficient, and the cuts in the third block impose the JQC and the baseline analysis criteria.
Because our analysis is based on the two hardest jets, there is an increase by a factor of about two between the first and second blocks.
}
\label{tab:workflow}
\end{center}
\end{table*}

\begin{figure*}[p]
\centering
\includegraphics[width=2\columnwidth]{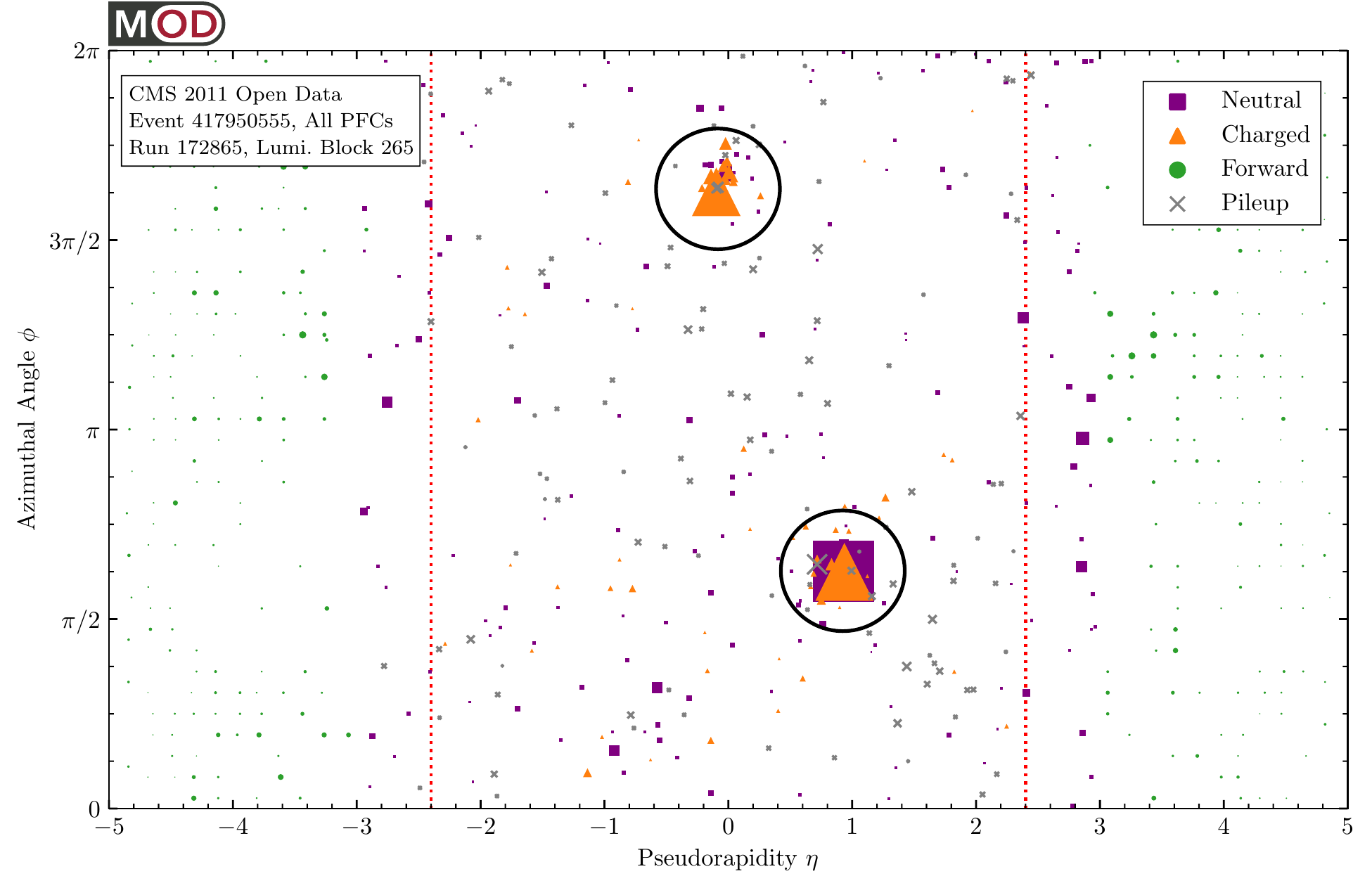}
\caption{\label{fig:eventdisplay}
Reconstructed PFCs in an example event from the CMS Open Data passing our jet selection criteria.
The size of the symbol indicates the PFC transverse momentum and the style indicates its charge, with purple squares for neutral PFCs, orange triangles for charged PFCs, and green circles for PFCs in the forward region where no charge information is available.
Charged pileup PFCs removed by CHS are indicated as gray crosses.
The leading two jets are shown as circles of radius $R = 0.5$, and the tracking region $|\eta|<2.4$ is within the dashed, vertical lines.
}
\end{figure*}

Our initial workflow is summarized in \Tab{tab:workflow}.
Because we consider the two hardest jets with $p_T^\text{jet}>10$~\text{GeV}, there are about twice as many jets in the analysis as the number of events.
In order to have a more homogenous jet sample, we impose the narrower $p_T^\text{jet} \in [375,425]~\text{GeV}$ range for our substructure and EMD studies below.
An example event from the CMS 2011 Open Data passing our kinematic jet selections is displayed in \Fig{fig:eventdisplay}, including information about the charges and vertices of the PFCs.

\section{Analyzing Jet Substructure}
\label{sec:jetsubstructure}

To validate the performance of the CMS detector for jet reconstruction, we present a variety of jet kinematics and jet substructure distributions derived from the CMS 2011 Open Data.
There are two main differences compared to a similar analysis performed in \Ref{Tripathee:2017ybi}.
First, we can now compare the open data distributions to detector-simulated MC samples to check for robustness.
Second, we have proper luminosity information~\cite{CMS:luminosity2011} such that we can plot (uncorrected) differential cross sections, instead of just normalized probability distributions.

\subsection{Overall Jet Kinematics}
\label{subsec:overalljet}

In \Fig{fig:pTspectrum}, we show the $p_T$ spectrum of the two hardest jets  (i.e.~two histogram entries per event), restricted to the region $|\eta^\text{jet}| < 1.9$ and $p_T^\text{jet} > \SI{375}{GeV}$.
Here, we compare the CMS Open Data in black to the simulated MC samples in orange.
We find very good agreement in the shape of the $p_T$ spectrum after including appropriate $K$-factors described below, though there are small disagreements and discontinuities for $p_T^\text{jet} > \SI{750}{GeV}$.
We also show the generation-level \textsc{Pythia} distribution without detector simulation in blue, which matches very well to the orange simulation-level distribution with detector response, indicating that the overall JEC factors have been chosen appropriately.
(Of course, the JEC factors also include data-driven corrections beyond just those captured by the detector simulation.)
Note that these distributions only include statistical uncertainties, without any estimate of systematic uncertainties.

Because \textsc{Pythia} is a leading-order generator, we have rescaled the MC events by a next-to-leading-order (NLO) $K$-factor.
This $p_T$-dependent $K$-factor is derived from \Ref{Kumar:2013hia} for $R = 0.5$ jets, with $K_{\rm NLO} \simeq 1.135$ in the vicinity of \SI{400}{GeV}.
As discussed further in \App{app:pileupreweighting}, we reweight the MC in order that the pileup level in the simulation matches the data.
Finally, we multiply by an additional factor of $K_{375} = 0.961$ to ensure that the lowest bin in the simulation has the same normalization as the actual data.
This factor partially accounts for effects like the efficiency of the medium JQC, which is difficult to extract reliably from the CMS simulation, as well as QCD corrections beyond NLO and uncertainties on the recorded luminosity.

\begin{table*}[t]
\begin{center}
\begin{tabular}{ r@{$\quad$} l @{$\quad$}|@{$\quad$}  r @{$\quad$} r @{$\quad$} r @{$\quad$}|@{$\quad$} r @{$\quad$} r @{$\quad$} r}
\hline
\hline
&& \multicolumn{3}{c @{$\quad$}|@{$\quad$}}{CMS 2011 Open Data} & \multicolumn{3}{c}{CMS 2011 Simulation} \\
PID & Candidate & Total Count & After CHS & $p_T > \SI{1}{GeV}$ & Total Count & After CHS & $p_T > \SI{1}{GeV}$ \\
\hline
\hline
$11$ & Electron ($e^-$) & 31,297 & 30,304 & 30,284 & 76,819 & 73,937 & 73,906\\
$-11$ & Positron ($e^+$) & 31,444 & 30,470 & 30,448 & 75,651 & 72,920 & 72,868\\
$13$ & Muon ($\mu^-$) & 16,779 & 14,957 & 14,912 & 47,871 & 42,604 & 42,511\\
$-13$ & Antimuon ($\mu^+$) &  17,453 & 15,373 & 15,310 & 50,009 & 44,256 & 44,149\\
$211$ & Positive Hadron (e.g.\,$\pi^+$) & 10,731,634 & 8,159,520 & 6,950,019 & 31,682,518 & 23,267,103 & 19,775,066\\
$-211$ & Negative Hadron (e.g.\,$\pi^-$) & 10,414,733 & 7,987,681 & 6,780,597 & 30,718,965 & 22,837,987 & 19,361,736\\
\hline
$22$ & Photon ($\gamma$) & 14,102,402 & 14,102,402 & 7,157,772 & 39,487,711 & 39,487,711 & 19,805,470\\
$130$ & Neutral Hadron (e.g.\,$K^0_L$) & 2,955,136 & 2,955,136 & 2,317,806 & 7,509,228 & 7,509,228 & 5,974,028\\
\hline
\hline
\end{tabular}
\end{center}
\caption{Counts of PFCs by PID code, considering the constituents of the two hardest jets with the restriction $|\eta^\text{jet}| < 1.9$ and $p_T^\text{jet} \in [375,425]~\text{GeV}$.
The MC simulation has a larger number of events than the CMS Open Data, and therefore more total PFCs.
Note that the PID code is based on the PDG MC numbering scheme, but a code like $\pm 211$ indicates any charged hadron candidate, not solely $\pi^\pm$.
}
\label{table:object_count}
\end{table*}

In \Fig{fig:etaspectrum}, we show the jet pseudorapidity spectrum.
After relaxing the $|\eta^\text{jet}| < 1.9$ requirement, we find a small number of jets at larger pseudorapidities.
Compared to the simulated data, the open data has more jets in the vicinity of $|\eta^\text{jet}| \simeq 1.2$ and fewer in the vicinity of $|\eta^\text{jet}| \simeq 0.0$, indicating a possible issue with the \textsc{Pythia} prediction or with the pseudorapidity dependence of the JEC factors.
That said, the overall agreement is very good, giving us confidence that we can make basic kinematic jet selections.
For completeness, the jet azimuth spectrum is shown in \Fig{fig:phispectrum} of \App{app:additionalplots}, which exhibits the expected flat spectrum with small fluctuations due to detector inhomogeneities.

\begin{figure*}[t]
  \subfloat[]{
	\includegraphics[height=0.475\textwidth]{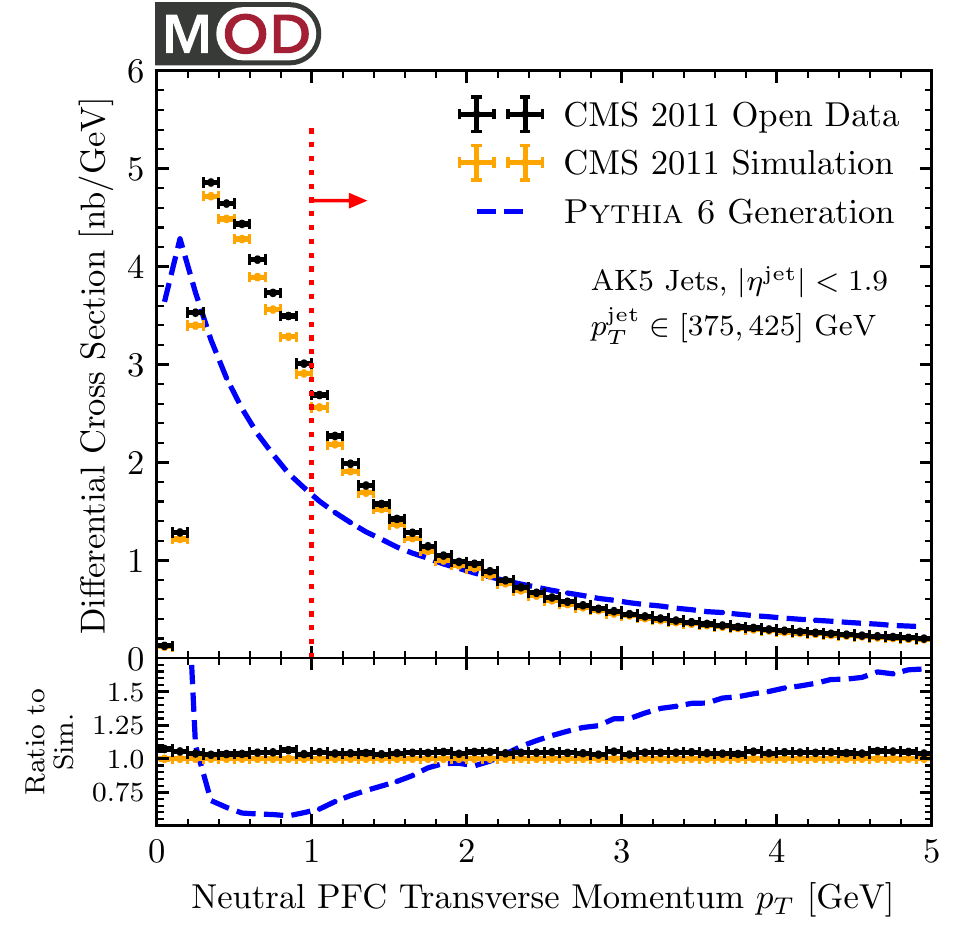}
	\label{fig:neutral_PFC_spectrum_zoom}
	}
\subfloat[]{
	\includegraphics[height=0.475\textwidth]{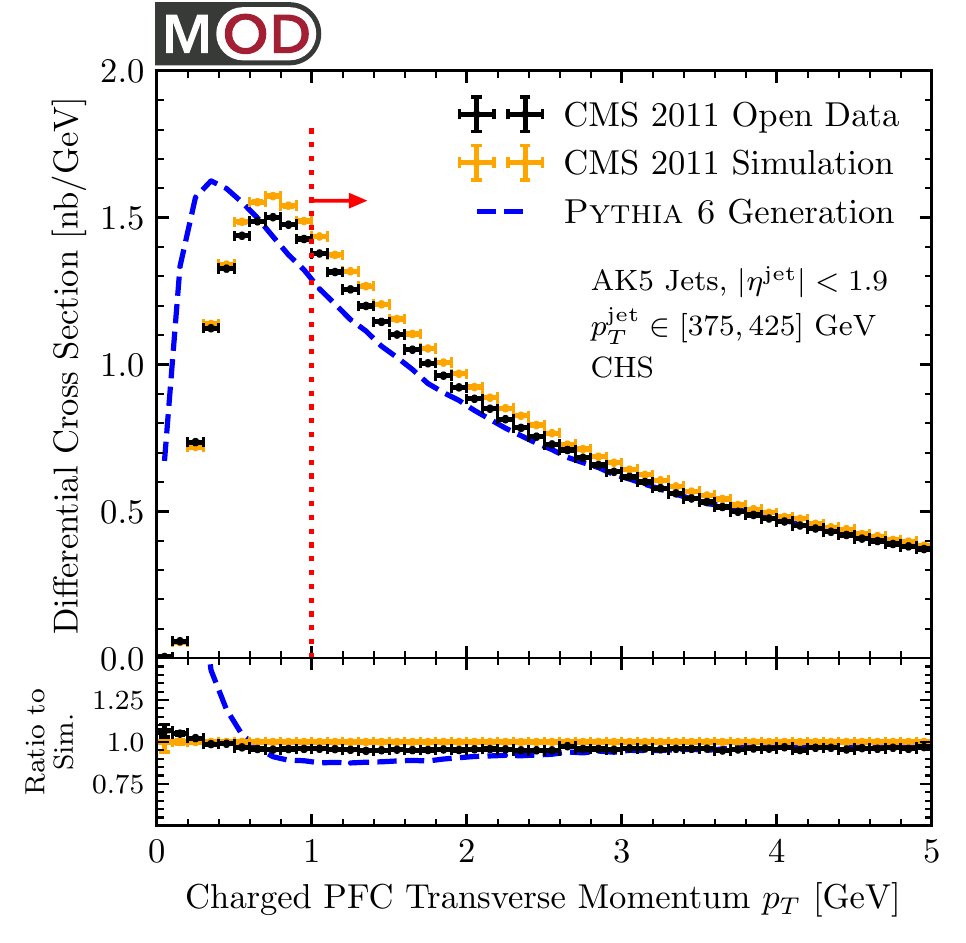}
	\label{fig:charged_PFC_spectrum_zoom}
	}
  \caption{Transverse momentum spectra for (a) neutral PFCs and (b) charged PFCs, including CHS to mitigate charged pileup, restricted to PFCs that are within the analyzed jets.
  The CMS simulation captures the key features of the CMS Open Data.
  Only for charged PFCs with $p_T^\text{PFC} > \SI{1}{GeV}$ is there reasonable agreement with the generation-level expectations from \textsc{Pythia}.
  The complete PFC $p_T$ spectrum is shown in \Fig{fig:PFC_spectrum} of \App{app:additionalplots}.  
  }
  \label{fig:PFC_spectrum_zoom}
\end{figure*}

\subsection{Jet Constituents}
\label{subsec:jetconstituents}

In addition to the reconstructed AK5 jets, the CMS Open Data contains the complete list of PFCs, which allows us to calculate a wide range of jet substructure observables.
Due to detector effects, one has to be careful when interpreting the PFC information.
Ultimately, we will focus on track-based observables which have better reconstruction performance as well as better pileup stability.

In \Tab{table:object_count}, we list the PID codes of the PFCs and their absolute counts in the jet sample with $|\eta^\text{jet}| < 1.9$ and $p_T^\text{jet} \in [375,425]~\text{GeV}$.
Note that there are more events in the MC samples than in the open data, so there is a corresponding increase in the number of total PFCs.
The PID codes indicate the most likely particle candidate, using the PDG MC numbering scheme~\cite{Tanabashi:2018oca}.
In particular, code 211 includes $\pi^+$, $K^+$, and proton candidates, code 22 includes photon and merged $\pi^0 \to \gamma \gamma$ candidates, and code 130 includes $K^0_L$ and neutron candidates.

The counts in \Tab{table:object_count} include contamination from pileup.
As shown in \Fig{figure:NPVhistogram} of \App{app:additionalplots}, there are typically $\sim 5$ pileup events per beam crossing.
While the CMS Open Data already includes a pileup correction for the jet $p_T$ via the JEC factors, this is insufficient to correct substructure distributions.
We have two ways to mitigate the effect of pileup.
First, we apply the CHS procedure~\cite{CMS:2014ata} to remove charged particles not associated with the primary vertex.
This is possible since \texttt{MODProducer} now stores vertex information (see \Sec{subsec:modformat} above), so we can remove charged jet constituents assigned to pileup vertices.
Though CHS cannot remove neutral particles from pileup, it does reduce the overall pileup contamination by a factor of $\sim$2/3.
Second, inspired by the SoftKiller procedure~\cite{Cacciari:2014gra}, we impose a $p_T^\text{PFC} > \SI{1}{GeV}$ cut on all PFCs, where this value is motivated by \Fig{fig:PFC_spectrum_zoom} below.
This helps control the level of neutral pileup, though we will still focus on track-based observables in our subsequent analyses.

The $p_T$ spectrum of neutral PFCs is shown \Fig{fig:neutral_PFC_spectrum_zoom}.
The neutral PFCs do not benefit from CHS, so there is a significant excess of neutral PFCs from pileup below around 2 GeV, compared to generation-level expectations.
That said, the CMS simulation appropriately captures this neutral pileup contamination.
Because of finite calorimeter granularity, there is a depletion of moderate $p_T$ neutral PFCs as a result of merging.
This merging results in an excess of higher $p_T$ neutral PFCs, which can be seen in \Fig{fig:neutral_PFC_spectrum} of \App{app:additionalplots}.

The $p_T$ spectrum of charged PFCs is shown in \Fig{fig:charged_PFC_spectrum_zoom}.
With CHS, the PFC $p_T$ spectrum is rather similar between the CMS Open Data and the MC event samples, even at the generator level and even going out to higher $p_T$ in \Fig{fig:charged_PFC_spectrum} of \App{app:additionalplots}.
The main difference is below $\SI{1}{GeV}$, where one sees the impact of tracking inefficiencies and momentum misreconstruction.
For this reason, we impose a cut of $p_T^\text{PFC} > \SI{1}{GeV}$ for all of our jet substructure studies, which results in better data/MC agreement for observables like track multiplicity that are sensitive to such effects. 
Note that this same $p_T^\text{PFC}$ cut was advocated for in \Ref{Tripathee:2017ybi}, though a looser cut of \SI{500}{MeV} is used by CMS in its track multiplicity study~\cite{Chatrchyan:2012mec}.

\subsection{Jet Substructure Observables}
\label{subsec:subobs}

\begin{figure*}[p]
\centering
\subfloat[]{\includegraphics[width=0.8\columnwidth]{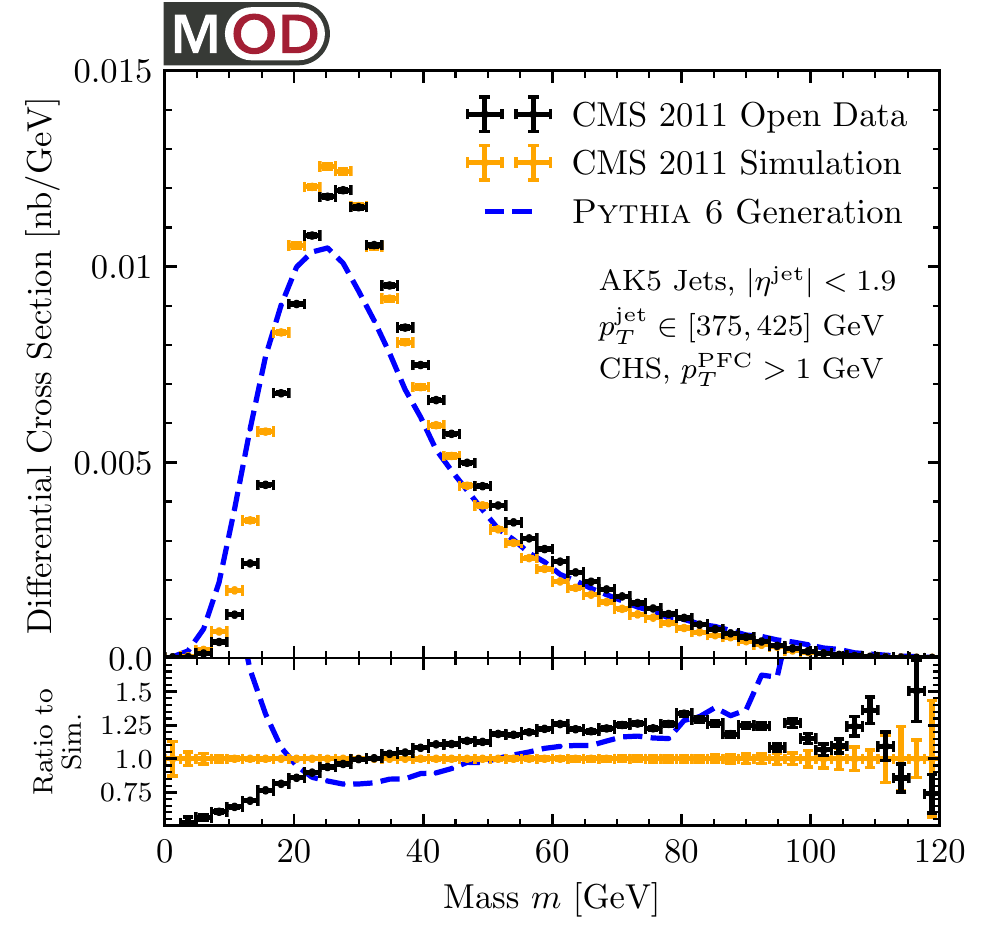}\label{fig:mass}}
\hspace{10mm}
\subfloat[]{\includegraphics[width=0.8\columnwidth]{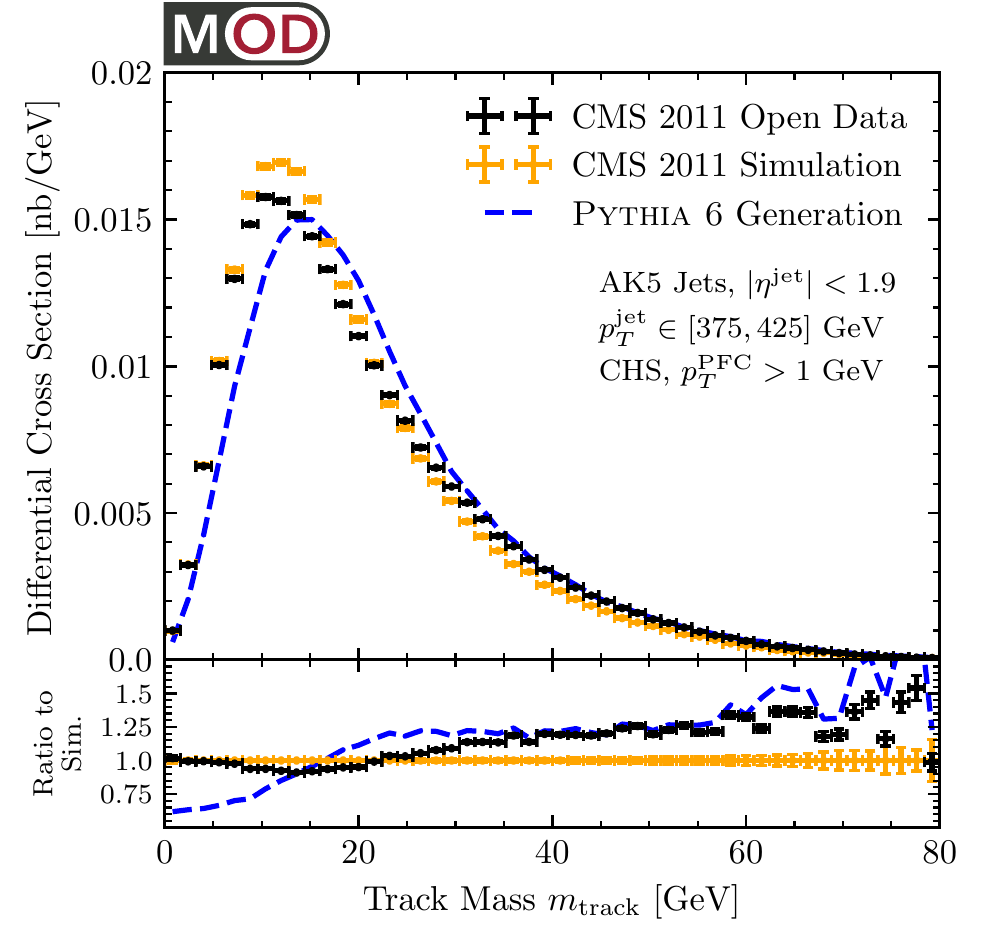}\label{fig:trmass}}\\
\subfloat[]{\includegraphics[width=0.8\columnwidth]{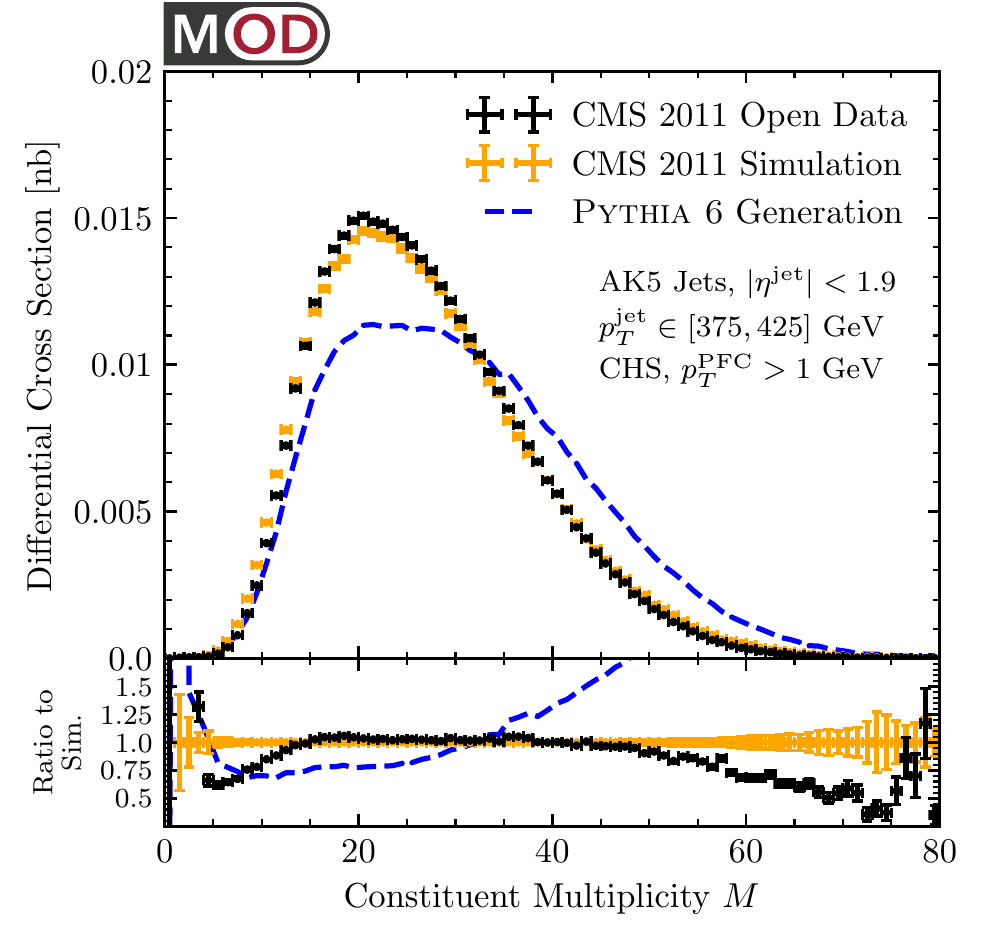}\label{fig:mult}}
\hspace{10mm}
\subfloat[]{\includegraphics[width=0.8\columnwidth]{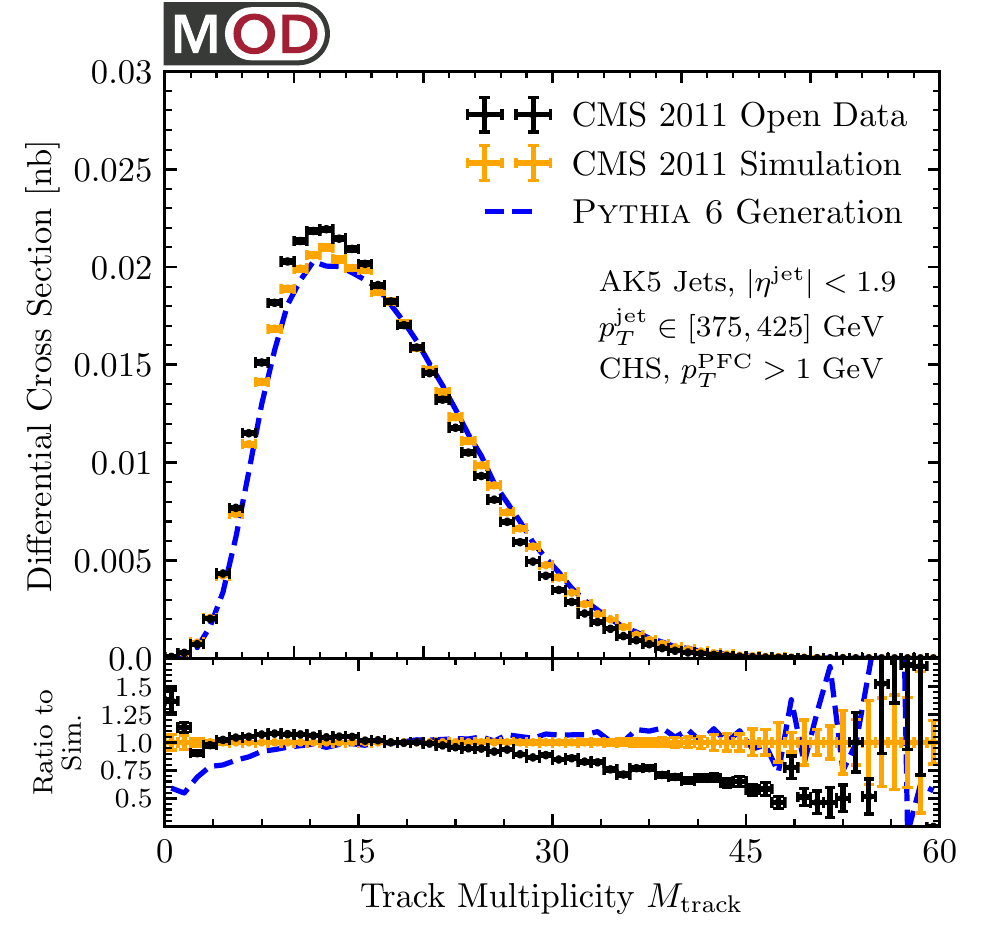}\label{fig:trmult}}\\
\subfloat[]{\includegraphics[width=0.8\columnwidth]{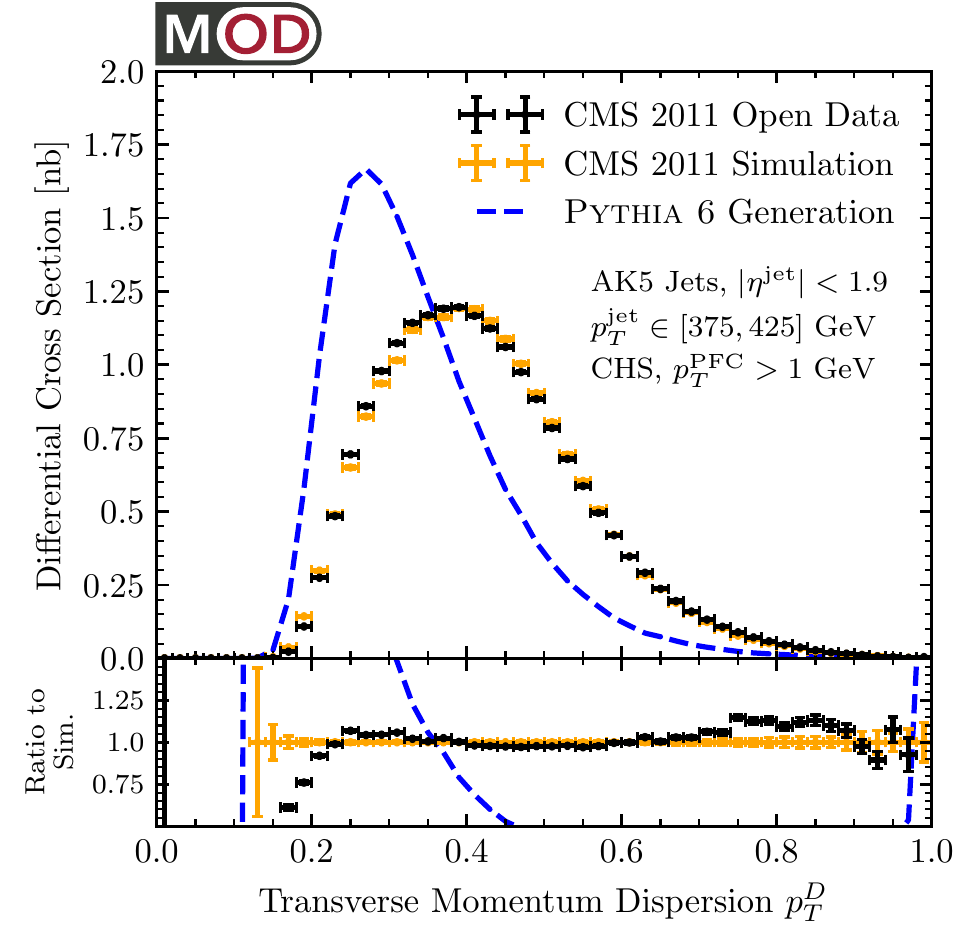}\label{fig:ptd}}
\hspace{10mm}
\subfloat[]{\includegraphics[width=0.8\columnwidth]{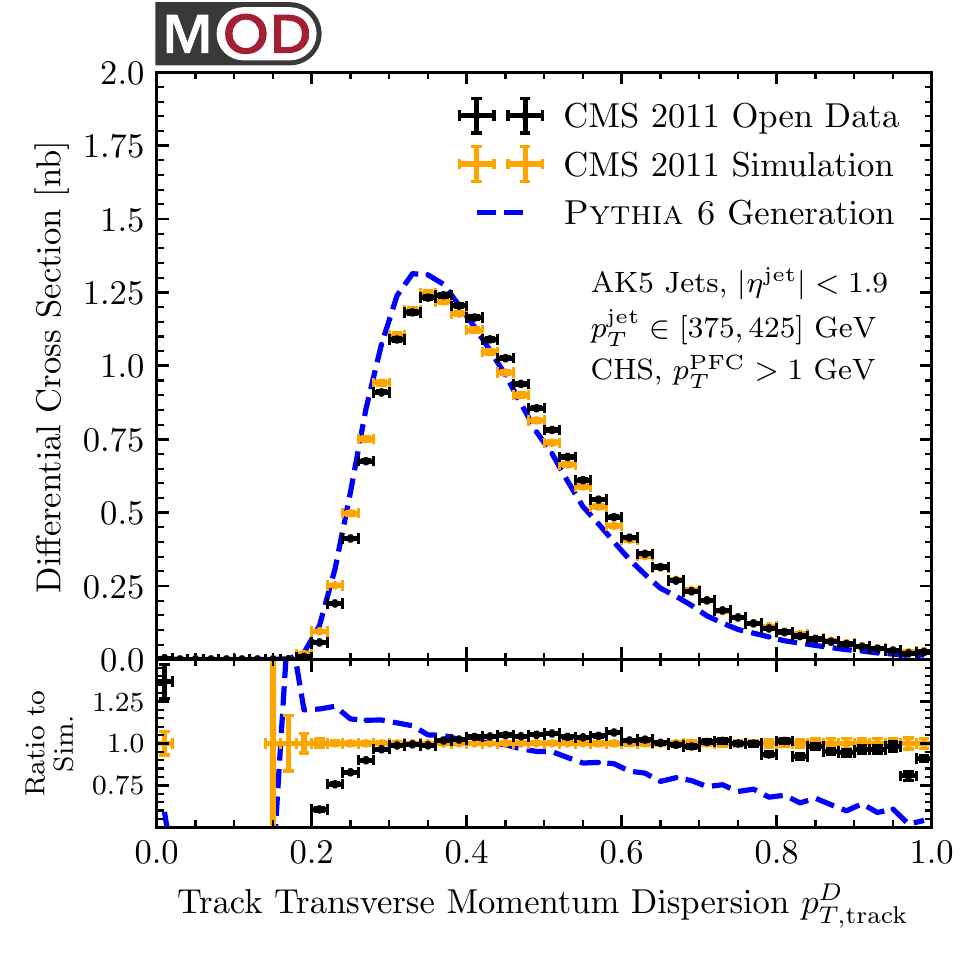}\label{fig:trptd}}
\caption{Jet substructure observables using (left column) all PFCs and (right column) charged PFCs.
In all cases, we apply CHS and enforce $p_T^\text{PFC} > \SI{1}{GeV}$.
The observables are (top row) jet mass, (middle row) constituent multiplicity, and (bottom row) transverse momentum dispersion ($p_T^D$).
}
\label{fig:substructureobs}
\end{figure*}

\begin{figure*}[p]
\subfloat[]{\includegraphics[width=0.8\columnwidth]{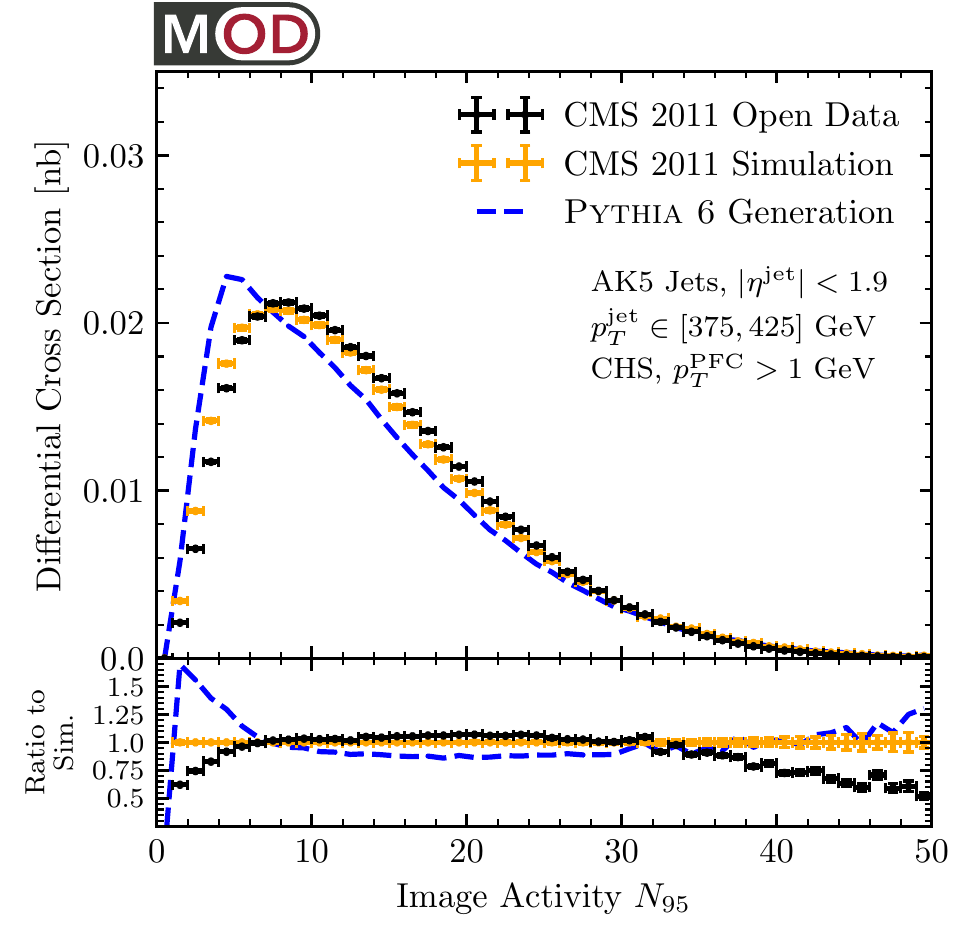} \label{fig:N95}}
\hspace{10mm}
\subfloat[]{\includegraphics[width=0.8\columnwidth]{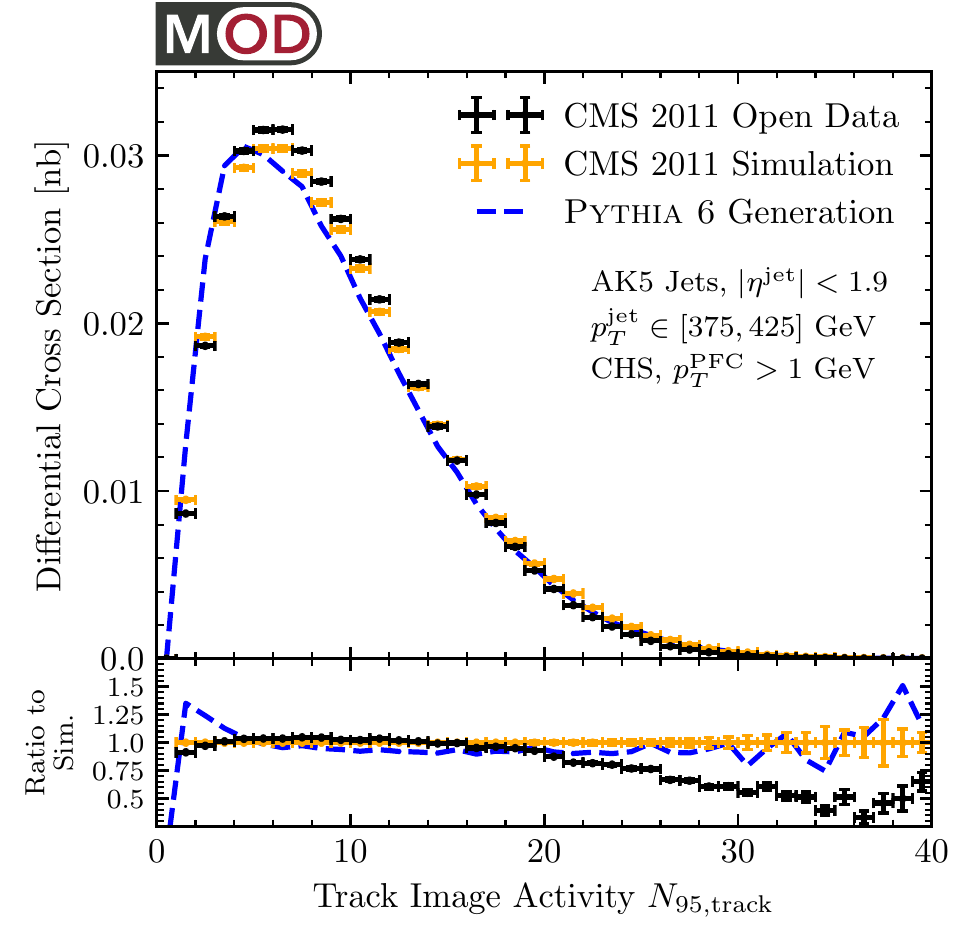} \label{fig:trN95}}\\
\subfloat[]{\includegraphics[width=0.8\columnwidth]{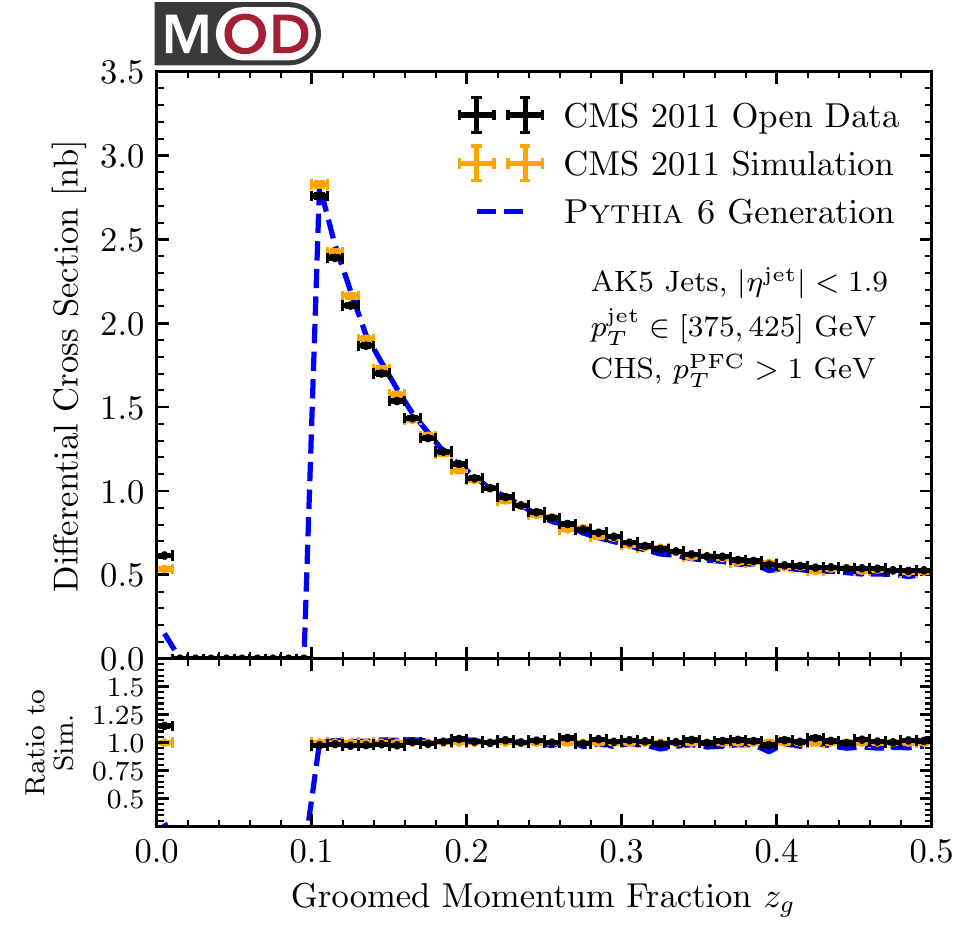} \label{fig:zg}}
\hspace{10mm}
\subfloat[]{\includegraphics[width=0.8\columnwidth]{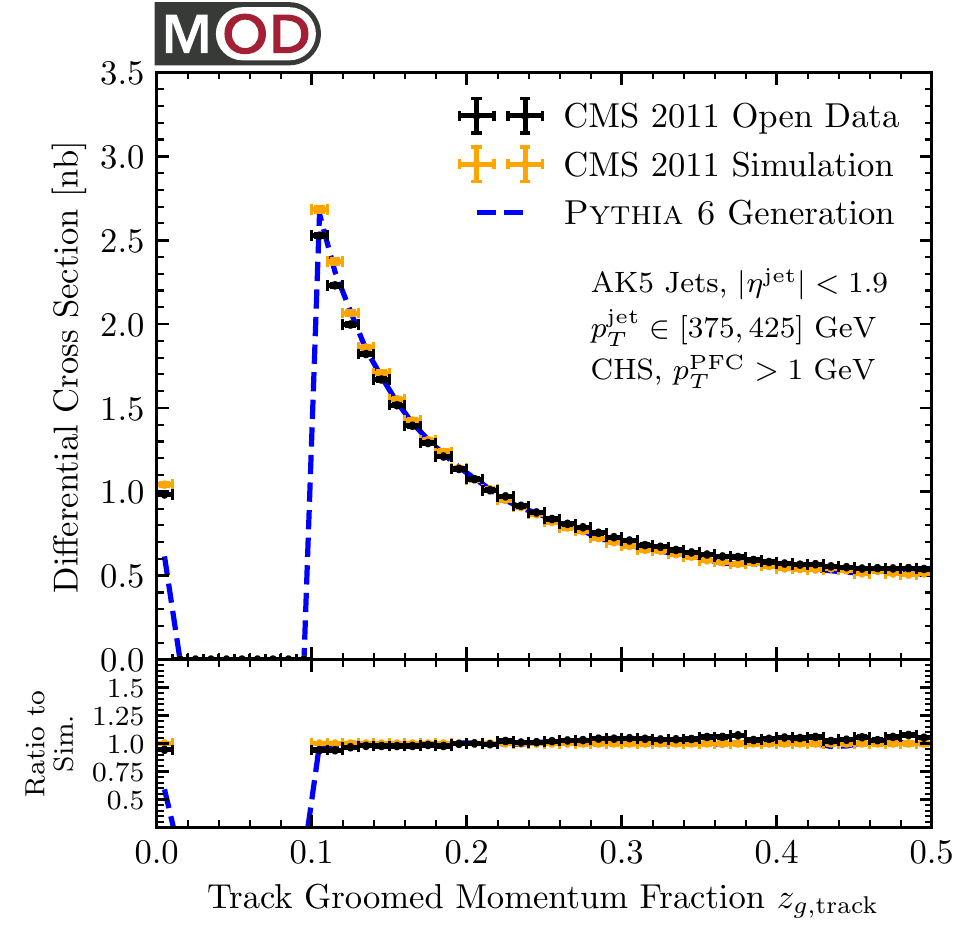} \label{fig:trzg}}\\
\subfloat[]{\includegraphics[width=0.8\columnwidth]{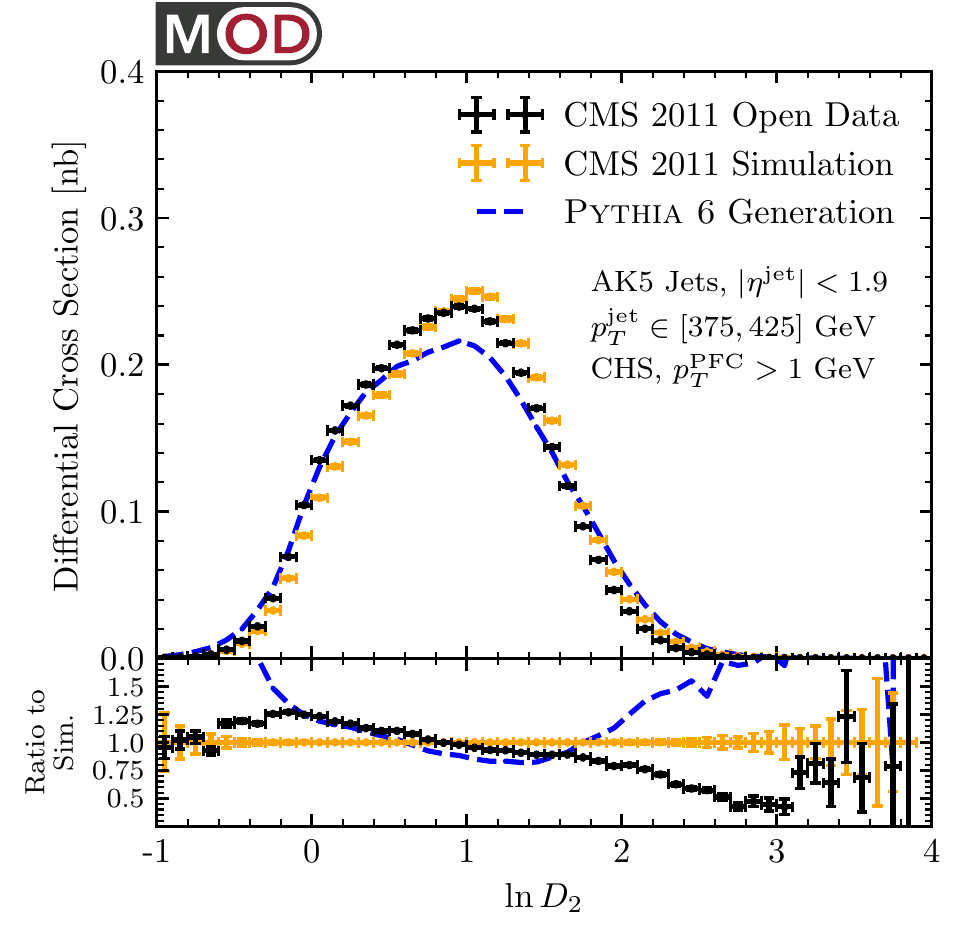} \label{fig:D2}}
\hspace{10mm}
\subfloat[]{\includegraphics[width=0.8\columnwidth]{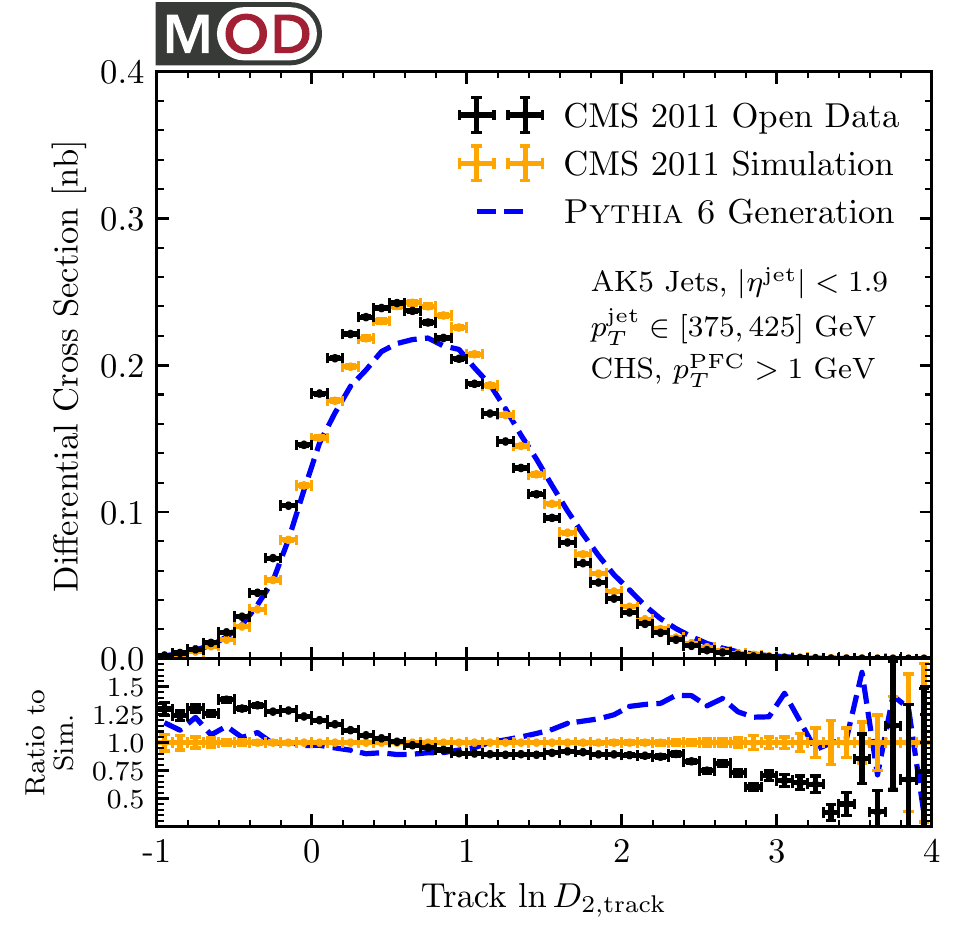} \label{fig:trD2}}
\caption{
Same as \Fig{fig:substructureobs} for three more jet substructure observables:  (top row) $N_{95}$, (middle row) $z_g$, and (bottom row) $D_2$.
}
\label{fig:substructureobs2}
\end{figure*}

We now plot a representative sample of jet substructure observables, comparing the CMS Open Data to the MC samples, both before and after detector simulation.
Based on the conclusions of \Sec{subsec:jetconstituents}, we always implement CHS and impose the $p_T^\text{PFC} > \SI{1}{GeV}$ cut.
In order to analyze jets with similar total $p_T$, we focus on the relatively narrow range of $p_T^\text{jet} \in [375,425]~\text{GeV}$.

In \Fig{fig:substructureobs}, we show three classic substructure distributions:  jet mass, constituent multiplicity, and $p_T^D$~\cite{CMS:2013kfa}.
Using all PFCs, shown in the left column of \Fig{fig:substructureobs}, there is good agreement between the CMS Open Data and the simulation-level MC events.
This suggests that \textsc{Pythia} 6 with tune Z2 has a reasonable model for jet fragmentation and that the CMS simulation provides a faithful characterization of the detector response; see related studies in \Ref{Chatrchyan:2012mec}, as well as \Ref{Aad:2011kq} for alternative \textsc{Pythia} tunes.

That said, there are significant differences when comparing the generation-level and simulation-level MC distributions, even after applying CHS for pileup mitigation.
Roughly speaking, the CMS detector reconstructs fewer PFCs than expected, which is consistent with merging of neutral PFCs due to finite calorimeter granularity.
On the other hand, the CMS detector reconstructs a larger jet mass than expected, which is consistent with residual neutral pileup contamination.

We can improve the generation-level and simulation-level agreement by restricting our analysis to just charged PFCs, as shown in the right column of \Fig{fig:substructureobs}.
The agreement improves most notably for the IRC-unsafe observables of multiplicity and $p_T^D$.
While the CMS detector reconstructs fewer charged PFCs than expected from \textsc{Pythia} at the generation level, the difference is well within the theoretical uncertainties in MC generation (see further discussion in \Ref{Gras:2017jty}).
Since we will not attempt to unfold the data in this paper, it is important for us to use observables that are robust to detector effects.
For this reason, the focus of our EMD studies will be on track-based observables.

It is worth remarking that the good agreement in the track multiplicity distribution in \Fig{fig:trmult} is due in part to using the medium JQC.
If we were to use the loose JQC, there would be an excess of events with very low track multiplicity in the CMS Open Data.
Most likely, these are prompt photons which barely pass the loose JQC, and to describe these properly, we would need to include photon-plus-jet MC samples.
This excess is removed by the medium JQC, with only a modest impact on other substructure distributions.

We investigate three additional jet substructure distributions in \Fig{fig:substructureobs2}: $N_{95}$~\cite{Pumplin:1991kc}, $z_g$~\cite{Larkoski:2015lea}, and $D_2$~\cite{Larkoski:2014gra} with $\beta=1$.
These observables probe, respectively, the uniformity of jet activity, the momentum sharing between subjets, and the two-prong substructure of jets.
We implement $N_{95}$ as the minimum number of pixels in a $33\times33$ jet image from $-R$ to $R$ required to account for at least 95\% of the total $p_T$.
The soft drop jet grooming~\cite{Dasgupta:2013ihk,Larkoski:2014wba} parameters used to define the groomed momentum fraction $z_g$ are $z_\text{cut}=0.1$ and $\beta=0$.
Jets with $z_g=0$ indicate that the grooming procedure results in just a single remaining particle.
Again, we find good agreement between the CMS Open Data and the simulation-level MC samples when using all PFCs, but the detector-level and simulation-level distributions agree somewhat better when restricted to track-based observables.
Using our released samples~\cite{MOD:ZenodoCMS,MOD:ZenodoMC170,MOD:ZenodoMC300,MOD:ZenodoMC470,MOD:ZenodoMC600,MOD:ZenodoMC800,MOD:ZenodoMC1000,MOD:ZenodoMC1400,MOD:ZenodoMC1800}, it is straightforward to plot a wide range of jet substructure observables~\cite{MODDemo}, a number of which have already been implemented in the \textsc{EnergyFlow} package~\cite{EnergyFlow}.

\section{Exploring the Space of Jets}
\label{sec:emd}

We now turn from considering individual substructure observables at the histogram level to studying the radiation pattern in jets more broadly.
In this section, we will use the energy mover's distance~\cite{Komiske:2019fks} as a metric to compare the energy flow of jets.
We perform a range of exploratory EMD studies on the CMS Open Data to universally probe jet modifications, explore the space of jets, and visualize the most representative jets.

\subsection{Review of the Energy Mover's Distance}
\label{subsec:emd_review}

The jet energy flow can be characterized by an energy density on a two-dimensional surface, corresponding to an idealized detector at infinity~\cite{Tkachov:1995kk,Sveshnikov:1995vi,Cherzor:1997ak}.
For proton-proton collisions, we typically use transverse momentum $p_{T}$ instead of energy and we indicate angular directions via rapidity $y$ and azimuth $\phi$.
In these coordinates, the energy flow (more precisely, the transverse momentum flow) is:
\begin{equation}
\label{eq:energyflow}
\rho(y,\phi) = \sum_{j \in \mathcal{J}} p_{Tj} \, \delta(y - y_j) \, \delta(\phi - \phi_j),
\end{equation}
where $j$ labels the constituents of the jet $\mathcal{J}$.

The expression in \Eq{eq:energyflow} is IRC safe by construction, since a particle with zero $p_T$ does not contribute to the sum and a collinear splitting does not change the sum.
The energy flow does not include any PID information, which is important to ensure IRC safety.
To handle constituent masses, one could include velocity information~\cite{Mateu:2012nk}, but that is beyond the scope of this paper.

Given two jets $\mathcal{I}$ and $\mathcal{J}$, the EMD is~\cite{Komiske:2019fks}:
\begin{equation}
\label{eq:EMD}
\text{EMD}(\mathcal{I}, \mathcal{J}) = \min_{\{f_{ij}\}} \sum_{i \in \mathcal{I}} \sum_{j \in \mathcal{J}} f_{ij} \frac{R_{ij}}{R} + \bigg| \sum_{i \in \mathcal{I}} p_{Ti} - \sum_{j \in \mathcal{J}} p_{Tj}  \bigg|,
\end{equation}
where $R_{ij}^2 = (y_i - y_j)^2 + (\phi_i - \phi_j)^2$ is the rapidity-azimuth distance, $R$ is the jet radius, and $f_{ij}$ is the amount of transverse momentum ``transported'' from particle $i$ in jet $\mathcal I$ to particle $j$ in jet $\mathcal J$, subject to the constraints:
\begin{align}
&f_{ij} \ge 0, \quad \sum_{j \in \mathcal{J}} f_{ij} \leq p_{Ti}, \quad \sum_{i \in \mathcal{I}} f_{ij} \leq p_{Tj},
\label{eq:EMDconstraints}\\
&\sum_{i \in \mathcal{I}} \sum_{j \in \mathcal{J}} f_{ij} = \min \bigg( \sum_{i \in \mathcal{I}} p_{Ti}, \sum_{j \in \mathcal{J}} p_{Tj}  \bigg).
\end{align}
Finding the minimum over $\{f_{ij}\}$ in \Eq{eq:EMD} is an optimal transport problem which can be solved efficiently using the network simplex algorithm~\cite{DBLP:journals/mp/Orlin97,DBLP:journals/mp/Tarjan97,DBLP:journals/mp/OrlinPT93}.

The expression in \Eq{eq:EMD} is non-negative, symmetric, and satisfies the triangle inequality:
\begin{align}
\text{EMD}(\mathcal{I}, \mathcal{J}) &\ge 0, \nonumber\\
\text{EMD}(\mathcal{I}, \mathcal{J}) &= \text{EMD}(\mathcal{J}, \mathcal{I}), \nonumber \\
\text{EMD}(\mathcal{I}, \mathcal{J}) &\leq \text{EMD}(\mathcal{I}, \mathcal{\mathcal{K}}) + \text{EMD}(\mathcal{K},\mathcal{J}) \label{eq:EMDproperties}.
\end{align}
Therefore, EMD is a proper metric on the space of energy flows, with units of energy (i.e.\ GeV).
If the EMD between two jets is zero, then they are treated as identical.
For this reason, it is often convenient to perform symmetry transformations on the jets prior to calculating the EMD.
(This transformation procedure is closely related to the tangent earth mover's distance~\cite{temd}.)
For all of the EMD studies in this paper, we longitudinally boost and azimuthally rotate each jet such that its four-vector is at the $(y,\phi)$ origin.

The second term in \Eq{eq:EMD} is a cost term when two jets have different values of their scalar sum $p_T$.
Because we are primarily interested in relative jet energy flows and not absolute jet energy scales, it is convenient to rescale the jets to make this cost term vanish.
For jets with $p_T^\text{jet} \in [375,425]~\text{GeV}$, we rescale the jet constituents uniformly such that
\begin{equation}
\label{eq:EMDrescaling}
\sum_{j \in \mathcal{J}} p_{Tj} \Rightarrow \SI{400}{GeV}.
\end{equation}
Since we are working in relatively narrow $p_T$ range and since QCD is a quasi-scale-invariant theory, this rescaling has only a mild impact on our results.
Experimentally, this rescaling has the nice feature of reducing the dependence of our results on the JEC factors and on any PFC selection criteria.
Theoretically, this rescaling has the nice feature of making the EMD identical (up to an overall energy scale) to the 1-Wasserstein metric between probability densities~\cite{wasserstein1969markov,dobrushin1970prescribing}.
Changing the baseline from $\SI{400}{GeV}$ to some other scale would just proportionally rescale all the results below.

\begin{figure*}
  \subfloat[]{
    \includegraphics[width=0.49\textwidth]{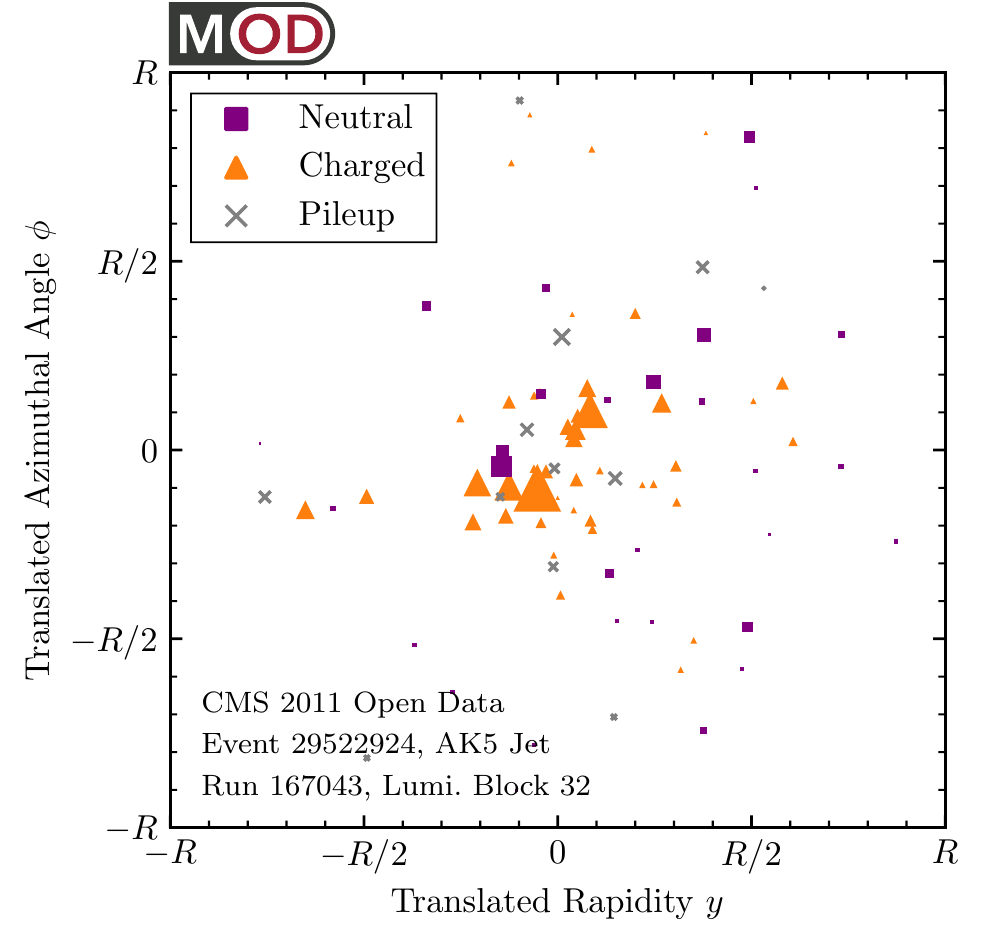}
	}
  \subfloat[]{
    \includegraphics[width=0.49\textwidth]{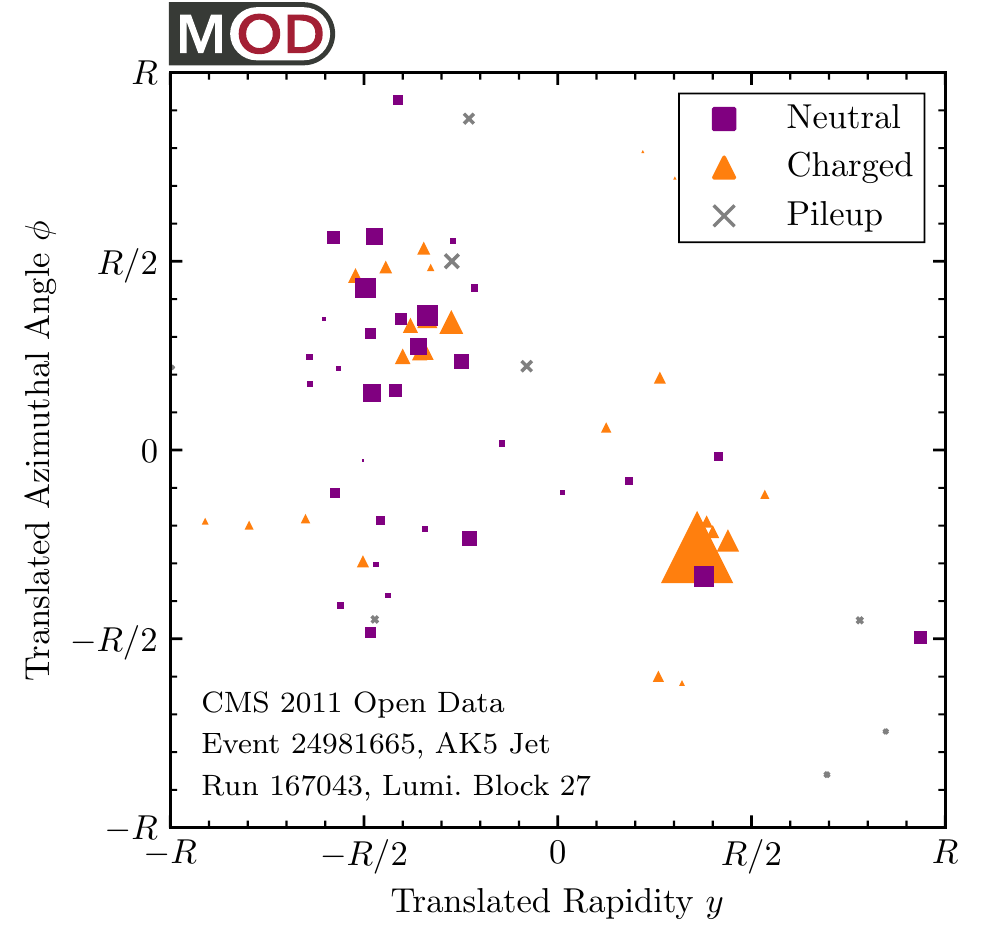}
	}
	\\
\subfloat[]{
    \includegraphics[width=0.65\textwidth]{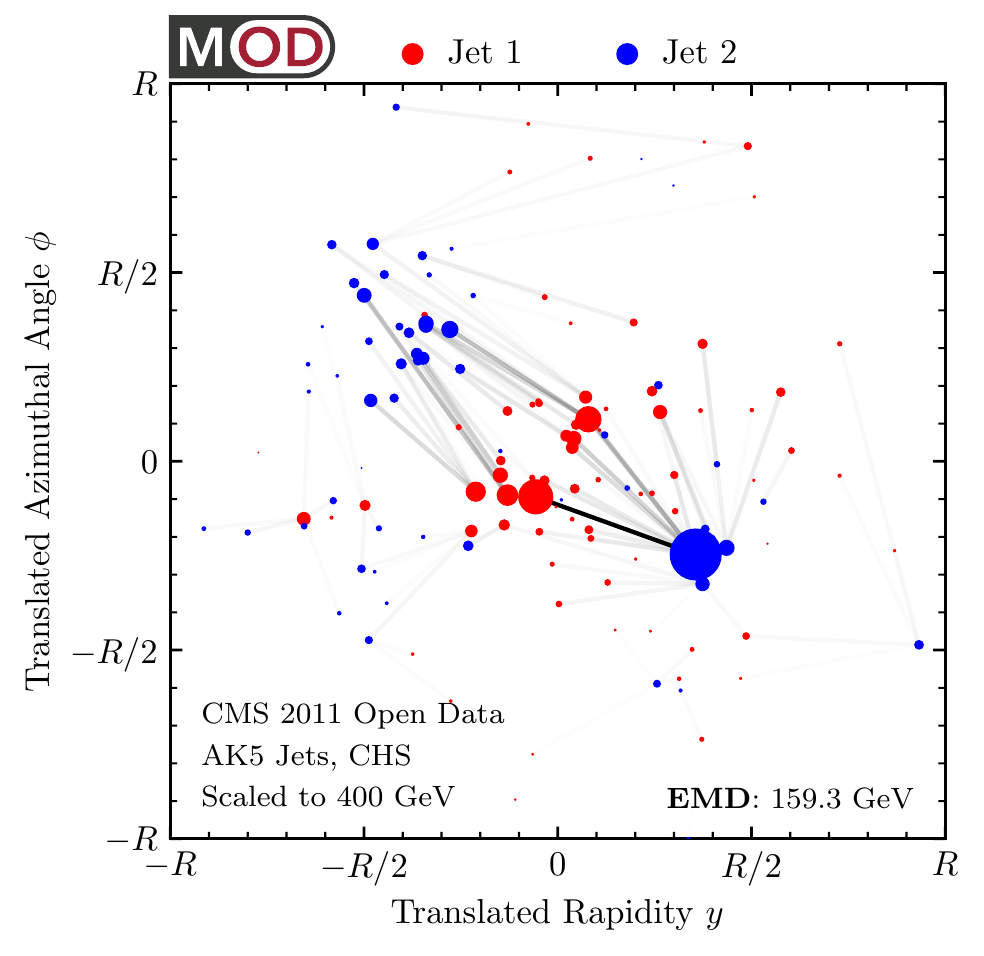}
	}
  \caption{
   Example EMD computation.
(top row) Two jets from the CMS Open Data shown in the style of \Fig{fig:eventdisplay}, with the size of each symbol indicating the particle transverse momentum and the style indicating the charge.
Pileup particles removed by CHS are indicated by gray crosses.
(bottom) Both jets represented as energy flow distributions via \Eq{eq:energyflow}, along with the optimal transportation plan to rearrange one jet into the other, with the intensity of each line corresponding to $\{f_{ij}\}$ of \Eq{eq:EMD}.
    \label{fig:EMDexample}
}
\end{figure*}

\begin{figure*}
  \subfloat[]{
    \includegraphics[width=0.49\textwidth]{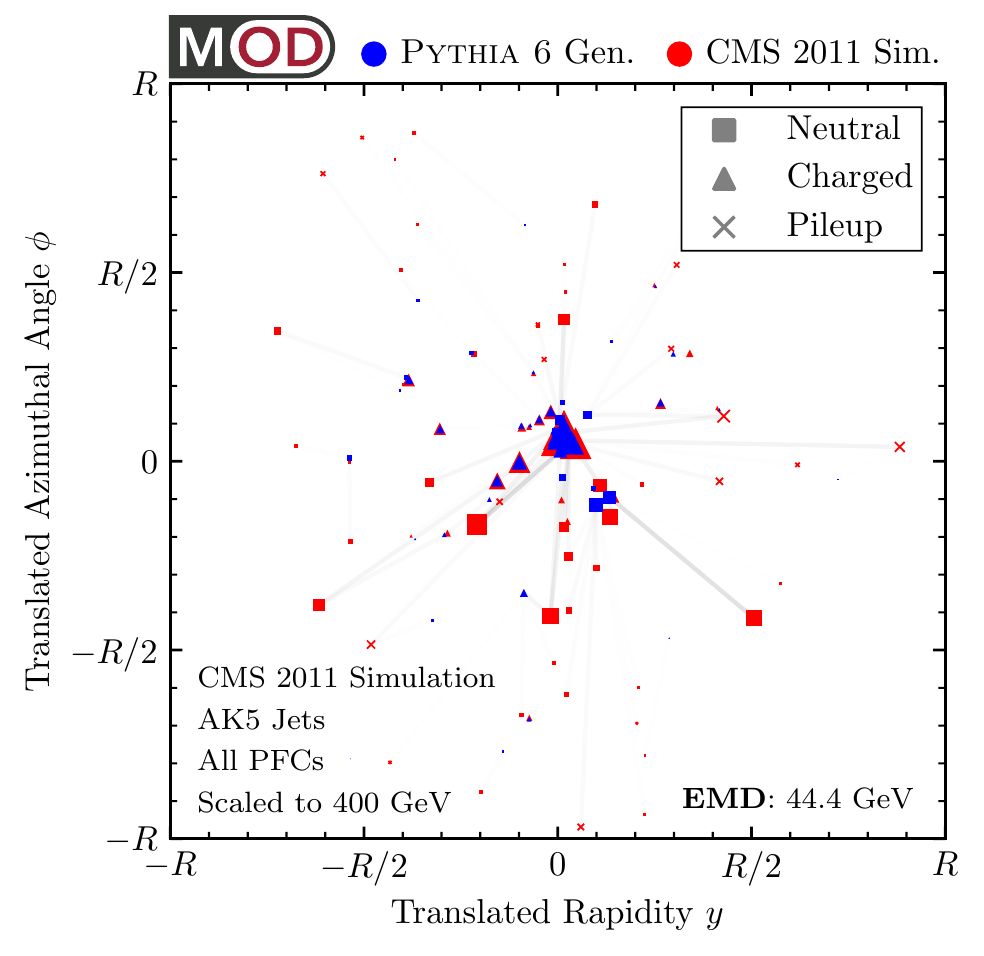}
	}
  \subfloat[]{
    \includegraphics[width=0.49\textwidth]{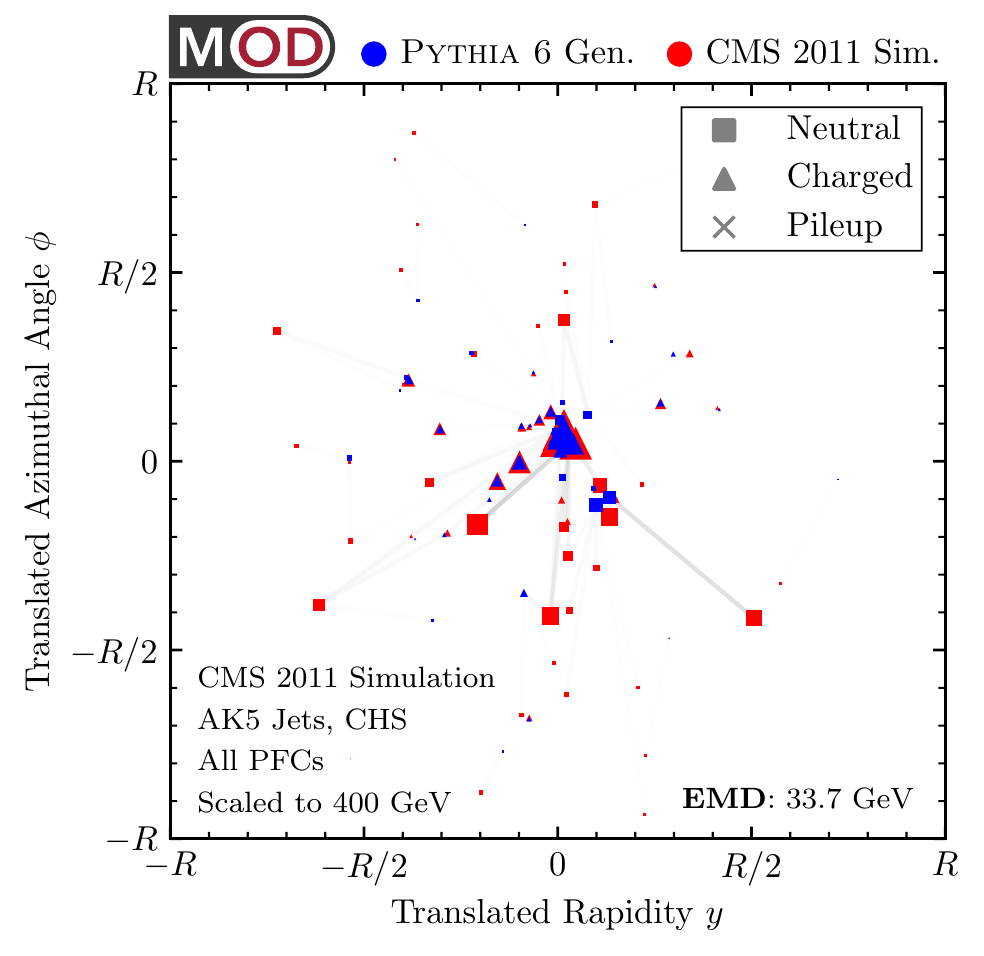}
	}\\
\subfloat[]{
    \includegraphics[width=0.49\textwidth]{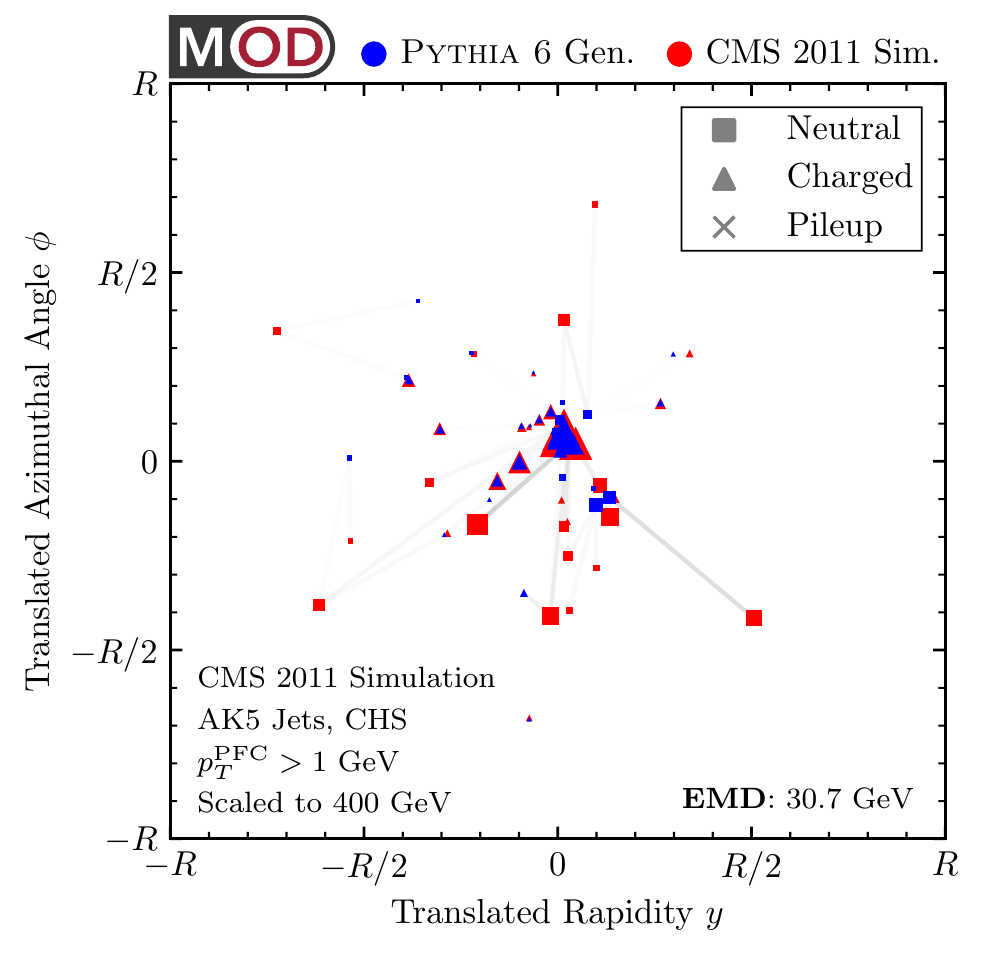}
	}
\subfloat[]{
    \includegraphics[width=0.49\textwidth]{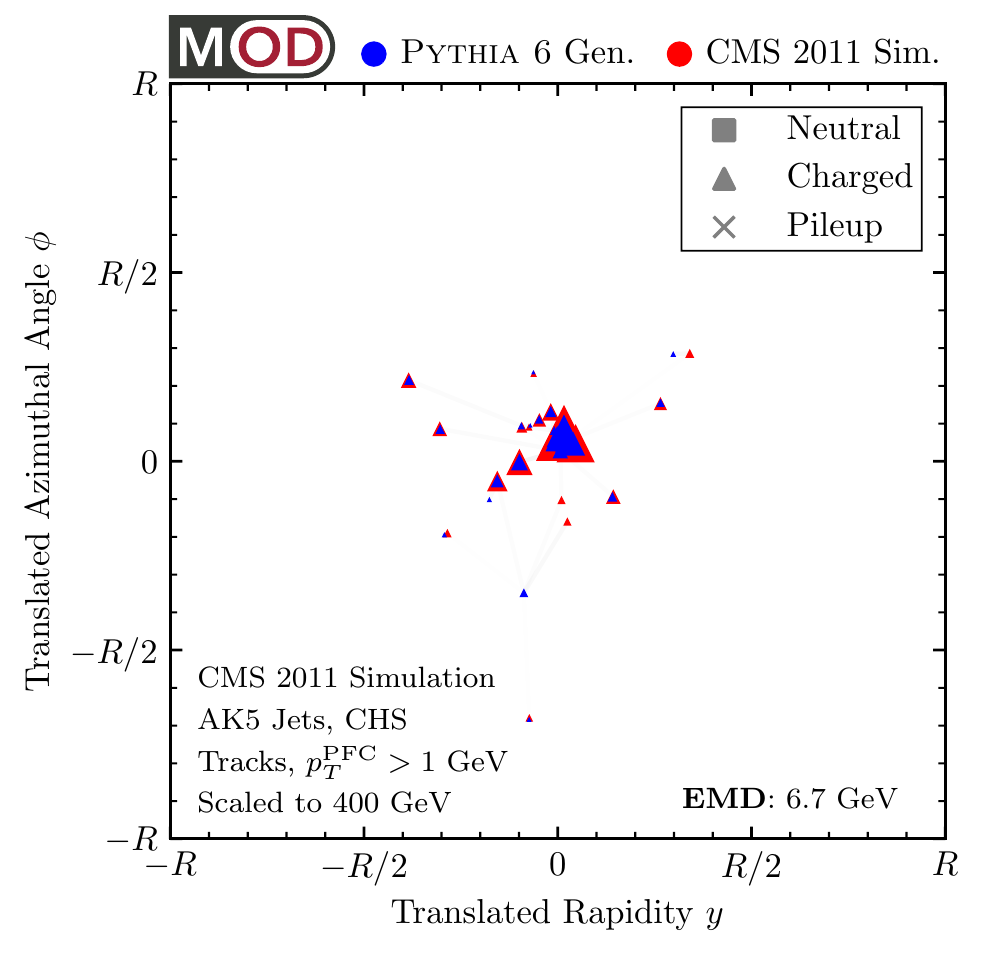}
	}
  \caption{
A jet from the \textsc{Pythia} hard QCD MC sample shown (blue) before and (red) after the \textsc{Geant}-based CMS detector simulation, with the size of each symbol indicating the particle transverse momentum and the shapes indicating the charge.
To improve visibility and clarity, the sizes of the symbols in the generator-level jet have been uniformly decreased.
Pileup particles removed by CHS are indicated by crosses, and the optimal transportation plans between the jets are shown as gray lines.
The jets are shown (a) with all PFCs, (b) after applying CHS to remove charged pileup, (c) after an additional $p_T^\text{PFC}>1$ GeV cut, and (d) after further restricting only to tracks.
The EMD between the jet before and after the detector simulation decreases as these cuts are applied, highlighting that these PFC cuts minimize the impact of detector effects.
    \label{fig:SIMGENexample}
}
\end{figure*}

As motivated by \Sec{sec:jetsubstructure} (and further motivated by \Sec{subsec:emd_detector} below), we often restrict our attention to charged particles with $p_T^\text{PFC}>1$ GeV.
Strictly speaking, such a PFC restriction breaks the collinear safety (though not the soft safety) of the EMD, though there are calculational strategies to account for this using track functions~\cite{Waalewijn:2012sv,Krohn:2012fg,Chang:2013rca,Chang:2013iba}.
Note that we always apply the rescaling in \Eq{eq:EMDrescaling} \emph{after} applying any PFC-level restrictions, such that our track-only results are similar in spirit to track-assisted observables~\cite{ATLAS:2016vmy,Elder:2018mcr}.
Crucially, the PFC restriction and overall rescaling still preserve the metric properties of the EMD in \Eq{eq:EMDproperties}.

An example EMD computation for two jets in the CMS Open Data is shown in \Fig{fig:EMDexample}.
In the top row, we show two jets plotted in the style of \Fig{fig:eventdisplay}.
Here, the size of the dots indicates the transverse momenta of the PFCs, the colors indicate whether the PFCs are neutral or charged, and the crosses indicate charged PFCs that have been removed by CHS.
In the bottom row, we drop the PID information and switch to the energy flow representation in \Eq{eq:energyflow}.
We overlay the two jets, with the red dots corresponding to the first jet, the blue dots corresponding to the second jet, and the gray lines indicating the optimal transport $\{f_{ij}\}$.
Because we have rescaled the jets by \Eq{eq:EMDrescaling}, all $p_T$ from the first jet can be transported to the second jet.

\begin{figure*}[t]
  \subfloat[]{
	\includegraphics[height=0.475\textwidth]{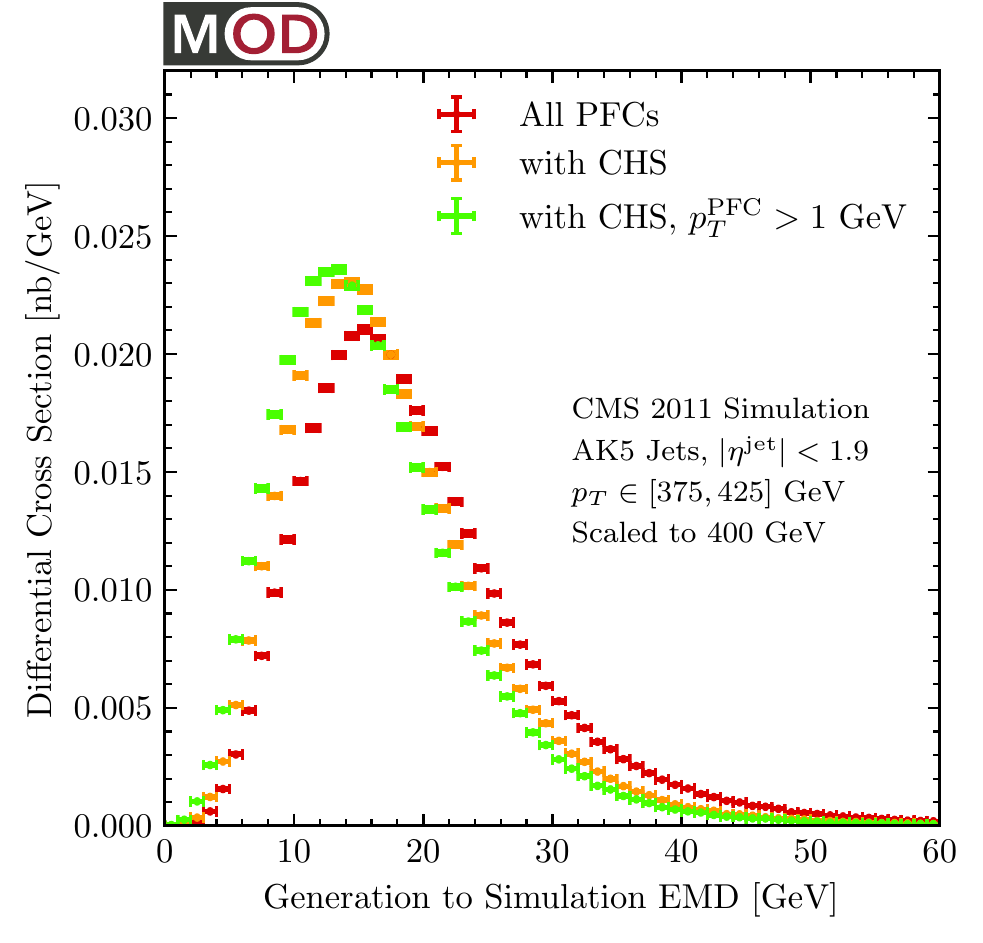}
	\label{fig:emdcmp_gensimEMDAll}
	}
\subfloat[]{
	\includegraphics[height=0.475\textwidth]{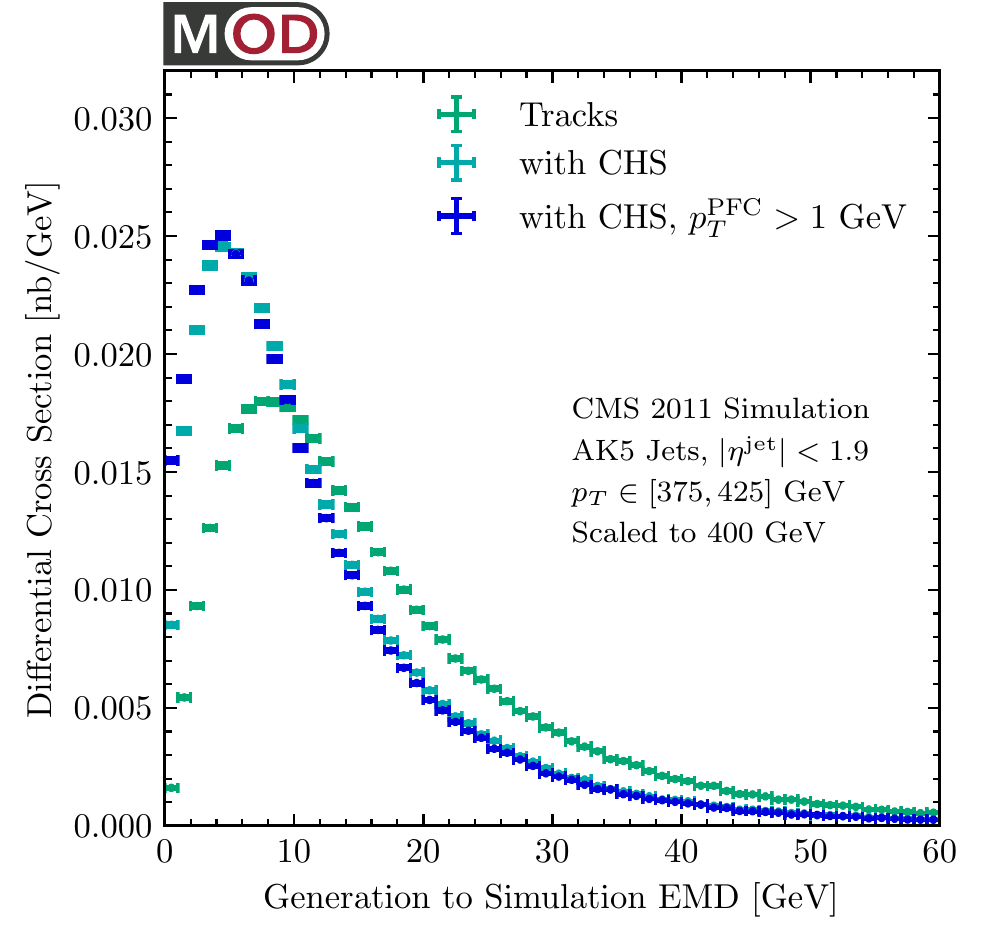}
	\label{fig:emdcmp_gensimEMDTrack}
	}
  \caption{
Quantifying detector effects through the distribution of generation-to-simulation EMDs.
Starting from the same jet generated by \textsc{Pythia}, we compute the EMD between the jet before and after the \textsc{Geant}-based CMS detector simulation.
These are shown for (a) all PFCs and (b) tracks only, with the subsequent application of CHS and the $p_T^\text{PFC} > \SI{1}{GeV}$ restriction.
The agreement between the generation-level and simulation-level radiation patterns (as quantified by EMD) can indeed be seen to improve as the selections tighten.
See \Fig{fig:pileupEMDs} in \App{app:pileupreweighting} for a study of the impact of CHS for different levels of pileup contamination.
See \Fig{fig:emdcmp_pT} in \App{app:additionalplots} for a study of the impact of the $p_T^\text{PFC}$ cut.
}
  \label{fig:emdcmp}
\end{figure*}

\subsection{Quantifying Detector Effects}
\label{subsec:emd_detector}

As a first application of the EMD, we investigate a novel way to quantify the impact of detector effects and pileup.
An example MC jet is shown in \Fig{fig:SIMGENexample}, where the EMD is computed between the same jet before and after detector simulation.
See \Sec{subsec:simulation} for how we associate simulation-level and generation-level jets.
Pileup is removed with CHS and a variety of PFC cuts are applied to improve the agreement between the particle-level and detector-level jets.
This is explicitly shown by the decreasing EMD as the cuts are applied, quantifying the fact that the radiation patterns within the jets are becoming more similar.

To see the impact of these cuts on the jet ensemble as a whole, in \Fig{fig:emdcmp} we histogram the EMDs between the same MC jet evaluated at generation level and simulation level.
Here, we impose $p_T^\text{jet} \in [375,425]~\text{GeV}$ on the simulation-level jet, while the generation-level jet could fall outside of this range.
We emphasize that these EMD calculations are performed \emph{after} the rescaling in \Eq{eq:EMDrescaling}, so this only quantifies the change in the radiation pattern, not the change in radiation intensity.
As emphasized in \Ref{Komiske:2019fks}, jets that are close in EMD are close in any (Lipschitz-bounded) IRC-safe measure, so small values of the generation-to-simulation EMD correspond to small differences between, for example, the generation- and simulation-level jet mass.
In this way, the EMD provides a universal bound on the impact detector effects can have on IRC-safe observables, which is a convenient alternative to studying the impact on specific observables individually.

Considering all PFCs in \Fig{fig:emdcmp_gensimEMDAll}, the generation-to-simulation EMD peaks at around \SI{17}{GeV}.
We can decrease the generation-to-simulation difference by sequentially applying CHS and the $p_T^\text{PFC} > \SI{1}{GeV}$ cut, though the impact is relatively modest.
In evaluating the EMD, the $p_T^\text{PFC} > \SI{1}{GeV}$ restriction is applied at both the generation and simulation levels.
Imposing the track-only restriction in \Fig{fig:emdcmp_gensimEMDTrack}, the generation-to-simulation EMD peak is shifted downward by a factor of about 2.
Now, CHS has a much more pronounced impact, since it decreases substantially the relative pileup contamination.
The $p_T^\text{PFC} > \SI{1}{GeV}$ cut has a modest, but non-negligible, impact.
As expected, the impact of detector effects and pileup is minimized for track-based observables after CHS.
In \Fig{fig:pileupEMDs} in \App{app:pileupreweighting}, we further investigate the performance of CHS for pileup mitigation.
In \Fig{fig:emdcmp_pT} in \App{app:additionalplots}, we investigate the impact of the $p_T^\text{PFC}$ cut in more detail.

From these studies, we conclude that our default selection (charged PFCs with $p_T^\text{PFC} > \SI{1}{GeV}$) is a reasonable compromise between reconstruction performance and substructure sensitivity.
More generally, we see that the EMD is an effective and intuitive way to quantify the impact of detector effects and pileup contamination.

\subsection{Visualizing the Space}
\label{subsec:emd_tsne}

It is interesting to directly visualize the metric space of jets defined by EMD.
There are a variety of techniques to visualize high-dimensional data in low dimensions, which provide a fascinating way to see the broad features of a dataset.
Here, we apply t-Distributed Stochastic Neighbor Embedding (t-SNE)~\cite{vanDerMaaten2008,DBLP:journals/jmlr/Maaten09,DBLP:journals/ml/MaatenH12,DBLP:journals/jmlr/Maaten14}, which finds a low-dimensional embedding of the data in a way that respects the distance between data points.
We run t-SNE with a two-dimensional embedding space, in which the procedure defines two axes and attempts to place data points in this two-dimensional plane in such a way that jets close in EMD are nearby and jets far in EMD are distant.

\begin{figure*}
  \subfloat[]{
  \label{fig:tSNE_DATA}
    \includegraphics[width=0.7\textwidth]{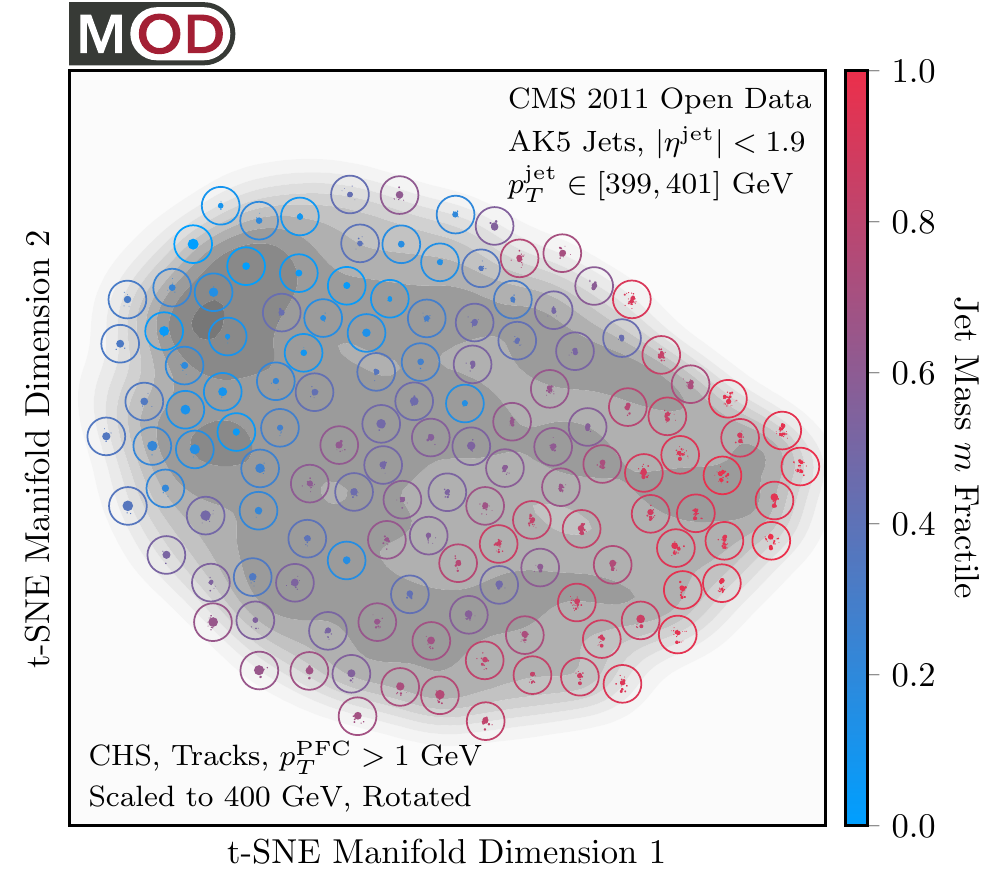}
	}\\
  \subfloat[]{
    \includegraphics[width=0.495\textwidth]{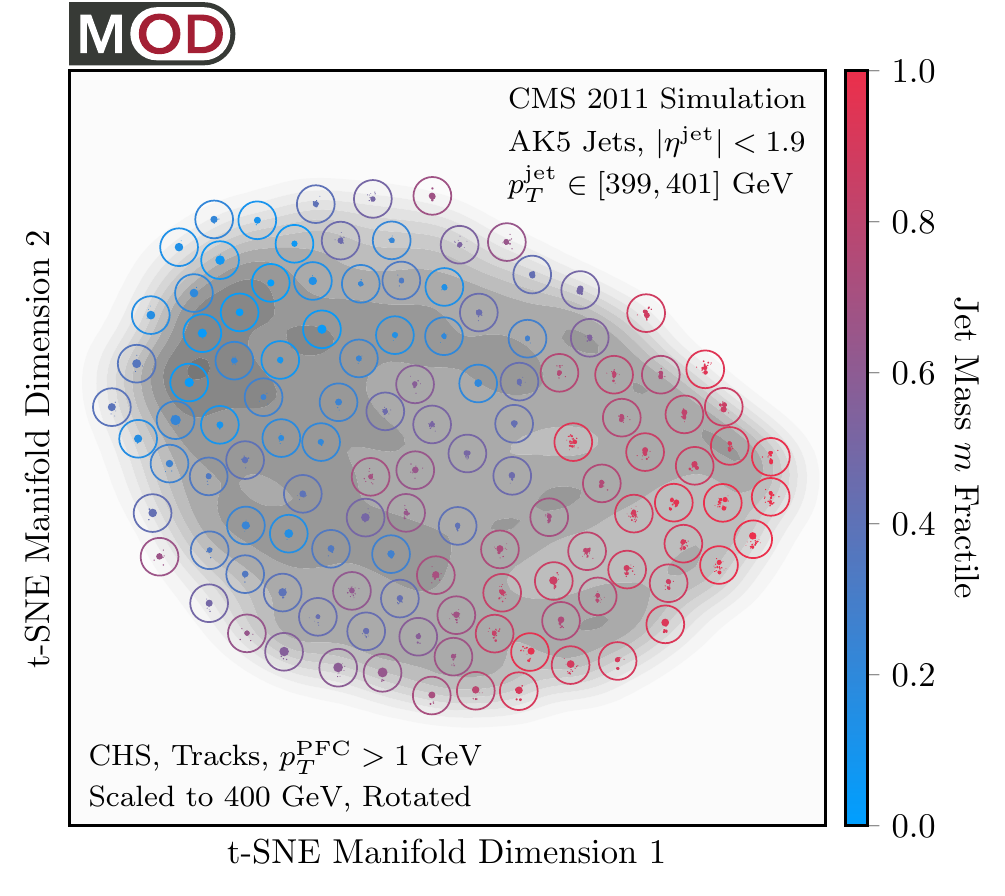}
  \label{fig:tSNE_SIM}
	}
\subfloat[]{
    \includegraphics[width=0.495\textwidth]{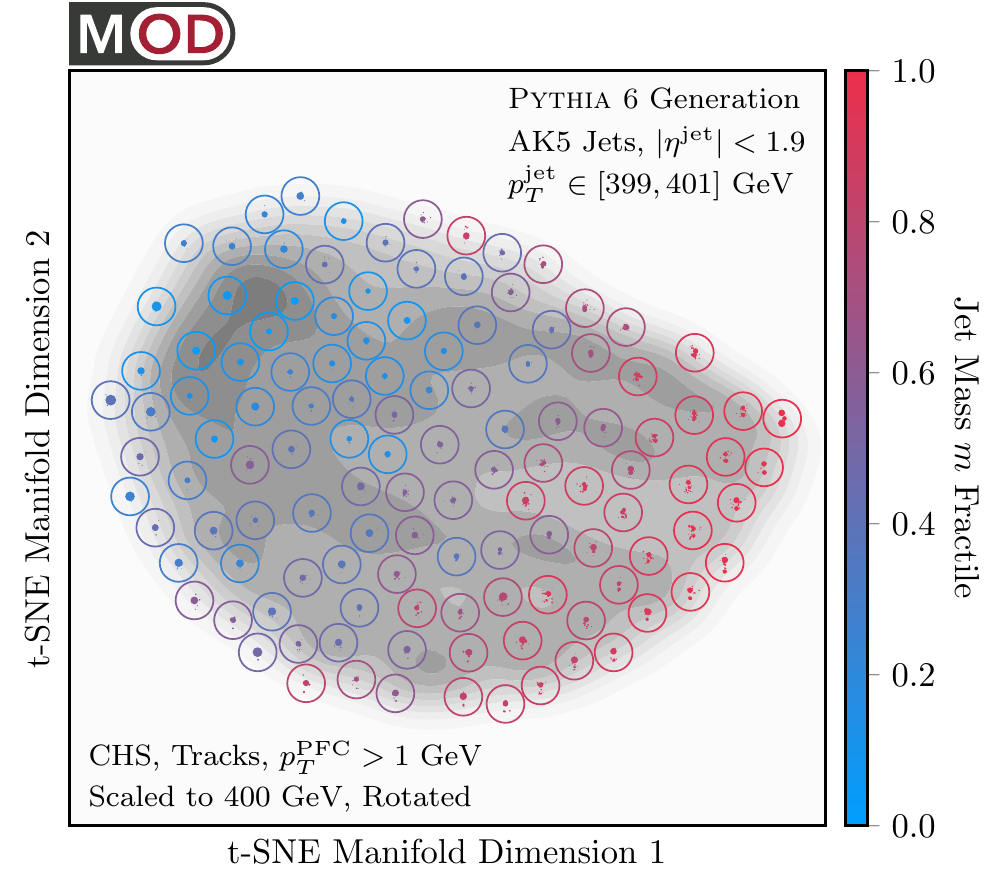}
\label{fig:tSNE_GEN}
	}
  \caption{
  Two-dimensional t-SNE embedding of jets in the $p_T^\text{jet}\in[399,401]~\text{GeV}$ range from the (a) CMS Open Data, (b) simulation-level MC, and (c) generation-level MC.
  The gray contours indicate the density of embedded jets, and the example jets are color coded by the jet mass fractile in the corresponding dataset.
  }
    \label{fig:tSNE}
\end{figure*}

Though there are techniques to implement t-SNE on $N$ data points in $\mathcal{O}(N \log N)$ runtime~\cite{DBLP:journals/jmlr/Maaten14}, due to current limitations in the {\tt scikit-learn}~\cite{scikit-learn} implementation that we use, we have to perform $\mathcal{O}(N^2)$ operations.
To make this computationally tractable, we restrict our attention to the $p_T^\text{jet}\in[399,401]\,\text{GeV}$ range, which yields approximately 40,000 jets in the CMS Open Data.
We also subsample and unweight the MC events to obtain around 40,000 generation-level and 40,000 simulation-level jets as well.
(Because there are insufficient events in the $\hat{p}_T \in [170,300]$\,GeV MC sample~\cite{CMS:QCDsim170-300}, we have to downweight them by a factor of around 10 to achieve an approximately unweighted sample.)
We apply CHS, the $p_T^\text{PFC} > \SI{1}{GeV}$ cut, and the track-only restriction on all jets.
To reduce the effective dimensionality of the dataset and remove a trivial isometry, we rotate the jets around the jet axis such that the principle component of the transverse radiation pattern is aligned vertically in the rapidity-azimuth plane, breaking the two-fold degeneracy by enforcing that the jet has more scalar sum $p_T$ at positive azimuth.
We also keep only the particles within a jet radius of the jet axis.

The results of t-SNE embedding into a two-dimensional space are shown in \Fig{fig:tSNE}, for the CMS Open Data and for the simulation-level and generation-level MC samples.
For visual clarity, we rotate the t-SNE manifold such that the three embeddings exhibit roughly the same large-scale structure.
The gray contours represent the density of the embedded jets.
Example jets are sprinkled throughout the space and color coded by their jet mass fractile (i.e.~fraction of events with smaller jet mass than the color coded value).

For the CMS Open Data in \Fig{fig:tSNE_DATA}, the t-SNE embedding exhibits a dominant cluster of jets with typically low jet mass, with a long slope extending out to typically higher jet masses.
The most exotic jets are furthest away from the dominant cluster.
The t-SNE embeddings of the MC samples in \Figs{fig:tSNE_SIM}{fig:tSNE_GEN} are qualitatively similar, though the specific density distributions differ.
Using smaller jets samples, we find that the variability between the data and MC t-SNE embeddings is comparable to the variability when running t-SNE multiple times on the same sample.
No obvious anomalies in the CMS Open Data appear visually, though we return to anomalous jet configurations in \Sec{subsec:emd_anomaly}.

\subsection{Correlation Dimension}
\label{subsec:emd_dimension}

\begin{figure*}[p]
\subfloat[]{
\includegraphics[width=0.5\textwidth]{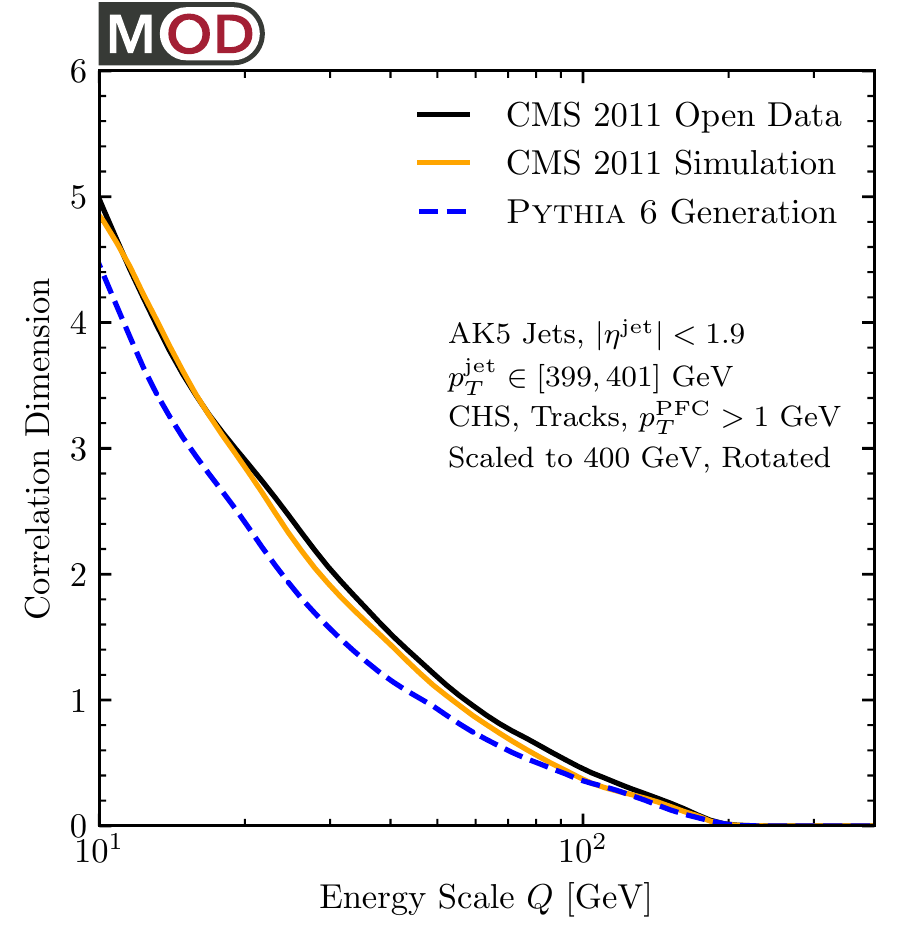}
\label{fig:cordim_comp}
}\\
\subfloat[]{
    \includegraphics[width=0.325\textwidth]{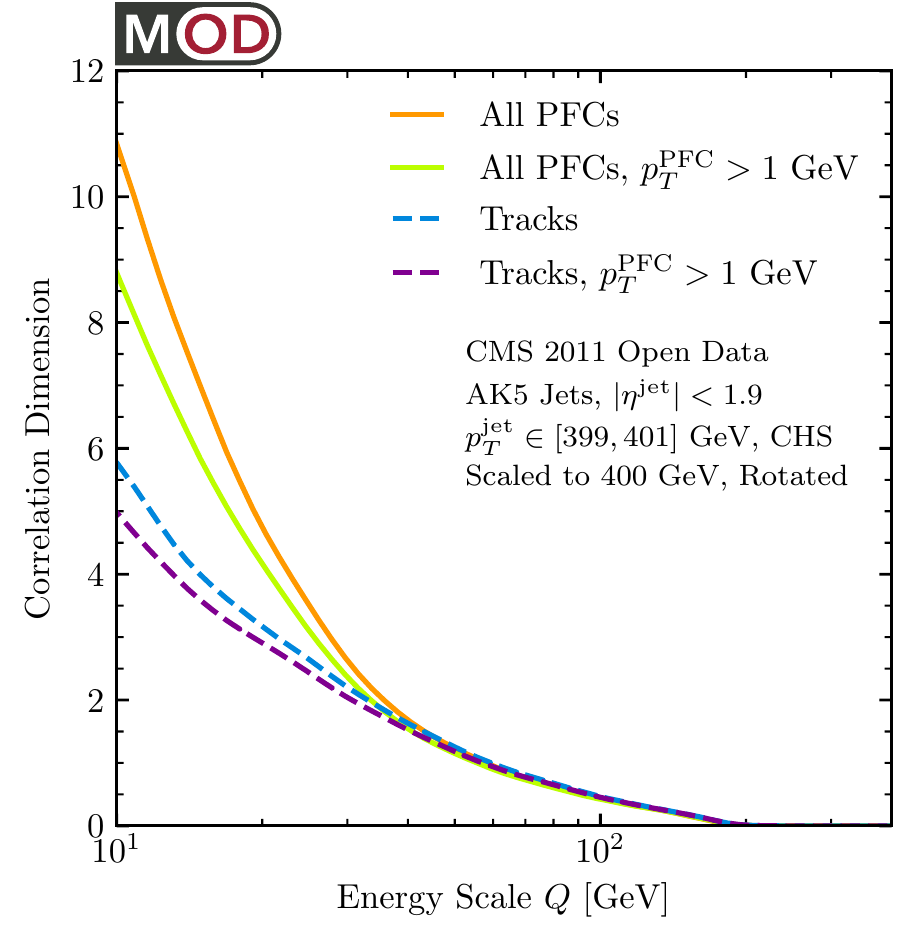}
  \label{fig:cordim_cms}
	}
\subfloat[]{
    \includegraphics[width=0.325\textwidth]{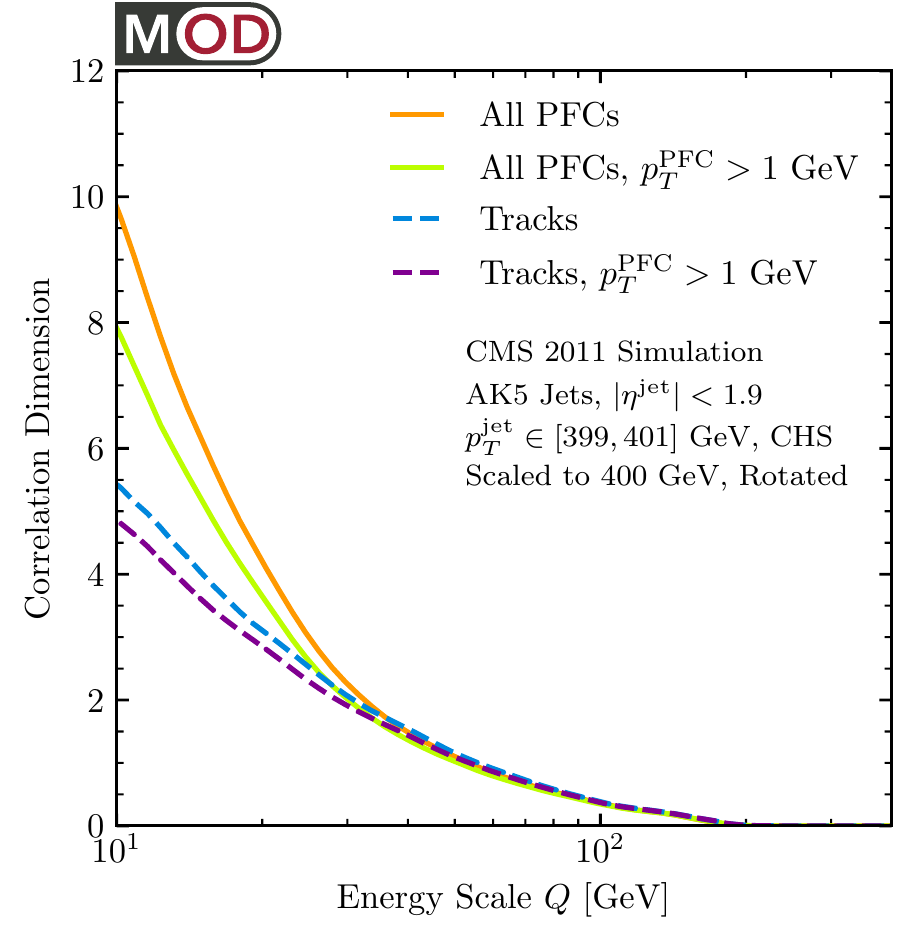}
  \label{fig:cordim_sim}
	}
\subfloat[]{
    \includegraphics[width=0.325\textwidth]{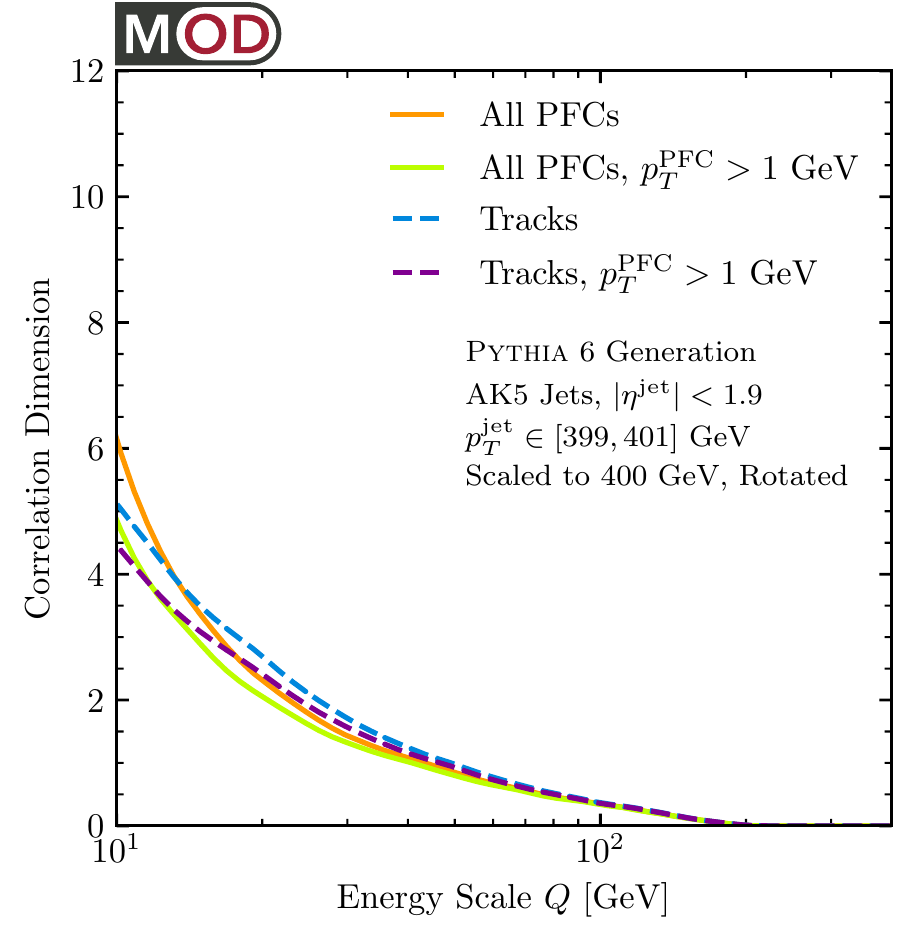}
  \label{fig:cordim_gen}
	}
\caption{The correlation dimension of the space of jets as a function of energy scale $Q$, (a) comparing the CMS Open Data to the generation-level and simulation-level MC samples.
There is good agreement between the MC simulation-level and the open data, while the MC generation-level jets have a systematically smaller correlation dimension over much of the energy range.
 Also shown are different PFC selections in the (b) CMS Open Data, (c) simulation-level MC, and (d) generation-level MC which either impose the $p_T^\text{PFC} > \SI{1}{GeV}$ cut or restrict to only tracks or both.
In all cases, the high-energy limit of the correlation dimension is robust to the PFC selection, with significant differences only appearing for $Q \lesssim \SI{40}{GeV}$.
}
\label{fig:corrdim_all}
\end{figure*}

To gain more quantitative insight into the space of jets, we can use the EMD to compute its dimensionality.
While a variety of definitions exist for intrinsic dimension, we use the correlation dimension~\cite{Grassberger:1983zz,DBLP:conf/nips/Kegl02}, which is a type of fractal dimension and was the measure used in \Ref{Komiske:2019fks}.
From a matrix of pairwise EMDs between jets, the correlation dimension is defined as:
\begin{equation}
\label{eq:corrdim}
\dim(Q) = Q\frac{\partial}{\partial Q}\ln\hspace{-0.5em}\sum_{1\le k < \ell \le N}\hspace{-0.5em}\Theta\big[\text{EMD}(\mathcal J_k, \mathcal J_\ell) < Q\big].
\end{equation}
Here, $N$ is the total number of jets in the sample and the Heaviside theta function indicates whether the jet $k$ is within an EMD $Q$ of jet $\ell$.
To gain an intuition for this formula, note that for a uniform data sample in $d$ dimensions, the expected number of neighbors $B$ within a ball of radius $Q$ scales like $Q^d$, such that $d \simeq \partial \ln B / \partial \ln Q$.
The expression in \Eq{eq:corrdim} has this same relation, where the number of neighbors $B$ is averaged over balls of radius $Q$ centered around each data point.

The computational cost of implementing \Eq{eq:corrdim} is $\mathcal{O}(N^2)$, so we restrict our attention to the same $p_T^\text{jet}\in[399,401]~\text{GeV}$ subsample as in \Sec{subsec:emd_tsne}.
(Because it is straightforward to compute $\dim(Q)$ using MC weights, this time we do not need to downweight the $\hat{p}_T \in [170,300]$\,GeV MC sample~\cite{CMS:QCDsim170-300}.)
We also perform the same jet rotation in \Sec{subsec:emd_tsne}.

After the rescaling in \Eq{eq:EMDrescaling}, the maximum possible value of the EMD is \SI{400}{\GeV}, so $\dim(Q)$ always equals zero for $Q > \SI{400}{\GeV}$.
Because we cluster jets with the anti-$k_T$ algorithm, though, the jet configurations that could in principle lead to this maximum EMD value are not present in our samples.
For example, consider two jets of equal scalar sum $p_T$: one consists of a single particle; the second consists of two particles, each with transverse momentum $p_T/2$, separated by $\Delta R$.
The EMD between these configurations is $\frac{1}{2} p_T \Delta R$.
Within a jet region of size $R$, $\Delta R$ could in principle be as large as $2R$ (i.e.~EMD as large as $p_T$), but anti-$k_T$ would split the second jet in two unless $\Delta R < R$ (i.e.~EMD of $p_T/2$).
In practice, we find that $\dim(Q)$ indeed goes to zero around $Q \simeq \SI{200}{\GeV}$.

In \Fig{fig:cordim_comp}, we compare the correlation dimension between the CMS Open Data and the MC samples, again with CHS and tracks only with $p_T^\text{PFC} > \SI{1}{GeV}$.
The agreement between the open data and the MC sample at simulation-level is very good, though the correlation dimension is roughly 0.5 above the generation-level curve for much of the plotted $Q$ range.
Naively, one might think that detector effects would decrease the correlation dimension, since finite granularity effects decrease the relative complexity of jet configurations.
Instead, the added half dimension suggests that the detector has more of a smearing effect, analogous to the way that smearing a zero-dimensional point generates a higher-dimensional manifold.

The fact that the correlation dimension in \Fig{fig:corrdim_all} increases logarithmically with decreasing $Q$ is expected from first principles QCD.
The number of jet constituents scales up logarithmically with decreasing energy scale (see e.g.~\cite{Bolzoni:2012ii,Bolzoni:2013rsa}), as does the entropy of a jet~\cite{Neill:2018uqw}, and both of these quantities are related to the effective dimensionality of the space of QCD jets.
We leave a QCD calculation of $\dim(Q)$ to future work, noting that the result will depend on the strong coupling constant $\alpha_s$ as well as on the relative fraction of quark and gluon jets in the sample.

The correlation dimension gives us an interesting handle to understand the impact of applying cuts on the PFCs, complementary to the studies in \Sec{subsec:emd_detector}.
In the bottom row of \Fig{fig:corrdim_all}, we show $\dim(Q)$ for all PFCs and just tracks, as well as the effect of the $p_T^\text{PFC} > \SI{1}{GeV}$ cut, always with CHS applied.
For the CMS Open Data in \Fig{fig:cordim_cms} and for the simulation-level MC in \Fig{fig:cordim_sim}, there is relatively little impact on the correlation dimension for $Q \gtrsim \SI{40}{GeV}$.
Below this scale, though, the correlation dimension is significantly smaller when restricting to just tracks and/or when imposing $p_T^\text{PFC} > \SI{1}{GeV}$.
Interestingly, for the generation-level curves in \Fig{fig:cordim_gen}, there is a much more modest impact from these restrictions.
In fact, restricting to charged PFCs can sometimes \emph{increase} the correlation dimension, since after applying the rescaling in \Eq{eq:EMDrescaling}, the charged PFC restriction acts like a kind of smearing.
From this we conclude that $\dim(Q)$ is a robust measure of dimensionality at high $Q$, and very sensitive to QCD fragmentation and detector effects at small $Q$.

\subsection{The Most Representative Jets}
\label{subsec:emd_representative}

\begin{figure}[t]
\includegraphics[width=\columnwidth]{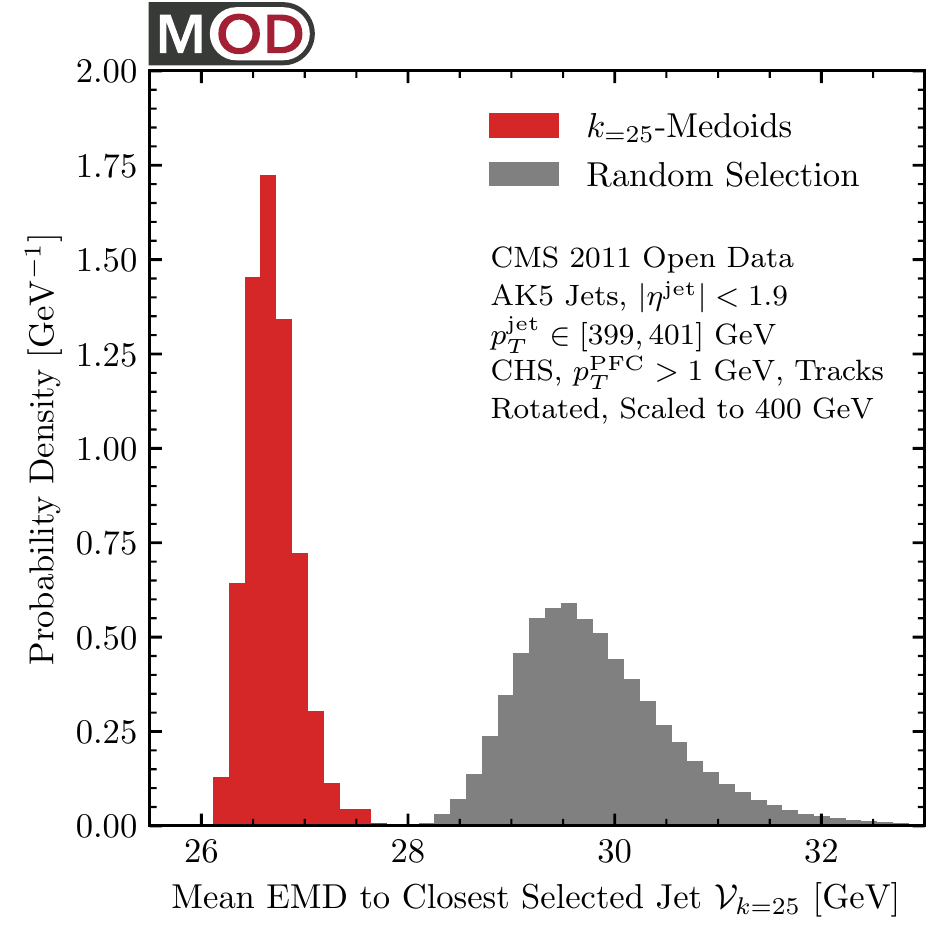}
\caption{
\label{fig:medoidsloss}
The distance of our jet dataset to a selection of 25 representative jets, shown for (red) jets selected with the $k$-medoids algorithm as well as (gray) randomly selected jets.
The $k$-medoids are systematically closer to the dataset, demonstrating that jets chosen in this way are significantly more representative than a random selection of jets.
}
\end{figure}

Computing the EMD also allows us to visualize the space of jets in such a way that observable values can be correlated with jet topologies.
Specifically, given a set of jets, we can find the $k$ jets $\{\mathcal K_1,\cdots,\mathcal K_k\}$ (called medoids) that minimize the sum of the distances of each jet to its closest medoid:
\begin{equation}
\label{eq:kmedoidsloss}
\hspace{-1.9mm}\mathcal{V}_k =\frac1N \sum_{i=1}^N \min\!\big\{\text{EMD}(\mathcal J_i, \mathcal K_1),\ldots,\text{EMD}(\mathcal J_i, \mathcal K_k)\big\}.
\end{equation}
The value of \Eq{eq:kmedoidsloss} provides a quantitative notion of how well approximated the dataset is by the $k$ jets.
Inspired by the $N$-subjettiness observables of \Ref{Thaler:2010tr,Thaler:2011gf}, this quantity can be thought of as the ``$k$-eventiness'' of the dataset.

While naively optimizing the choice of the medoids takes $\mathcal O(N^{K+1})$ runtime, we use a fast iterative approximation techniques from the {\tt pyclustering} \textsc{Python} package~\cite{andrei_novikov_2018_1491324}.
This $k$-medoids procedure provides a significantly more representative selection of jets than a random subsample, as quantified by the $\mathcal{V}_k$ distribution in \Fig{fig:medoidsloss} for the case of $k=25$.
Along these lines, one might also consider clustering the full dataset of jets, for instance using iterative reclustering similar to techniques used to cluster particles into jets~\cite{Ellis:1993tq,Catani:1993hr,Dokshitzer:1997in,Wobisch:1998wt,Cacciari:2008gp}, though we leave further explorations in this direction to future work.

\begin{figure*}[p]
\centering
\subfloat[]{
\includegraphics[width=0.7\textwidth]{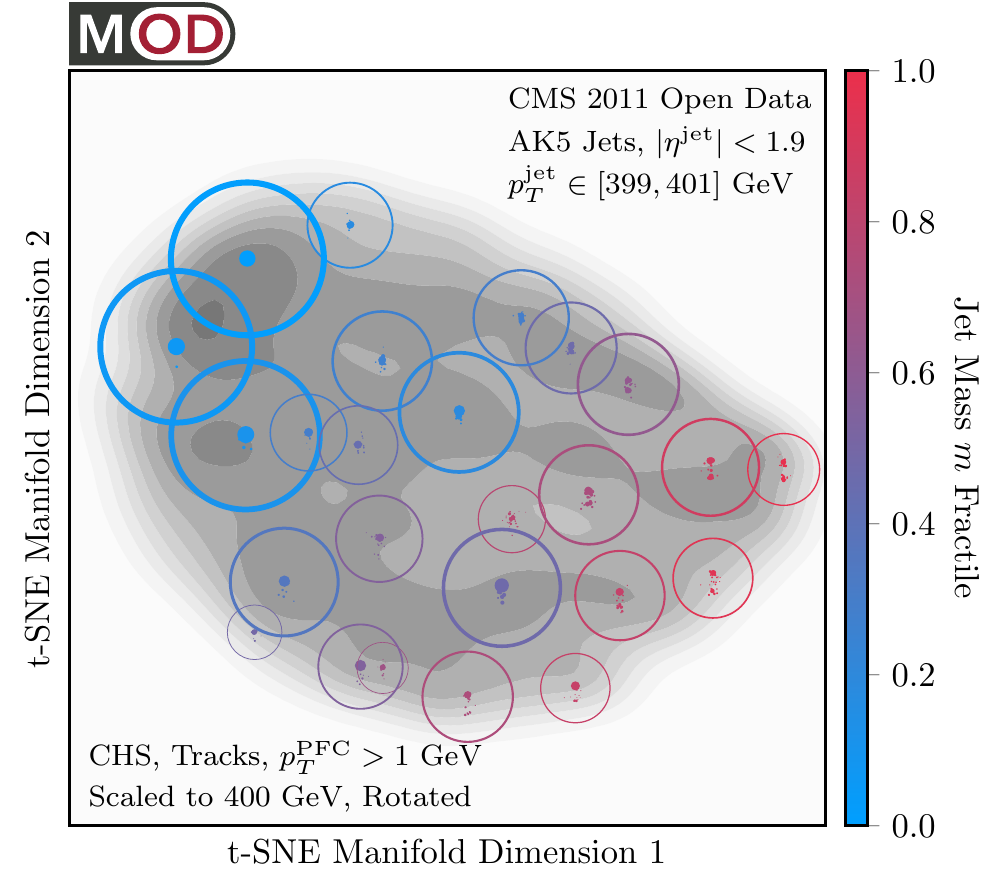}
}\\
\subfloat[]{
\includegraphics[width=0.495\textwidth]{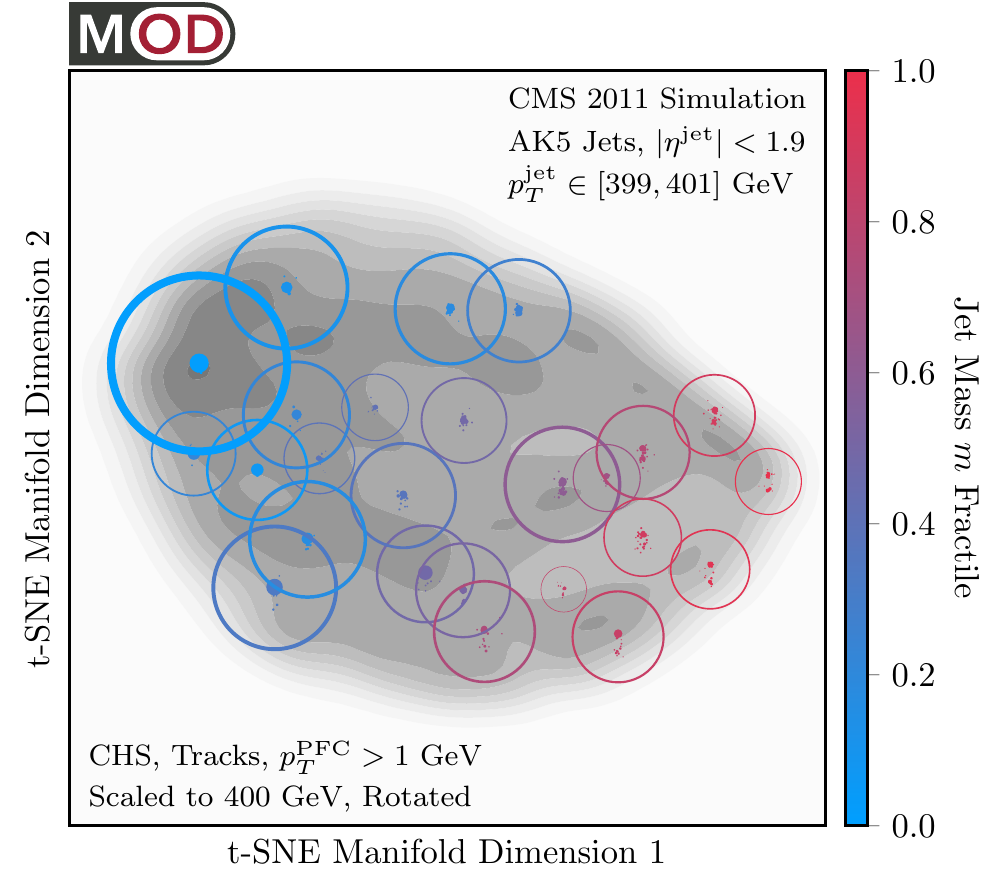}
}
\subfloat[]{
\includegraphics[width=0.495\textwidth]{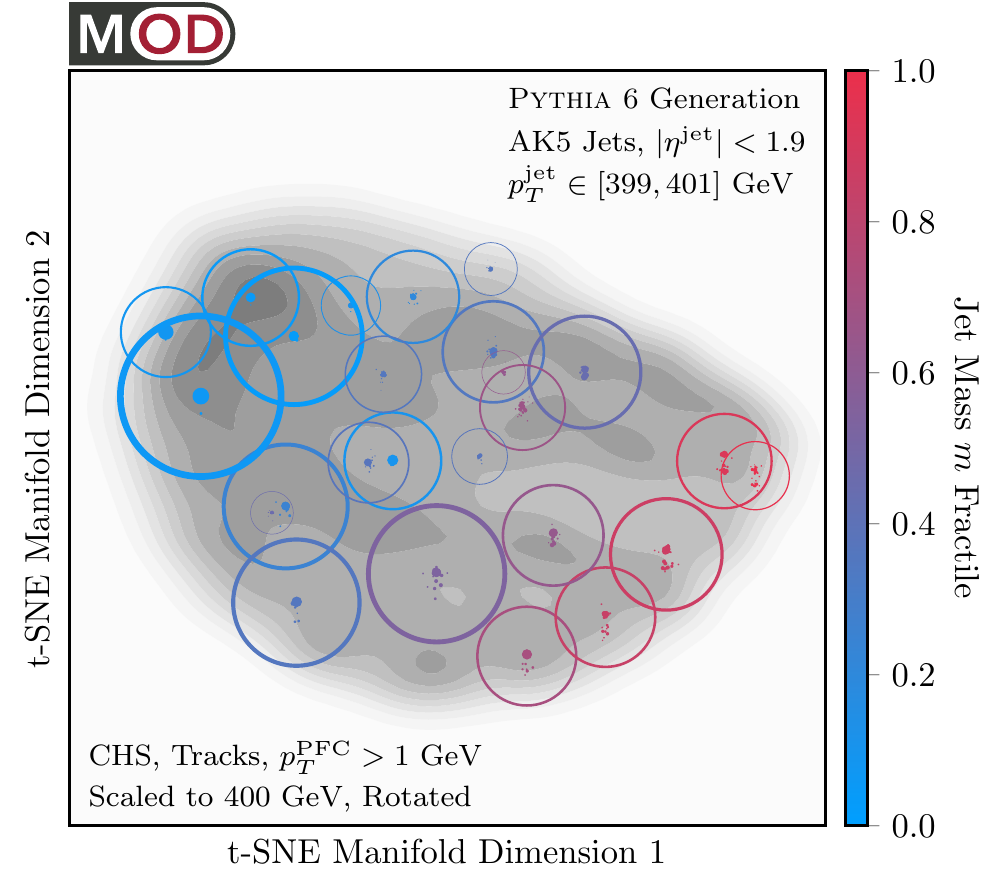}
}
\caption{
The 25 most representative jets (medoids) in the (a) CMS Open Data, (b) simulation-level MC, and (c) generation-level for $p_T^\text{jet}\in[399,401]~\text{GeV}$.
The jets are arranged according the t-SNE algorithm as in \Fig{fig:tSNE} and their area is proportional to the number of jets nearest to them.
The medoid jets try to ``tile'' the space in a rigorous sense.
}
\label{fig:kmedoids_tSNE_wholesample} 
\end{figure*}

\begin{figure*}[p]
  \subfloat[]{
    \includegraphics[width=0.725\columnwidth]{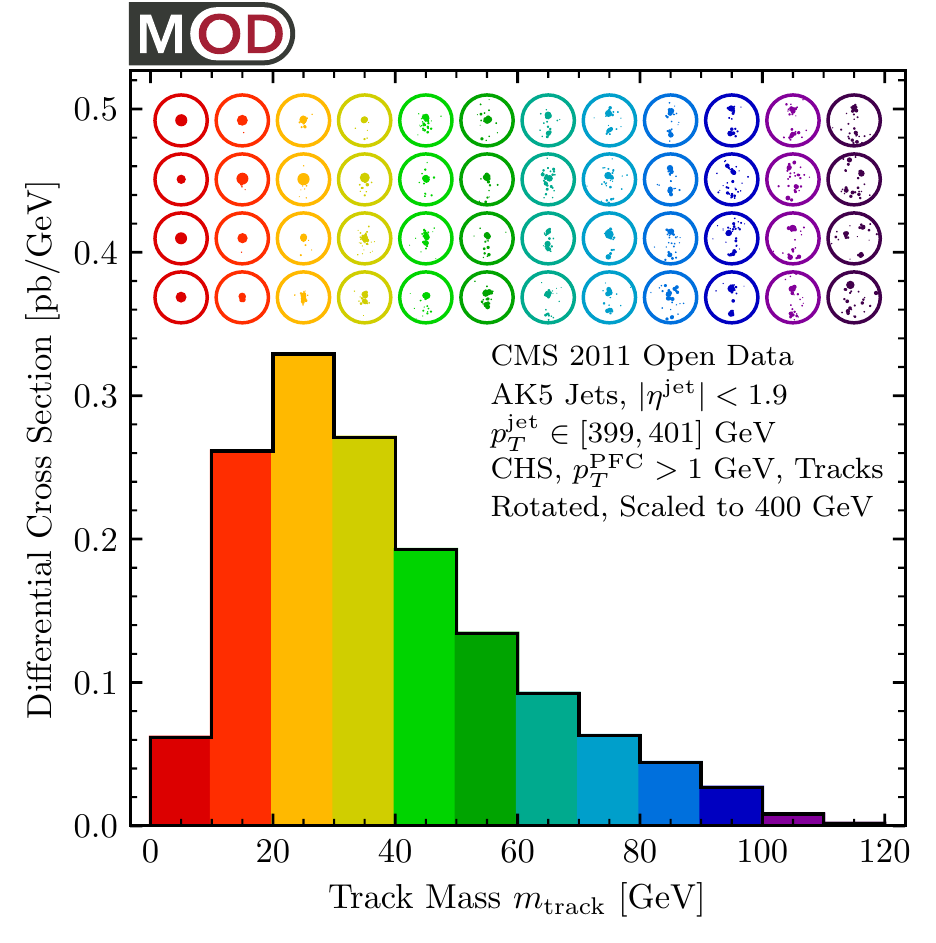}
  \label{fig:kmedoids_cms_mass}
	}
\subfloat[]{
  \label{fig:kmedoids_cms_D2}
    \includegraphics[width=0.725\columnwidth]{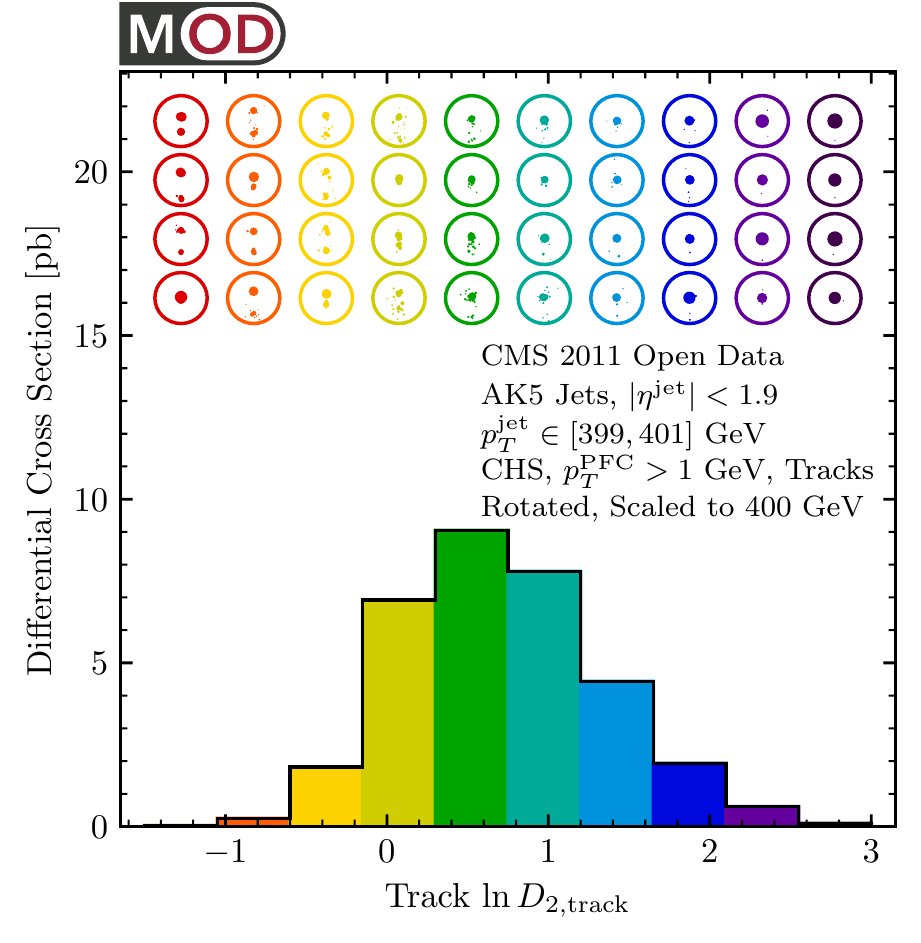}
	}\\
\subfloat[]{
  \label{fig:kmedoids_cms_track}
    \includegraphics[width=0.725\columnwidth]{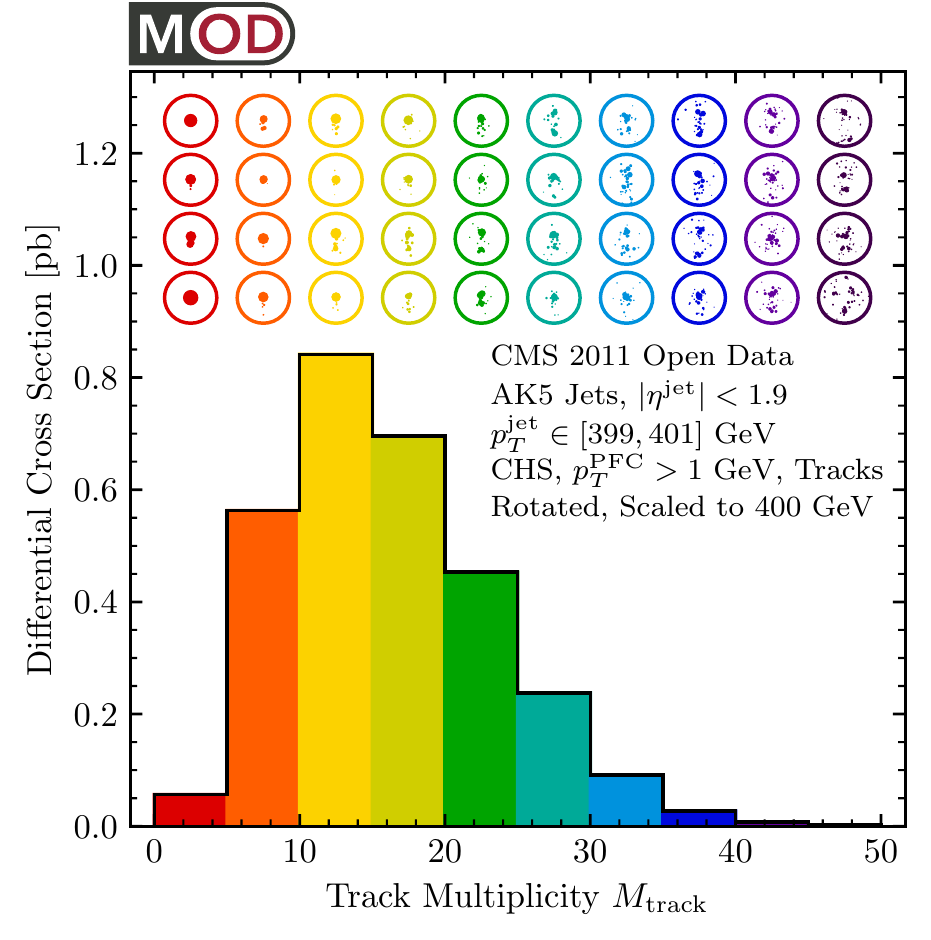}
	}
\subfloat[]{
  \label{fig:kmedoids_cms_N95}
    \includegraphics[width=0.725\columnwidth]{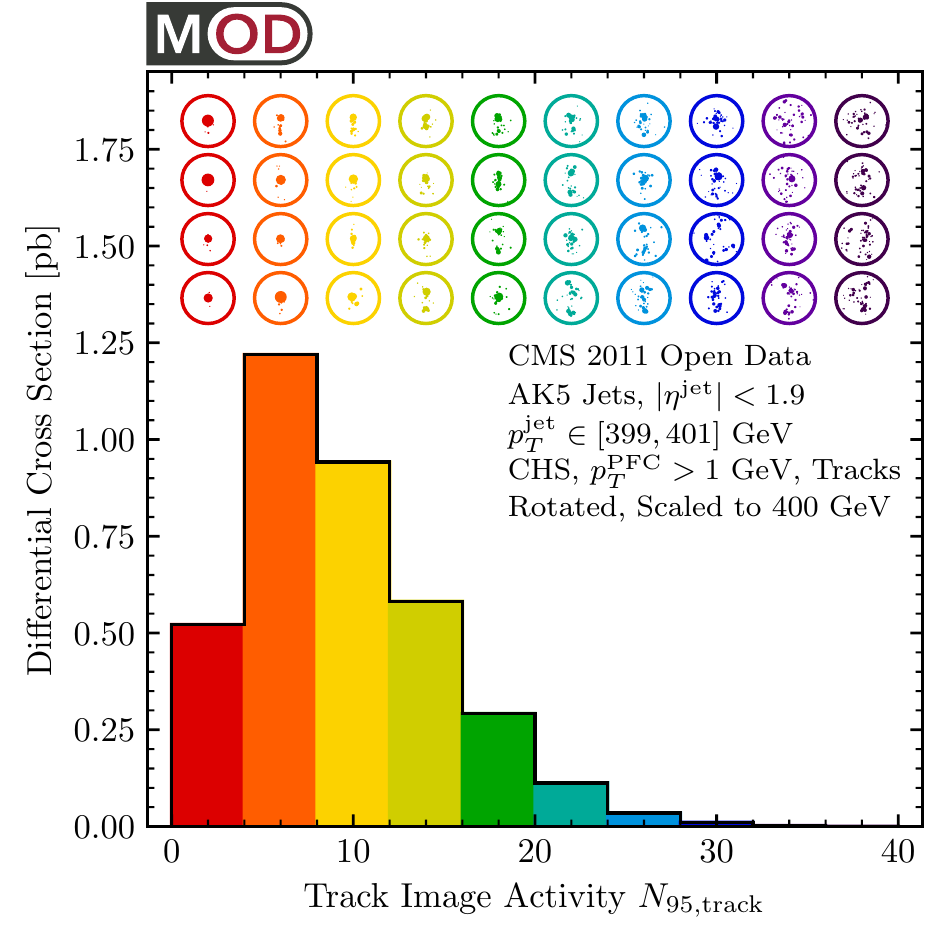}
	}\\
\subfloat[]{
  \label{fig:kmedoids_cms_ptd}
    \includegraphics[width=0.725\columnwidth]{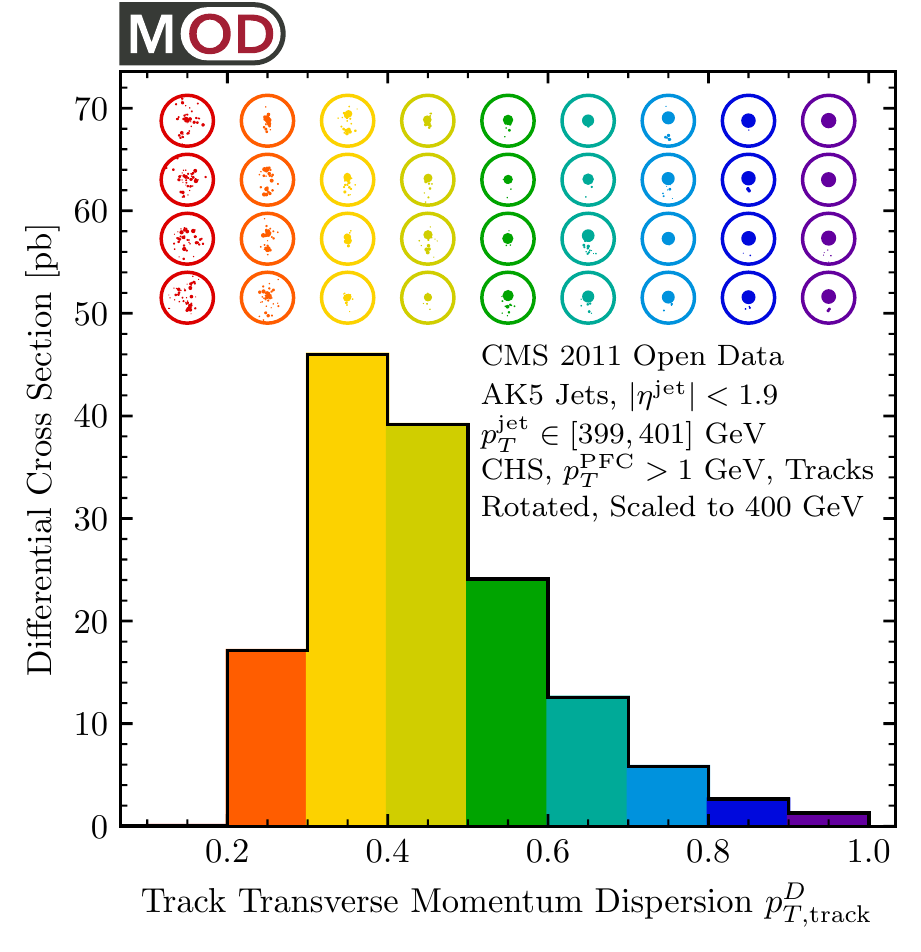}
	}
\subfloat[]{
  \label{fig:kmedoids_cms_zg}
    \includegraphics[width=0.725\columnwidth]{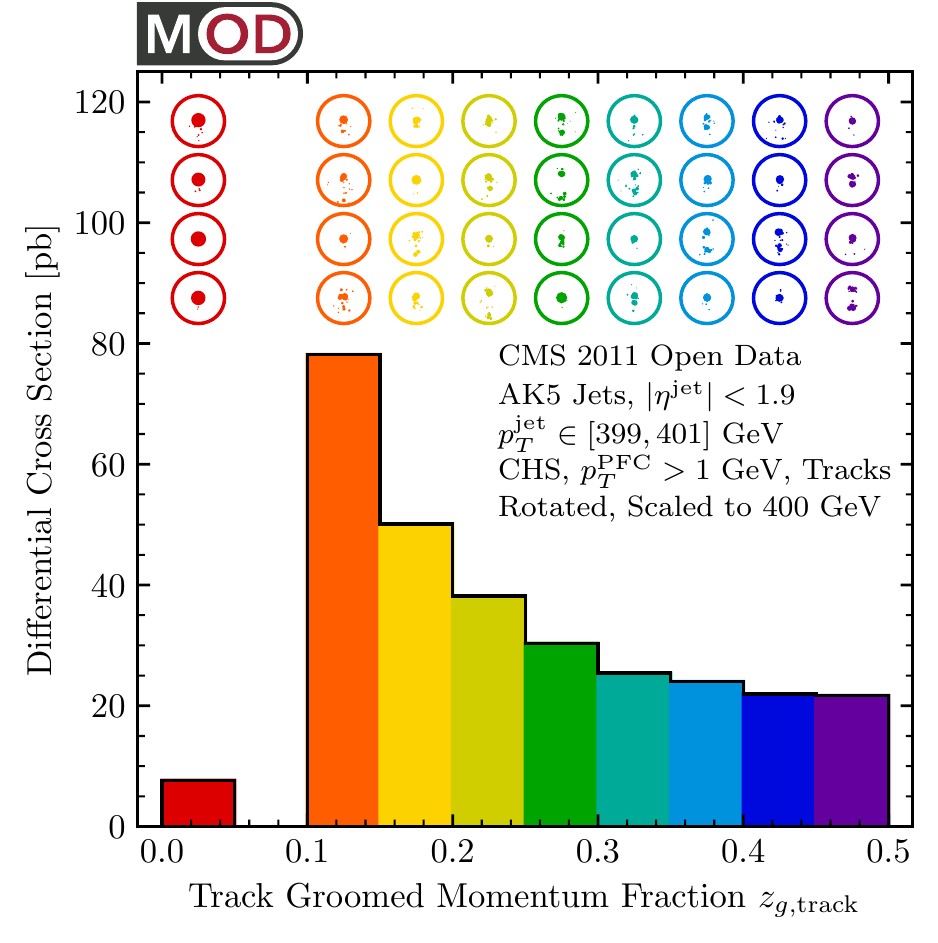}
	}
  \caption{The same jet substructure observables from \Sec{subsec:subobs}, but now showing the four most representative jets (medoids) in each histogram bin.
  These distributions are obtained from the CMS Open Data after applying CHS, the $p_T^\text{PFC} > \SI{1}{GeV}$ cut, the track-only restriction, as well as the rotation and rescaling in \Eq{eq:EMDrescaling}.
  As in  \Fig{fig:substructureobs}, we show (a) jet mass, (c) track multiplicity, and (e) $p_T^D$.
  As in  \Fig{fig:substructureobs2}, we show (b) $D_2$, (d) $N_{95}$, and (f) $z_g$.
Track multiplicity and $p_T^D$ are IRC-unsafe observables, and hence are not fully described by the energy flow in the jet.
}
  \label{fig:kmedoids_cms_all}
\end{figure*}

\begin{figure*}[t]
\centering
  \subfloat[]{
\includegraphics[height=0.55\textwidth]{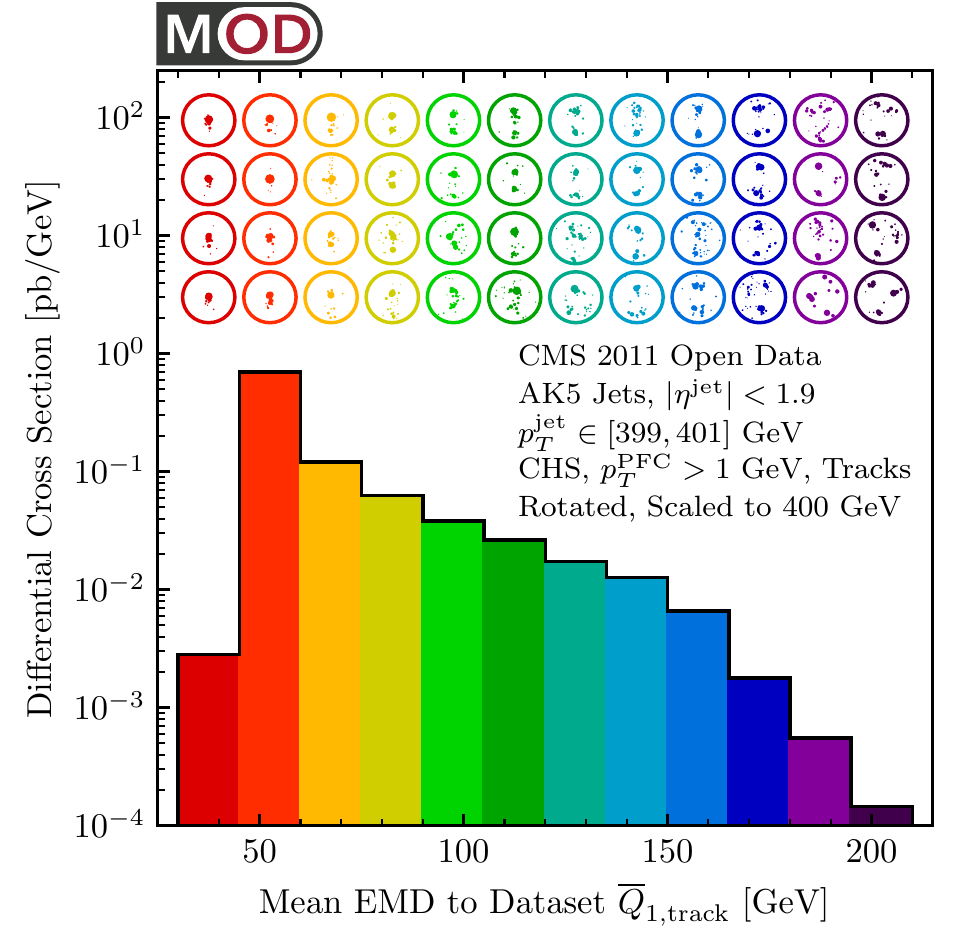}
}
\\
  \subfloat[]{
\includegraphics[height=0.3\textwidth]{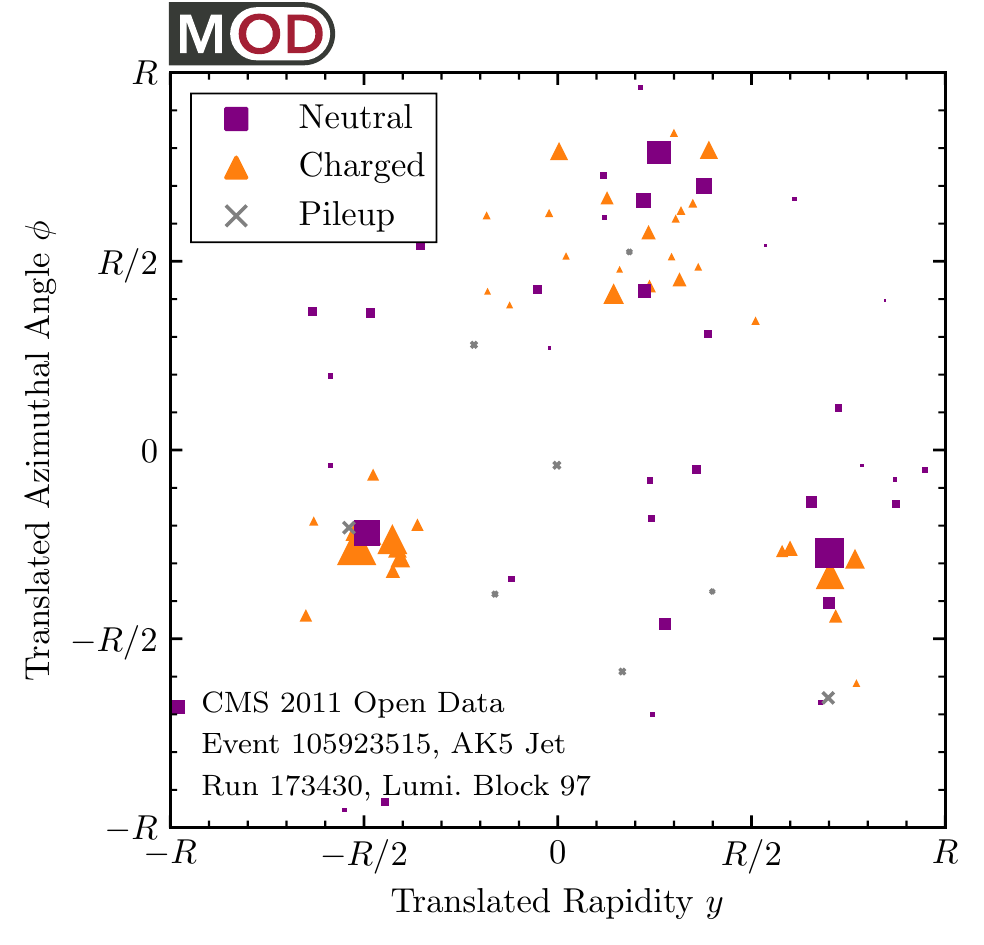}
}
  \subfloat[]{
\includegraphics[height=0.3\textwidth]{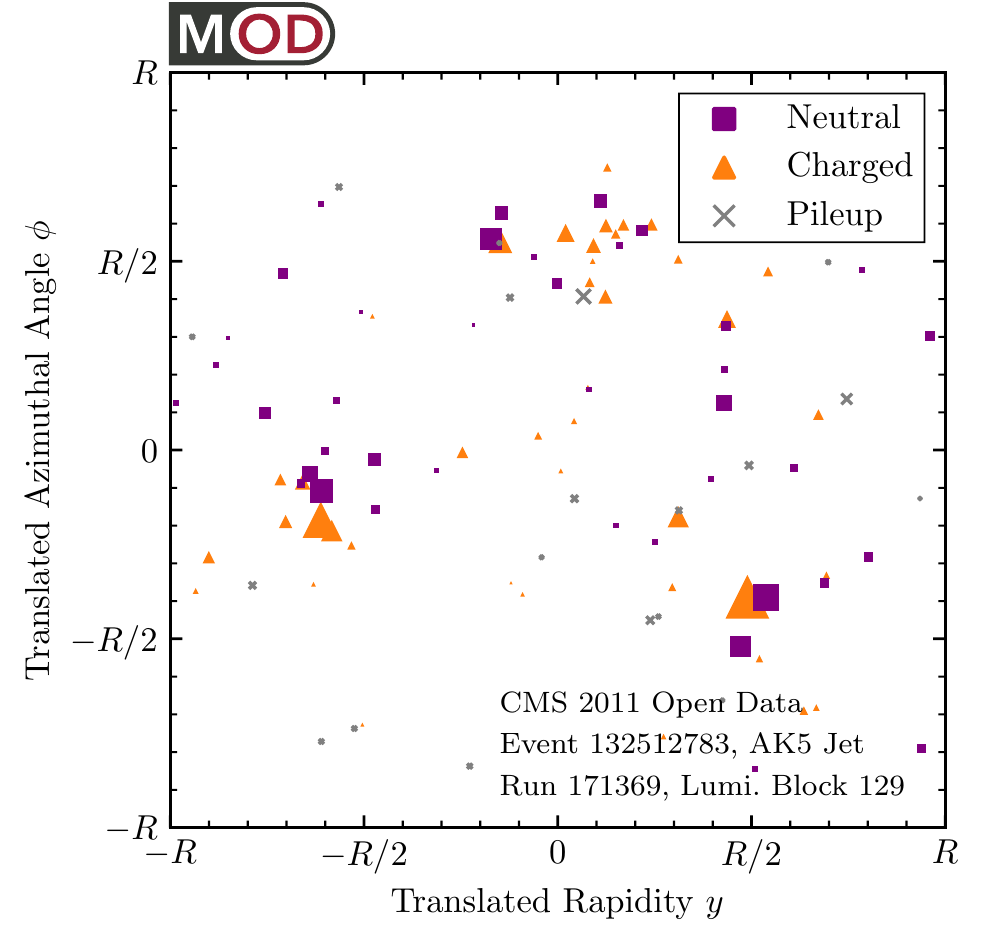}
}
  \subfloat[]{
\includegraphics[height=0.3\textwidth]{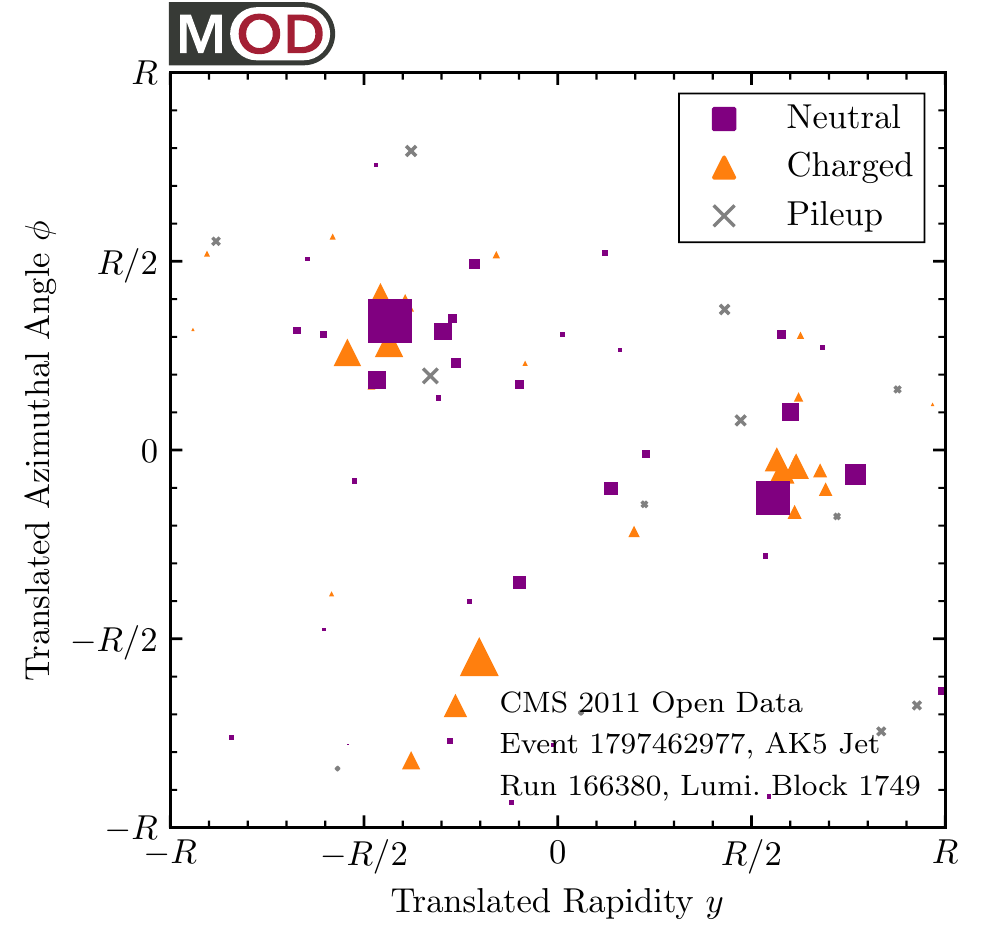}
}
\caption{
(a) Distribution on the CMS Open Data of $\overline{Q}_1$ from \Eq{eq:barQdef} along with the 4-medoids in each histogram bin.
The most typical (atypical) jets in the dataset have small (large) values of $\overline{Q}_1$.
Event displays are shown for the (b) most, (c) second most, and (d) third most anomalous jets in our CMS Open Data sample.
}
\label{fig:kmedoidsemd} 
\end{figure*}

In \Fig{fig:kmedoids_tSNE_wholesample}, we show the 25 most representative jets in the $p_T^\text{jet}\in[399,401]~\text{GeV}$ subsample from \Sec{subsec:emd_tsne}, arranged according to t-SNE and sized according the number of closest neighbors.
Because these medoids are representative (and not just randomly selected) in that they try to minimize $\mathcal{V}_k$, there is a rigorous sense in which understanding the structure of these 25 jets captures the structure of the CMS Open Data jet ensemble as a whole.

If we apply the $k$-medoid procedure to jets occupying the same histogram bins of a specific observable, we can then visualize how the jet topology changes as observable values change.
In \Fig{fig:kmedoids_cms_all}, we show histograms for the six substructure observables from \Sec{subsec:subobs}, using the CMS Open Data with CHS and only tracks with $p_T^\text{PFC} > \SI{1}{GeV}$.
In each histogram bin, we show the four most representative jets, as determined by the $4$-medoids procedure.
For jet mass in \Fig{fig:kmedoids_cms_mass}, we see a steady evolution from one-prong topologies to two-prong topologies.
The reverse behavior is shown for $D_2$ in \Fig{fig:kmedoids_cms_D2}, with two-prong topologies evolving into one-prong ones.
One low-$D_2$ medoid jet consists of two highly overlapping prongs, distinct from the one-prong high-$D_2$ configurations, highlighting the Sudakov safety of $D_2$~\cite{Larkoski:2014gra,Larkoski:2015lea}.
For the IRC-unsafe observables of track multiplicity in \Fig{fig:kmedoids_cms_track} and $p_T^D$ in \Fig{fig:kmedoids_cms_ptd}, we see evolutions between simple topologies and jets with more complex substructure.
For $N_{95}$ in \Fig{fig:kmedoids_cms_N95}, there is a progression from narrow jets to diffuse jets.
Finally, for $z_g$ in \Fig{fig:kmedoids_cms_zg}, there is an evolution from unbalanced subjets to balanced subjets, with its Sudakov safety apparent from the one-prong configurations throughout.
While all of these behaviors can be understood from the definition of these observables, the $k$-medoids procedure offer an intuitive visualization of the jet configurations that contribute to each observable value.

\subsection{Towards Anomaly Detection}
\label{subsec:emd_anomaly}

As the last application of the EMD in this paper, we present a first step towards using it for anomaly detection.
Instead of finding the most representative jets as in \Sec{subsec:emd_representative}, we can find the least representative jets.
As one way to quantify this, we can find the $n$-th moment of the EMD distribution of one jet to the rest of the dataset,
\begin{equation}
\label{eq:barQdef}
\overline{Q}_n(\mathcal{I}) = \sqrt[n]{
\frac{1}{N} \sum_{k=1}^N \Big( \text{EMD}(\mathcal{I},\mathcal{J}_k) \Big)^n,
}
\end{equation}
where we applied the $n$-th root such that $\overline{Q}_n$ has units of GeV.
Small values of $\overline{Q}_n$ indicate a common jet configuration.
Large values of $\overline{Q}_n$ indicate a jet which is far from the rest of the dataset, and therefore anomalous.

In \Fig{fig:kmedoidsemd}, we show the distribution of $\overline{Q}_n$ for $n =1$ (i.e.~mean EMD) along with the four medoids in each histogram bin.
As expected from the t-SNE visualization in \Fig{fig:tSNE}, the most typical jet configurations have a single hard prong, while the least typical configurations have multi-prong or diffuse topologies.
In \Fig{fig:kmedoidsemdapp} of \App{app:additionalplots}, we show a simular plot for $n=\frac12$ and $n=2$.
The most anomalous jets isolated by $\overline{Q}_n$ for $n=\frac12$, $1$, and $2$ agree for the six most anomalous jets, with the top three such jets shown in the bottom row of \Fig{fig:kmedoidsemd}.
The most anomalous jets are all highly complex three-prong topologies, hinting at a close relationship between this measure of anomalousness and observables such as $N$-subjettiness~\cite{Thaler:2010tr,Thaler:2011gf}.

The anomalousness of a jet, quantified by $\overline{Q}_n$, is nontrivially correlated with the jet mass, which is easily confirmed by observing the medoids in each bin in \Fig{fig:kmedoidsemd}.
While this is expected and understandable from QCD, this correlation can complicate searches for resonant new physics by sculpting the background.
To circumvent this correlation in the case of these searches, the EMD-based approach can be combined with mass decorrelation techniques~\cite{Dolen:2016kst,Shimmin:2017mfk,Moult:2017okx} or with ideas such as CWoLa hunting~\cite{Collins:2018epr} to look for anomalies within mass bins compared to sidebands.

\section{Conclusions}
\label{sec:conclusion}

The CMS Open Data is an exciting resource for performing exploratory studies in collider physics.
In this paper, we performed the first ever exploration of the metric space of QCD jets on real collider data, using the EMD~\cite{Komiske:2019fks} as our measure of jet similarity.
The EMD provides complementary information to traditional histogram-based analyses, and it also provides new strategies for data visualization in particle physics.
In terms of quantitative measures, we showed how to use the EMD to characterize the impact of detector effects and to calculate the intrinsic dimension of a jet ensemble.
For qualitative studies, we showed how to use the EMD to identify the most representative jets in a histogram bin and the least representative jets in the ensemble as a whole, where the latter analysis is particularly interesting in the context of anomaly detection for new physics searches~\cite{Aguilar-Saavedra:2017rzt,Collins:2018epr,Hajer:2018kqm,Heimel:2018mkt,Farina:2018fyg,Collins:2019jip,Roy:2019jae}.

Beyond the specific EMD studies here, a key outcome of this research is a processed and validated jet sample for use in future jet studies consisting of jets in the CMS 2011 Open Data with a $p_T$ above \SI{375}{GeV}.
This processed single-jet dataset is available on the \textsc{Zenodo} platform~\cite{MOD:ZenodoCMS,MOD:ZenodoMC170,MOD:ZenodoMC300,MOD:ZenodoMC470,MOD:ZenodoMC600,MOD:ZenodoMC800,MOD:ZenodoMC1000,MOD:ZenodoMC1400,MOD:ZenodoMC1800} along with the analysis tools needed to make the bulk of plots in this paper~\cite{EnergyFlow,MODRepo}.
This sample is ready to use out-of-the-box by future users, since JEC factors and JQC are available and easy to apply, and baseline event selection criteria have been chosen to ensure that the \texttt{Jet300} trigger is fully efficient.
Because we apply the same processing pipeline to corresponding simulated MC events, one can assess the impact of detector effects on new jet analysis strategies.
While we have not performed detector unfolding or estimation of systematic uncertainties in this exploratory study, our dataset contains sufficient information to implement these important elements, which we leave to future work.
As an important stress test of this archival strategy, we plan to perform our next jet physics analysis directly on the released datasets without ever accessing the underlying CMS AOD files.

There are a number of future directions to pursue using the EMD.
We focused on a narrow $p_T$ range of [375,425]\,GeV in this paper in order to have a more uniform jet sample, but it would be interesting to perform EMD studies on higher $p_T$ jets.
This is particularly relevant in the context of the intrinsic dimension; in a preliminary QCD calculation of the correlation dimension as a function of jet $p_T$, we find non-trivial dependence both on $Q$ and on the quark/gluon composition of the sample.
One application suggested in \Ref{Komiske:2019fks} is using EMD for jet classification, and it would be interesting to do a data/simulation classification study in the spirit of \Refs{DAgnolo:2018cun, DeSimone:2018efk} to identify regions of phase space that are not well modeled by the current generation/simulation tools.
In this study, we focused on applying the EMD to individual jets, but it could also be applied to events as a whole, which would be a novel strategy to explore the \texttt{MinimumBias} Primary Dataset.
It would also be interesting to explore alternative EMD definitions that incorporate PID information.

Finally, we applaud the commitment shown by the CMS experiment to releasing research-grade public data.
The inclusion of simulated datasets in the 2011 release was essential for us to gain confidence in the robustness of track-based observables for jet substructure studies.
Even without the actual data files, the simulated datasets are a valuable resource for phenomenological studies, since they cover a wide range of final states with fully realistic detector information.
As CMS continues to release research-grade data, we hope that more researchers take advantage of this unique resource for particle physics.

\begin{acknowledgments}
We thank CERN, the CMS collaboration, and the CMS Data Preservation and Open Access (DPOA) team for making research-grade collider data available to the public.
We specifically thank Edgar Carrera, Kati Lassila-Perini, and Tibor Simko for help processing the CMS Open Data, and Salvatore Rappoccio for help implementing CHS.
We thank Maximilian Henderson, Edward Hirst, and Ziqi Zhou for collaboration in the early stages of this work.
We benefitted from additional feedback from Cari Cesarotti, Kyle Cranmer, Achim Geiser, Matthew LeBlanc, David Miller, Benjamin Nachman, Jennifer Roloff, and Yotam Soreq.
This work was supported by the Office of Nuclear Physics of the U.S. Department of Energy (DOE) under Grant No. DE-SC0011090 and the DOE Office of High Energy Physics under Grant Nos. DE-SC0012567 and DE-SC0019128.
RM is additionally supported by a fellowship from the Heising-Simons Foundation.
JT is additionally supported by the Simons Foundation through a Simons Fellowship in Theoretical Physics.
We benefited from the hospitality of the Harvard Center for the Fundamental Laws of Nature and the Fermilab Distinguished Scholars program.
Cloud computing resources were provided through a Google Cloud allotment from the MIT Quest for Intelligence.
\end{acknowledgments}

\appendix

\section{Missing and Zeroed Luminosity Blocks}
\label{app:missinglumi}

As mentioned in \Sec{subsec:trigger_and_lumi}, there are 89 valid LBs tabulated in \Ref{CMS:luminosity2011} that do not appear anywhere in the \texttt{Jet} Primary Dataset~\cite{CMS:JetPrimary2011A}.
There is of course the possibility that we made a mistake in processing the data, though we verified that \texttt{MODProducer} recovers the total number of events (both valid and not) quoted in \Ref{CMS:JetPrimary2011A}.
Also, the missing LBs do not appear to represent a missing AOD file, which was an issue that had to be resolved for \Ref{Tripathee:2017ybi}.
In particular, the missing LBs do not appear to be linked in time, whereas a given AOD file typically has consecutive sequences of LBs.
Moreover, there are strange characteristics of the missing LBs that suggest that there might be more systematic issues at play.

We can classify the missing LBs into two main categories:
\begin{enumerate}
\item \textit{Near zero luminosity}.  For 17 missing LBs, the recorded luminosity was less than \SI{0.03}{\mu b^{-1}}. It is plausible that none of the jet triggers fired during these LBs, in which case they should count (negligibly) toward the integrated luminosity of the run.
\item \textit{Large delivered/recorded discrepancy}.  For 71 missing LBs, the recorded luminosity was at least an order of magnitude smaller than the delivered luminosity.  It is plausible that these LBs should not have been classified as  valid, in which case it is consistent to ignore them.
\end{enumerate}
Curiously, there was one missing LB where the discrepancy between the delivered and recorded luminosities was only 2.3\%.
This is consistent with the typical delivered/recorded mismatch for the valid LBs in the \texttt{Jet} Primary Dataset, which is around 3\%.

Another issue raised in \Sec{subsec:trigger_and_lumi} is that there are 201 valid LBs present in \Ref{CMS:luminosity2011} which have zero recorded luminosity.
The 164 such LBs in Run A can be categorized as follows:
\begin{enumerate}
\item \textit{Exactly zero delivered luminosity}.  For 3 zeroed LBs, the delivered luminosity was also zero. Of these, 1 LB contained 0 events; the other 2 contained a total of 3 events that were triggered in the \texttt{Jet} Primary Dataset.
\item \textit{Near zero delivered luminosity}.  For 20 zeroed LBs, the delivered luminosity was less than \SI{0.05}{\mu b^{-1}}, so it is expected that the recorded luminosity could be zero.  Of these, 11 LBs contained 0 events; the other 9 contained a total of 23 events that were triggered on in the \texttt{Jet} Primary Dataset, so we can safely ignore these as well.
\item \textit{Sizable delivered luminosity}.  For 141 zeroed LBs, the delivered luminosity was greater than \SI{2.7}{nb^{-1}}, so one expects at least one of the \texttt{Jet} triggers to have fired.  Of these, 9 LBs contained 0 events; the other 132 contained a total of 20,850 events, even though the recorded luminosity was zero.  Most likely, these were misclassified as valid LBs.
\end{enumerate}
Tallying these together, there are 21 zeroed LBs that have zero events, which are already counted as missing LBs above.
The remaining 143 zerored LBs have a total of 20,876 events, which is the number listed in \Tab{table:full_list_of_triggers}.
Following the recommendation of CMS, we omit all of the zeroed LBs from our analysis.

While these missing and zeroed LBs do not affect the conclusions of our physics studies, they do highlight the importance of stress-testing archival data strategies to make sure that there is validated information available to future generations of collider enthusiasts~\cite{strassler2019slow}.

\begin{figure}[t]
  \centering
    \includegraphics[width=\columnwidth]{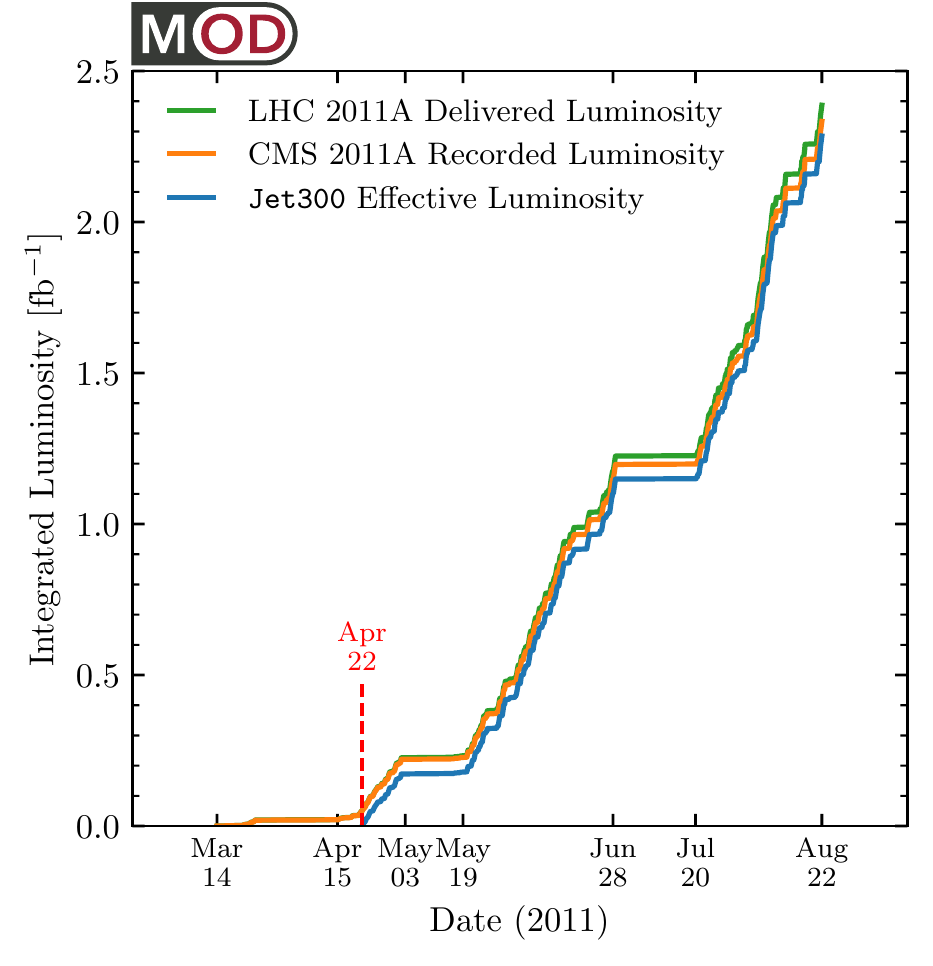}
    \caption{The delivered and integrated luminosity for the Run 2011A dataset over time.
    Also shown is the effective luminosity of the {\tt Jet300} trigger, which was activated on April 22, 2011.}
    \label{figure:del_rec_300}
\end{figure}

\begin{figure*}[t]
  \centering
    \subfloat[]{
    \includegraphics[height=0.45\textwidth]{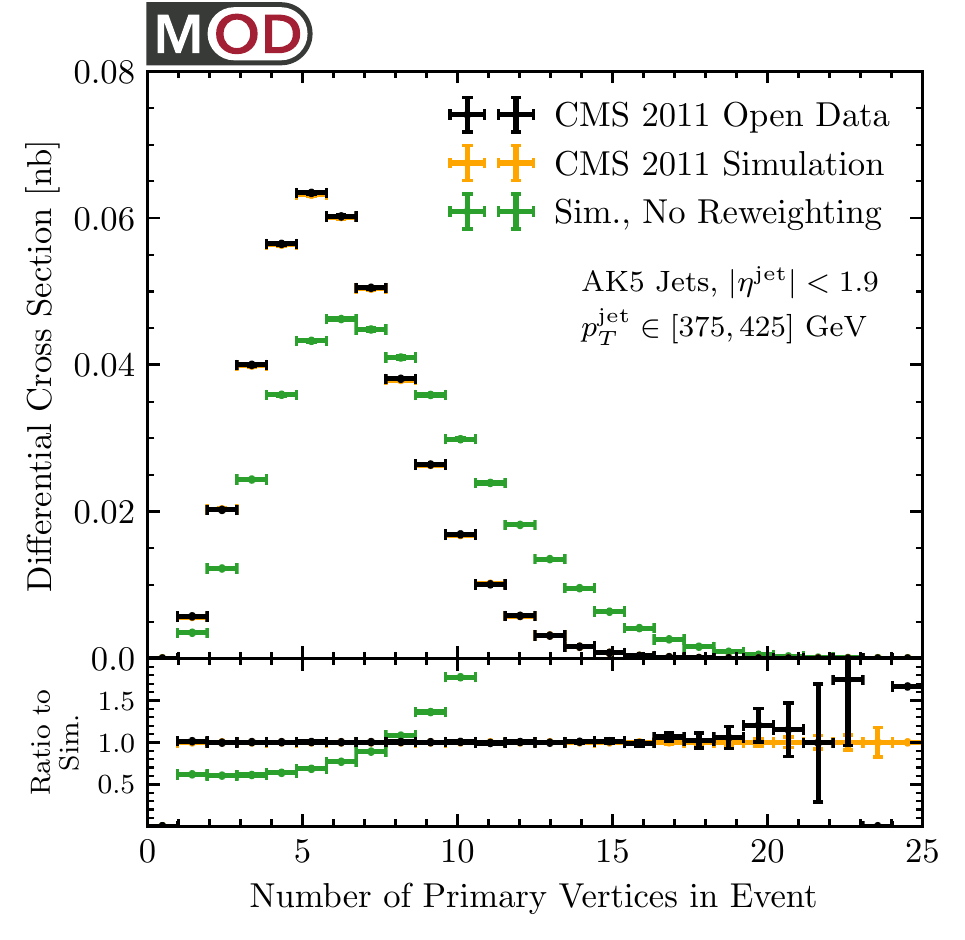}
        \label{figure:NPVhistogram}
}
    \subfloat[]{
    \includegraphics[height=0.45\textwidth]{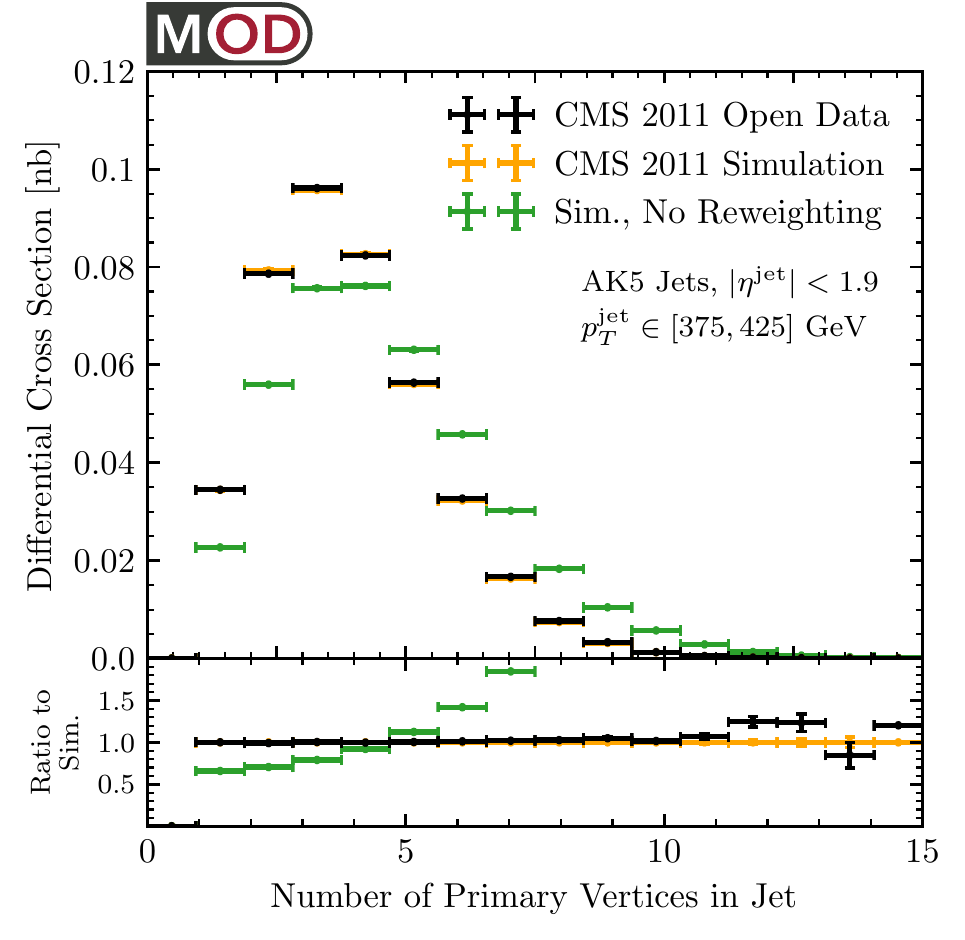}
        \label{figure:NPVhistogram_injet}
}
    \caption{
    Level of pileup contamination in the actual and simulated CMS datasets, with and without the pileup reweighting.
    Shown are the number of primary vertices (a) in the event as a whole and (b) associated with the reconstructed jets of interest. 
    Larger values of $N_{\rm PV}$ correspond to more pileup.
    }
    \label{figure:NPVhistogram_both}
\end{figure*}

For completeness, in \Fig{figure:del_rec_300}, we plot the total delivered and recorded luminosities for Run 2011A as a function of date, along with the effective luminosity for the \texttt{Jet300} trigger.
Note that the loss of luminosity due to the late turn-on of the \texttt{Jet300} trigger has a negligible effect on our analyses.

\section{Aspects of Pileup}
\label{app:pileupreweighting}

The CMS simulated MC samples include the effect of pileup, but the number of overlapping events differs from what is observed in the CMS 2011 Open Data.
To correct for this, we reweight the MC events to match the observed number of primary vertices ($N_{\rm PV}$).
Note that a larger number of primary vertices is associated with a larger amount of pileup contamination.

The effect of this reweighting is shown in \Fig{figure:NPVhistogram}, where we plot the number of primary vertices associated with each event in the CMS Open Data compared to the MC simulation, both before and after reweighting.
The reweighting factor is derived from all ``medium'' quality jets with $p_T^\text{jet}> 375$ GeV and $|\eta^\text{jet}|<1.9$, though the plot only shows the $p_T^\text{jet} \in [375,425]\,\text{GeV}$ range.
As a cross check, in \Fig{figure:NPVhistogram_injet}, we plot the number of primary vertices with at least one track associated with the reconstructed jet of interest.
From this, we conclude that the event-wide reweighting does indeed correct the in-jet pileup contamination level.

\begin{figure*}
\subfloat[]{
	\includegraphics[height=0.475\textwidth]{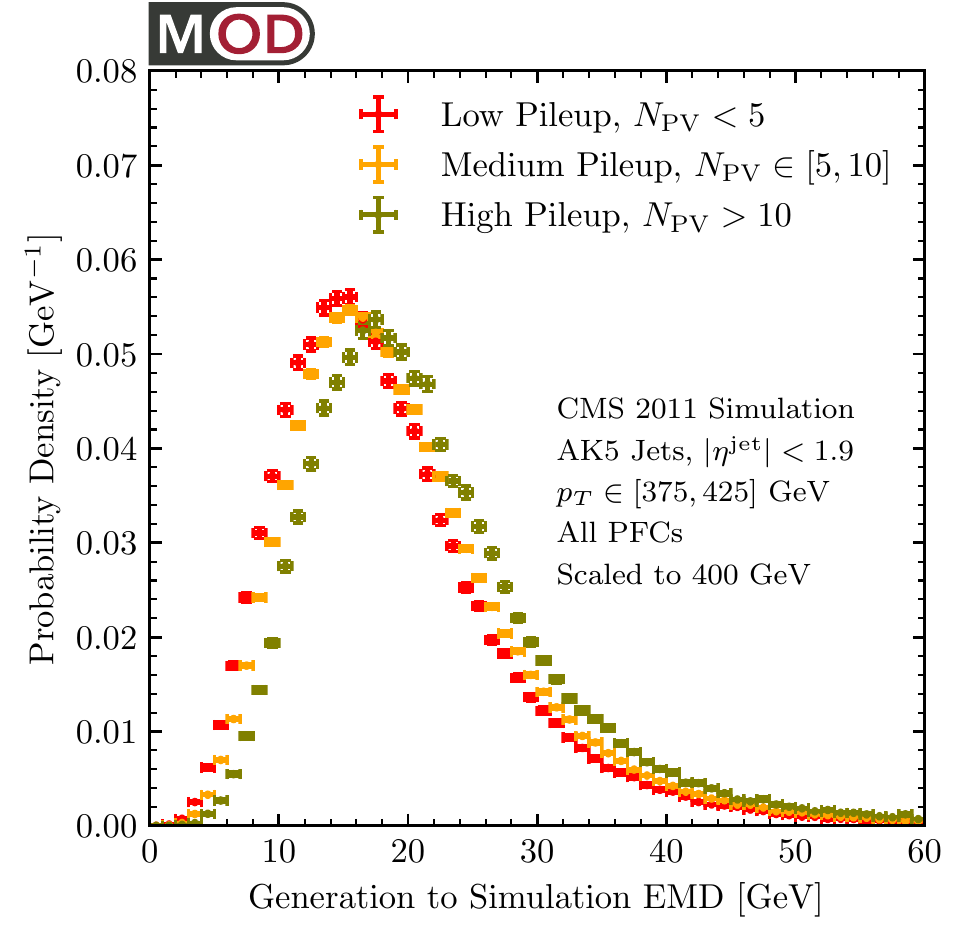}
}
\subfloat[]{
	\includegraphics[height=0.475\textwidth]{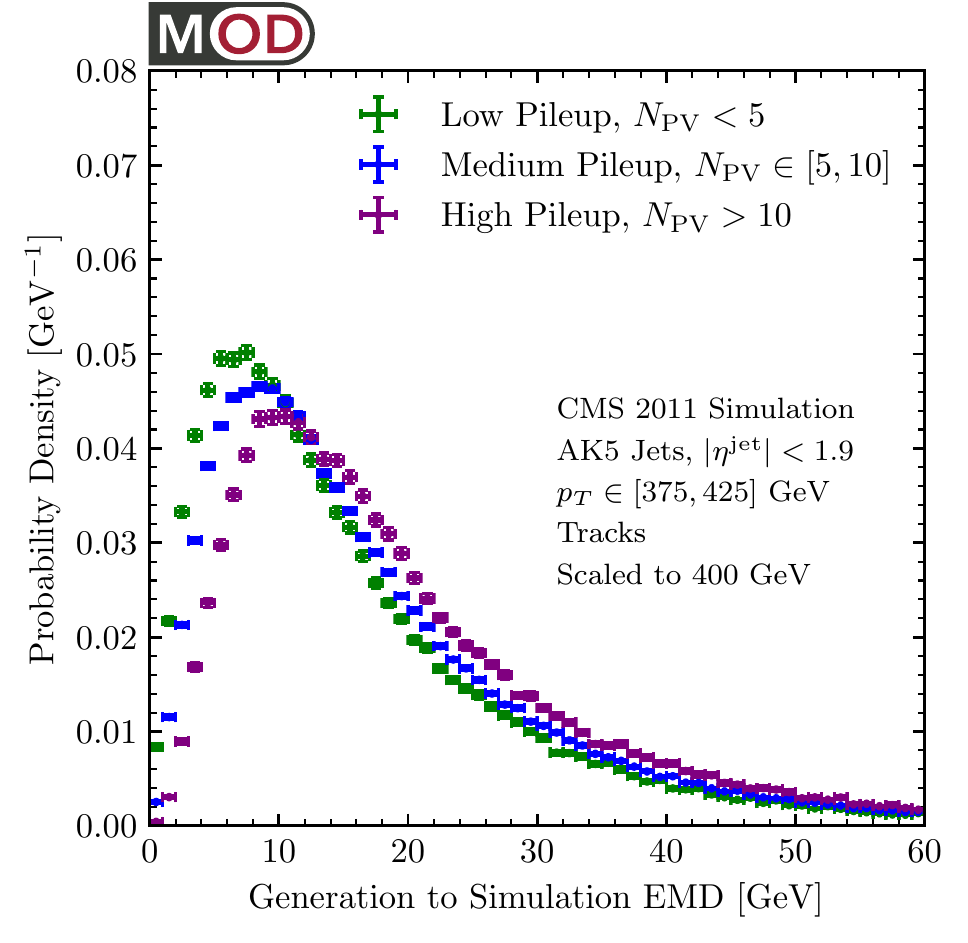}
}\\
\subfloat[]{
	\includegraphics[height=0.475\textwidth]{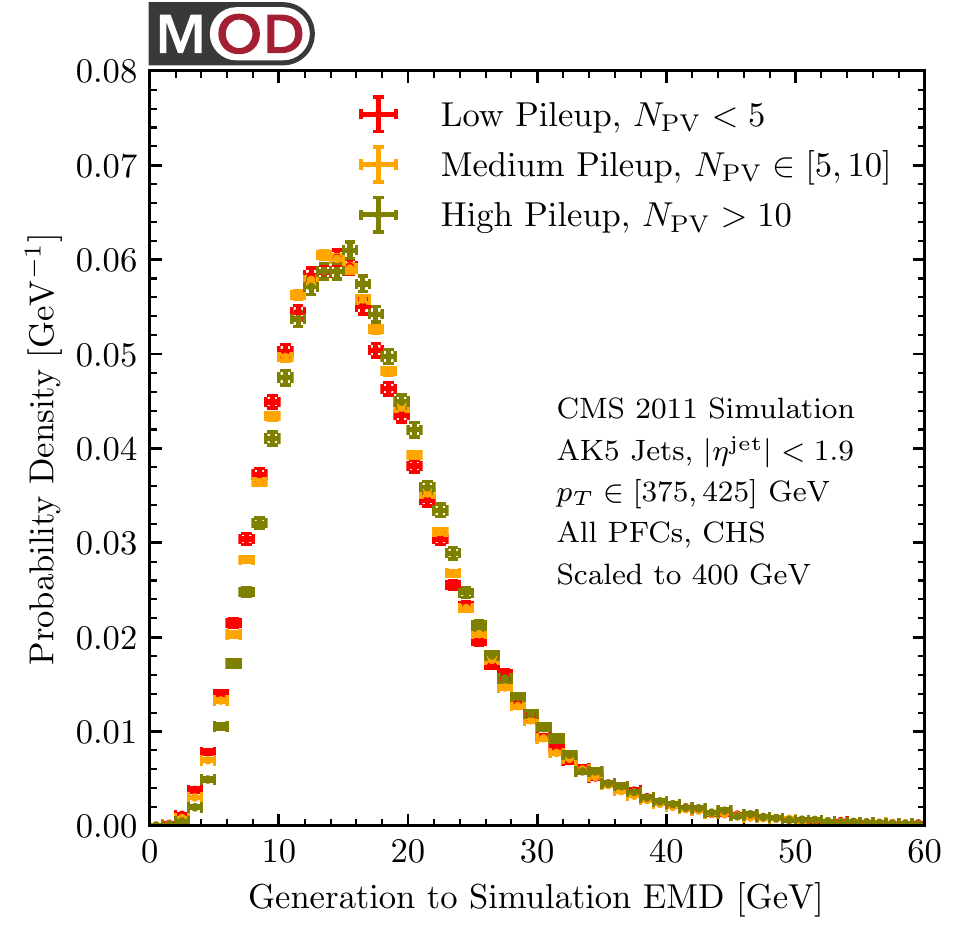}
}
\subfloat[]{
	\includegraphics[height=0.475\textwidth]{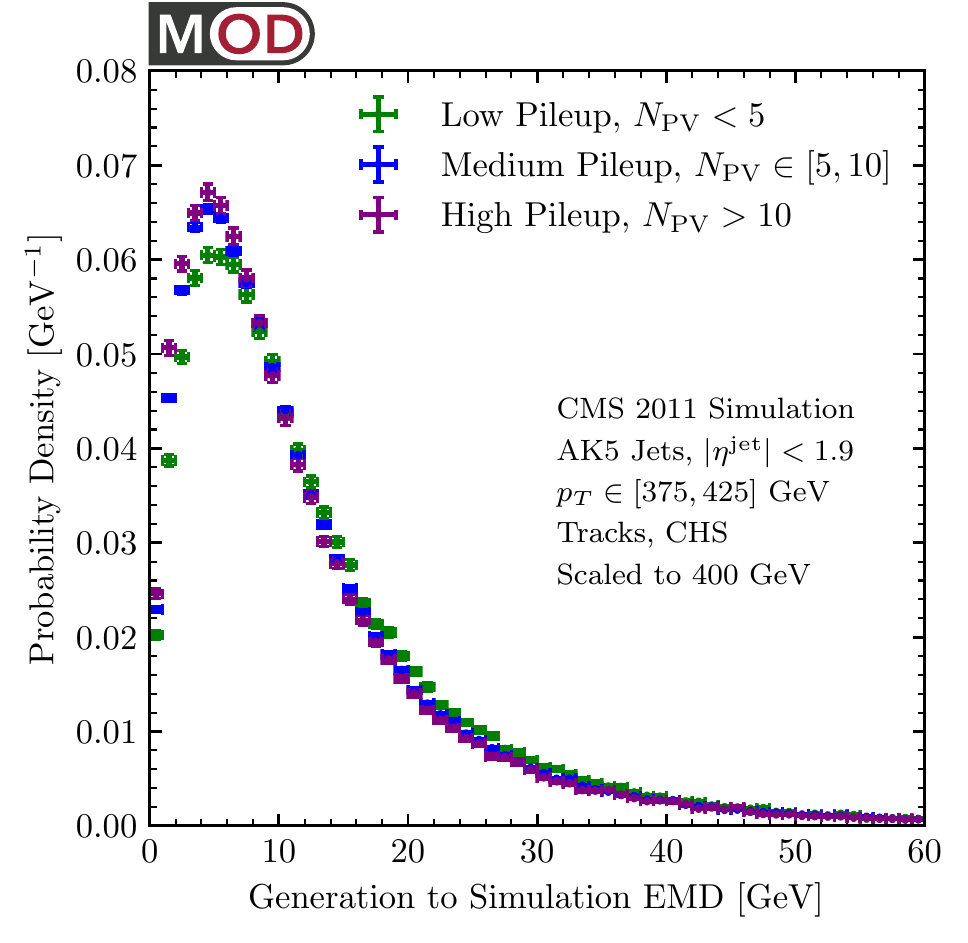}
	\label{fig:pileupEMDs_tracks_CHS}
}
\caption{
The generation-to-simulation EMD in the style of \Fig{fig:emdcmp} for different levels of pileup contamination, as quantified by the number of primary vertices ($N_{\rm PV}$) in the event.
Distributions are for (left column) all PFCs and (right column) just tracks, shown (top row) before and (bottom row) after CHS is applied.
}
\label{fig:pileupEMDs}
\end{figure*}

We can quantify the performance of CHS for pileup mitigation by performing an EMD analysis analogous to \Sec{subsec:emd_detector}.
In \Fig{fig:pileupEMDs}, we show the generation-to-simulation EMD before and after CHS is applied, split into low ($N_{\rm PV} < 5$), medium ($N_{\rm PV} \in [5,10]$), and high ($N_{\rm PV} > 10$) levels of pileup contamination.
First, we see that the EMD grows (i.e.~reconstruction degrades) as the pileup levels increase, though for these modest levels of pileup, the distortions are not so large.
As already shown in \Fig{fig:emdcmp}, CHS does mitigate the impact of pileup, with better performance when considering just tracks.

One surprise in \Fig{fig:pileupEMDs_tracks_CHS} is that the track-only EMD gets \emph{smaller} as the pileup contamination increases.
We are not sure of the origin of this behavior.
It might be related to the use of the rescaling factors in \Eq{eq:EMDrescaling}, or it might indicate a bias where low $N_{\rm PV}$ events often have unreconstructed primary vertices, so CHS does not removes tracks that it should.
Regardless, we see that the EMD is a useful way to quantify the performance of pileup mitigation strategies.

\section{Additional Plots}
\label{app:additionalplots}

In this appendix, we provide additional plots to complement the ones in the text.

\begin{figure*}[t]
\centering
  \subfloat[]{
	\includegraphics[width=0.5\textwidth]{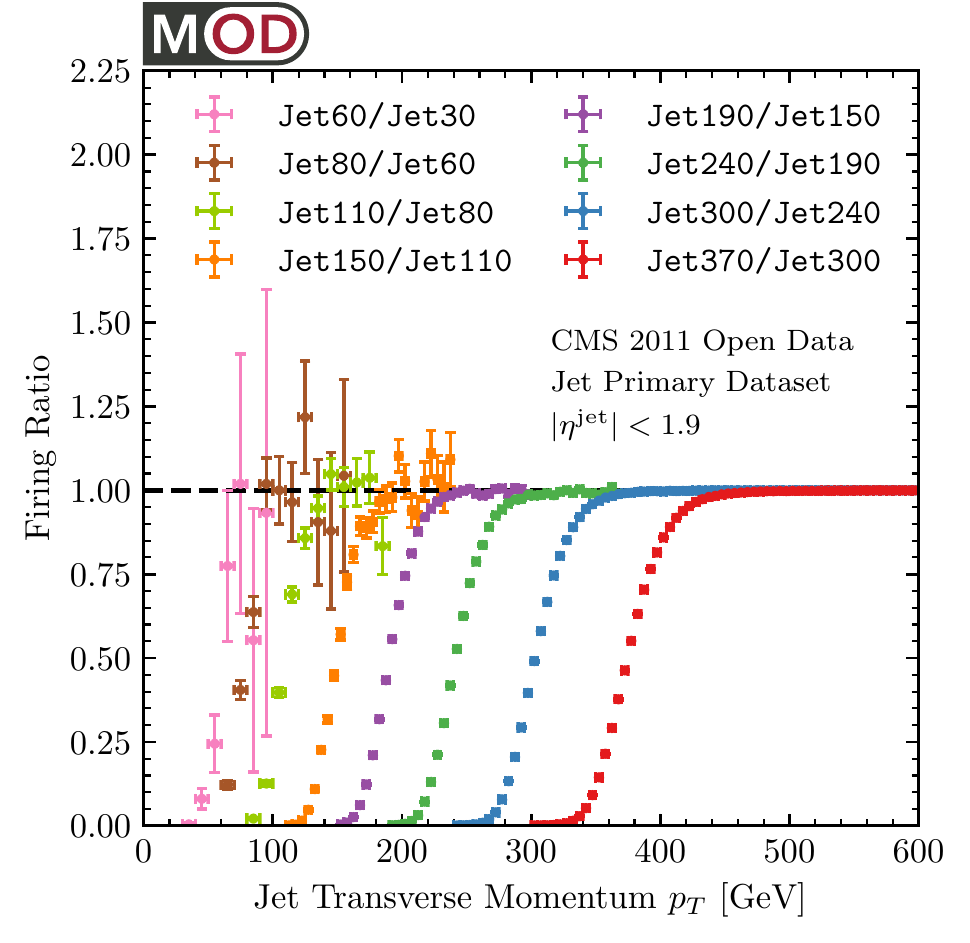}
	\label{fig:data_trigger_ratio_all}
	}
  \subfloat[]{
  \includegraphics[width=0.5\textwidth]{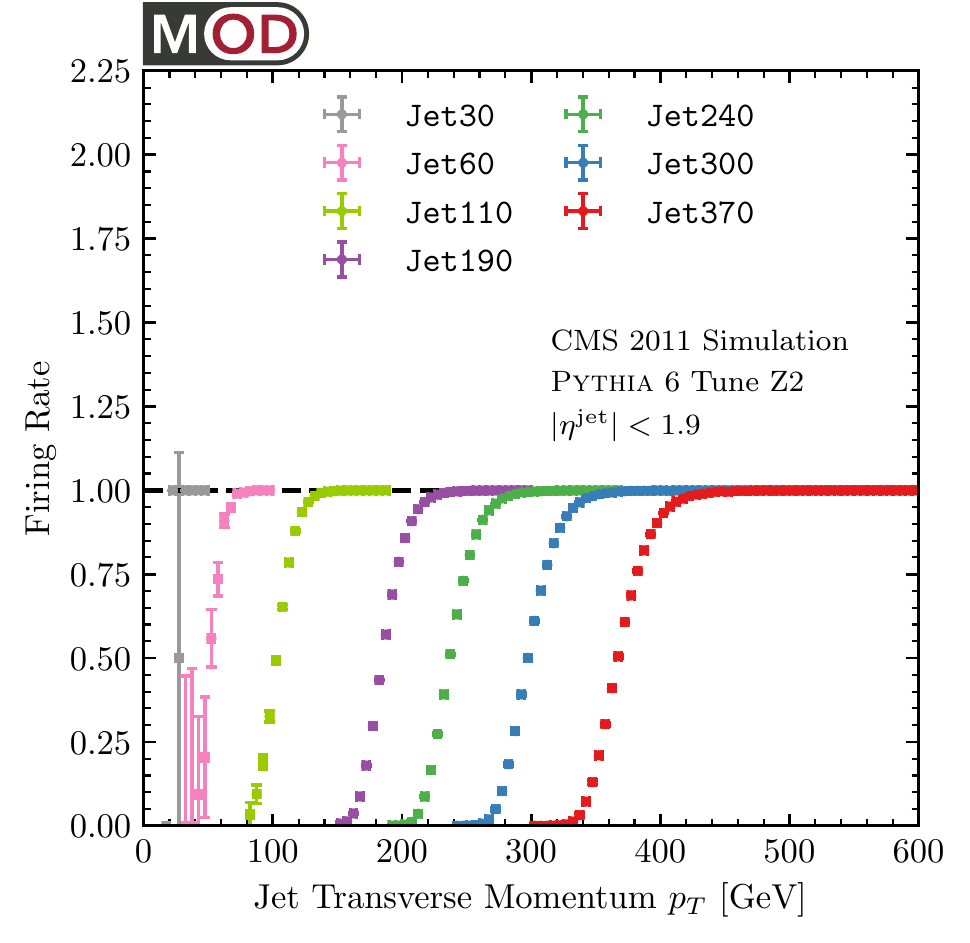}
  \label{fig:sim_trigger_efficiency_all}
  }
  \caption{(a) Relative trigger efficiency in the CMS Open Data, for 8 single-jet triggers compared to the adjacent trigger with lower $p_T$ threshold.
  Up to statistical fluctuations, the firing ratio approaches 1 in all cases, after correcting for the L1 trigger correlation subtlety in \Eq{eq:L1triggerissue}.
  (b) Absolute trigger efficiency in the MC simulation for seven single-jet triggers.
  The \texttt{Jet80} and \texttt{Jet150} triggers are not present in the simulated datasets, which are the two triggers that were turned off prior to the end of Run 2011A, as can be seen in \Fig{figures:integrated_effective_xsec}.
  Efficiency information for the \texttt{Jet300} trigger is highlighted in \Fig{fig:trigger_turn_on_sim}.
  }
\label{fig:trigger_turn_on_all_triggers}
\end{figure*}

In \Fig{fig:trigger_turn_on_all_triggers}, we plot the turn-on behavior for all of the relevant single-jet triggers, to compare to the \texttt{Jet300} study in \Fig{fig:trigger_turn_on_sim}.
In making this plot, we have to address the fact that some of the triggers share the same L1 trigger seed and their firing rates are therefore correlated.
For uncorrelated triggers, if trigger $A$ has prescale factor $p^{\rm trig}_A$ and trigger $B$ has prescale factor $p^{\rm trig}_B$ and both triggers are fully efficient, then the probability of $B$ firing given that $A$ fired is:
\begin{equation}
\mathcal{P}_{\rm uncorr}(B_{\rm fired} | A_{\rm fired}) = \frac{1}{p^{\rm trig}_B},
\end{equation}
which is independent of $p^{\rm trig}_A$ since the triggers are uncorrelated.
On the other hand, if two triggers have the same L1 seed, then the probability of $B$ firing given that $A$ fired is:
\begin{equation}
\label{eq:L1triggerissue}
\mathcal{P}_{\rm corr}(B_{\rm fired} | A_{\rm fired}) = \frac{\text{gcd}(p^{\rm trig}_A,p^{\rm trig}_B)}{p^{\rm trig}_B},
\end{equation}
where gcd is the greater common divisor.
This can be understood since if trigger $A$ ($B$) fires deterministically every $p^{\rm trig}_A$ ($p^{\rm trig}_B$) events, then they will overlap a factor of $\text{gcd}(p^{\rm trig}_A,p^{\rm trig}_B)$ more often than if the triggers fired randomly and independently.
For example, if $\text{gcd}(p^{\rm trig}_A,p^{\rm trig}_B) = p^{\rm trig}_B$, then the only time trigger $B$ can fire is if trigger $A$ has also fired, so $\mathcal{P}_{\rm corr}(B_{\rm fired} | A_{\rm fired}) = 1$.
In our case, this affects the \texttt{HLT\_Jet150}, \texttt{HLT\_Jet190}, \texttt{HLT\_Jet240}, \texttt{HLT\_Jet300}, and \texttt{HLT\_Jet370} triggers, which are all seeded by the same \texttt{L1\_SingleJet92} trigger~\cite{CMS:JetPrimary2011A}.

\begin{figure*}
  \centering
	\includegraphics[width=0.5\textwidth]{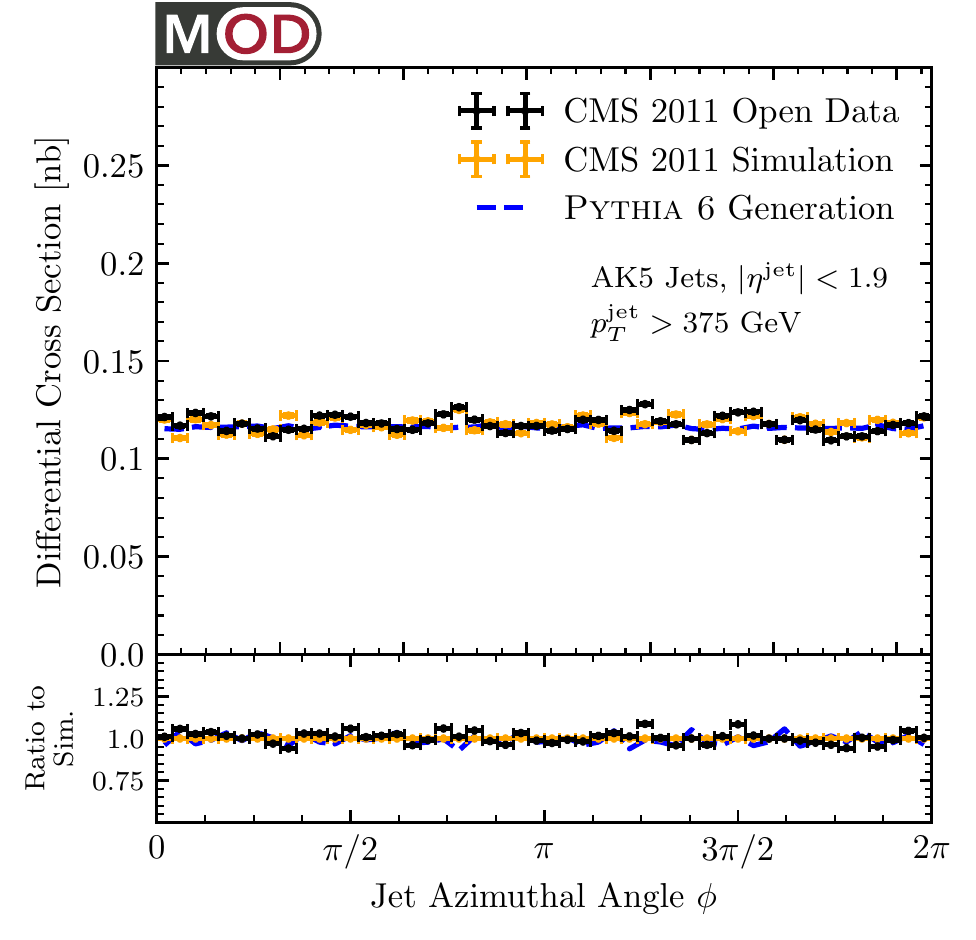}
 \caption{
 Jet azimuthal angle ($\phi$) distribution for the two hardest jets, comparing the CMS Open Data to MC event samples at the simulation level and generation level.
 See \Fig{fig:etaspectrum} for the pseudorapidity spectrum.}
    \label{fig:phispectrum}
\end{figure*}

In \Fig{fig:phispectrum}, we plot the azimuthal angle ($\phi$) distribution for the two hardest jets.
As expected, we observe a flat spectrum in both the CMS Open Data and the MC simulation, though the bin-to-bin fluctuations in the open data are larger than one would expect from statistics alone, possibly indicating an issue with the lack of $\phi$-dependence of the JECs.

\begin{figure*}
  \subfloat{
	\includegraphics[height=0.475\textwidth]{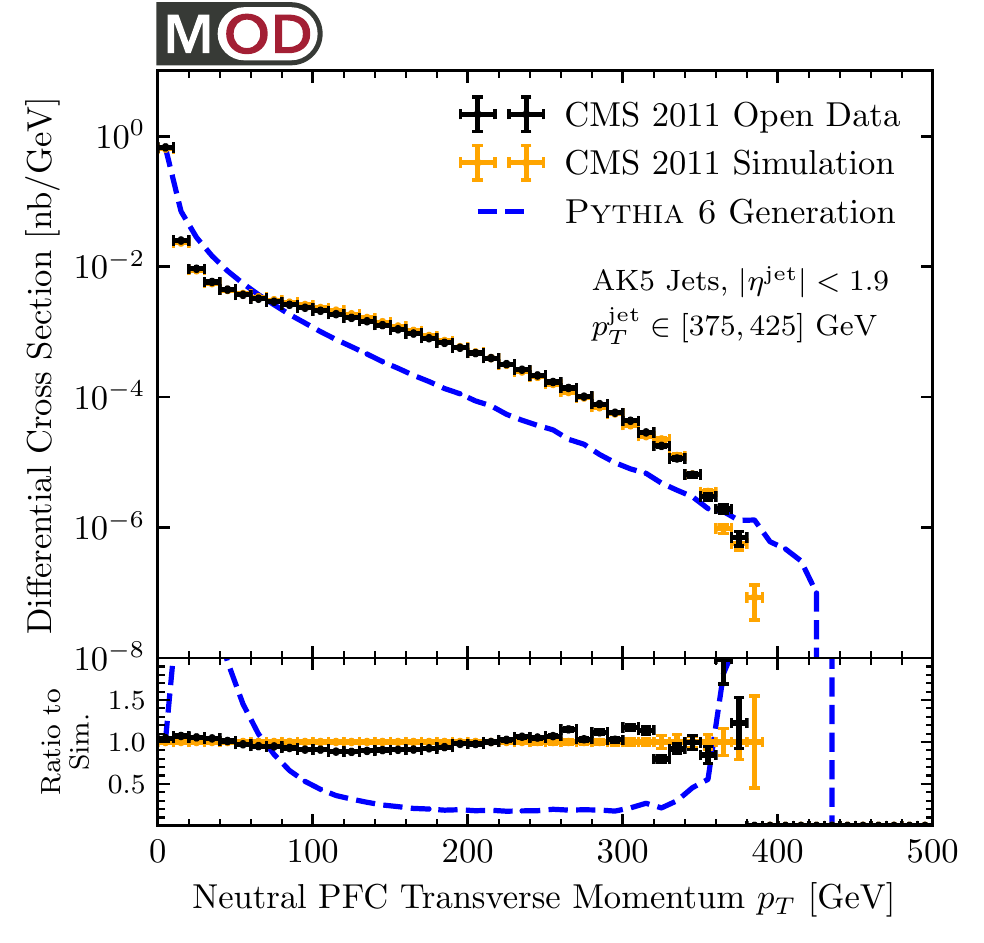}
	\label{fig:neutral_PFC_spectrum}
	}
\subfloat{
	\includegraphics[height=0.475\textwidth]{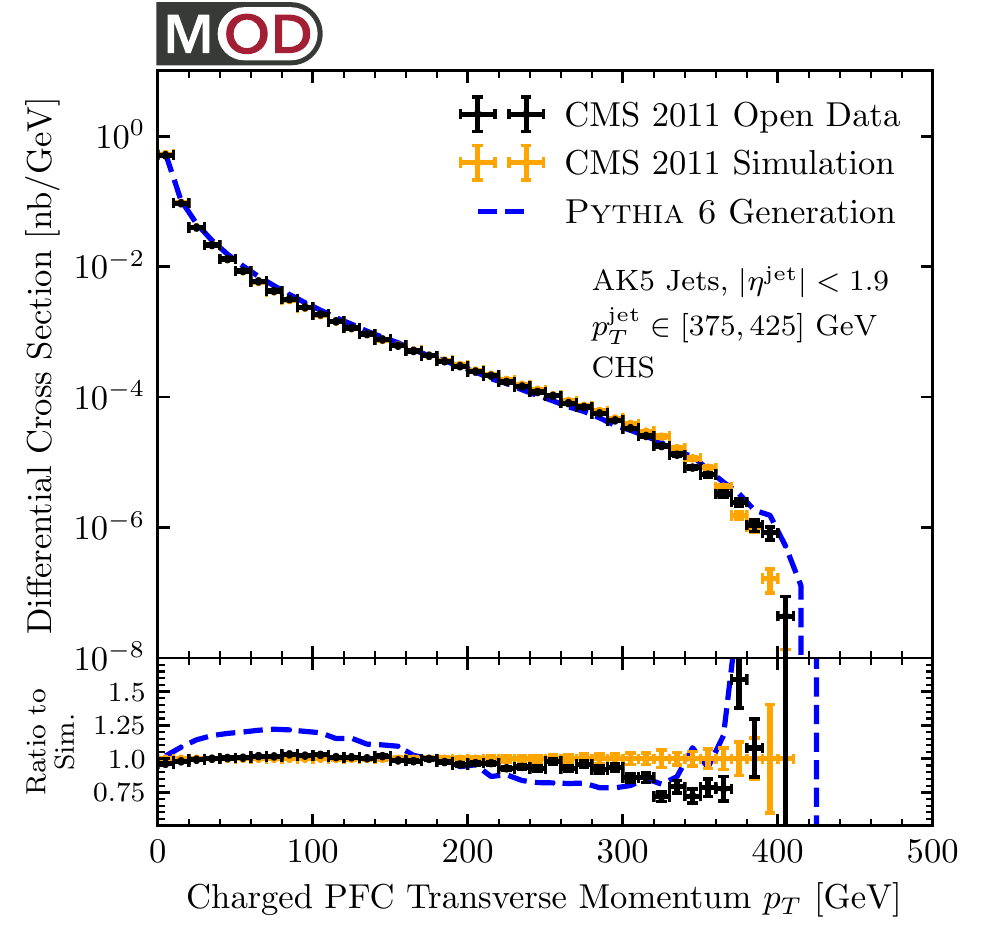}
	\label{fig:charged_PFC_spectrum}
	}
  \caption{Transverse momentum spectra for (a) neutral PFCs and (b) charged PFCs after CHS.  A zoomed version of these plots highlighting the region below 5 GeV is shown in \Fig{fig:PFC_spectrum_zoom}.}
  \label{fig:PFC_spectrum}
\end{figure*}

In \Fig{fig:PFC_spectrum}, we plot the complete PFC $p_T$ spectra for both neutral and charged constituents, going beyond the limited range shown in \Fig{fig:PFC_spectrum_zoom}.
This highlights the tighter relationship between generation-level and simulation-level information when using charged particles alone.
Though not shown, we used this plot when deciding to impose the medium JQC, since with only the loose JQC, there was an excess of events with high-$p_T$ neutral PFCs, most likely from photon-plus-jet events.

\begin{figure*}
  \subfloat[]{
	\includegraphics[height=0.47\textwidth]{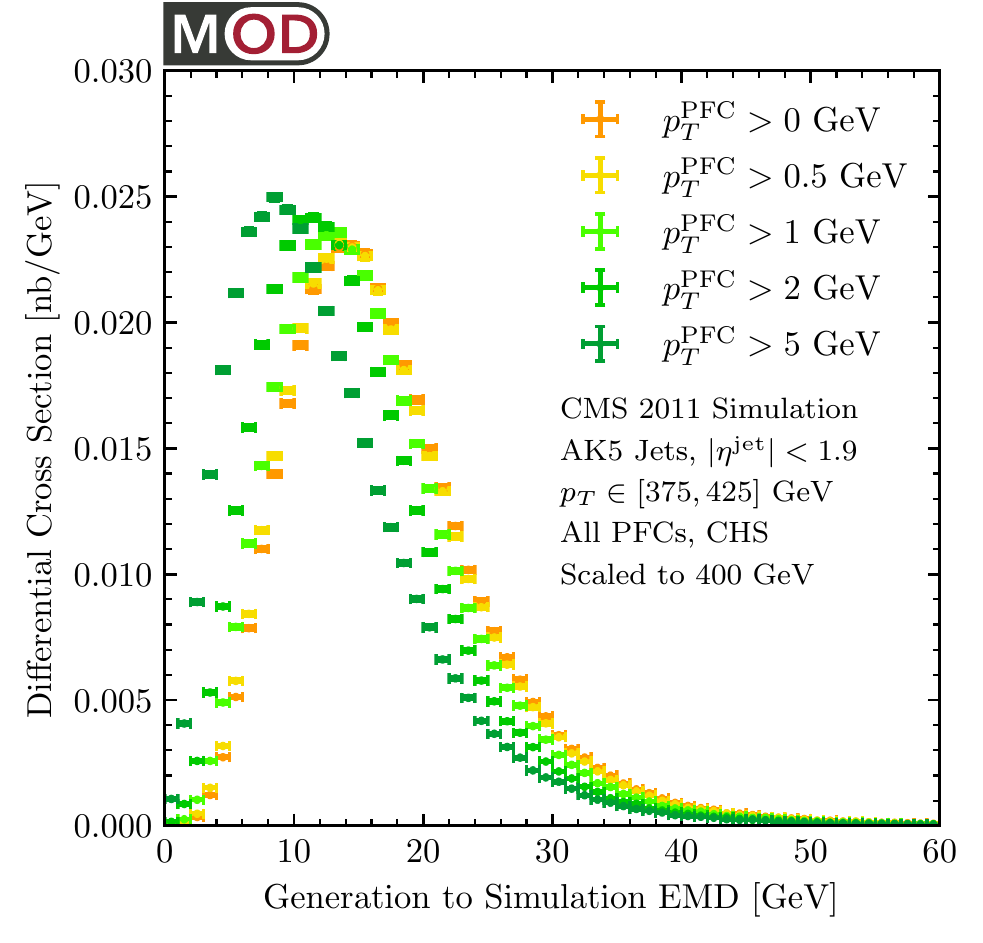}
	\label{fig:emdcmp_gensimEMDpTAll}
	}
\subfloat[]{
	\includegraphics[height=0.47\textwidth]{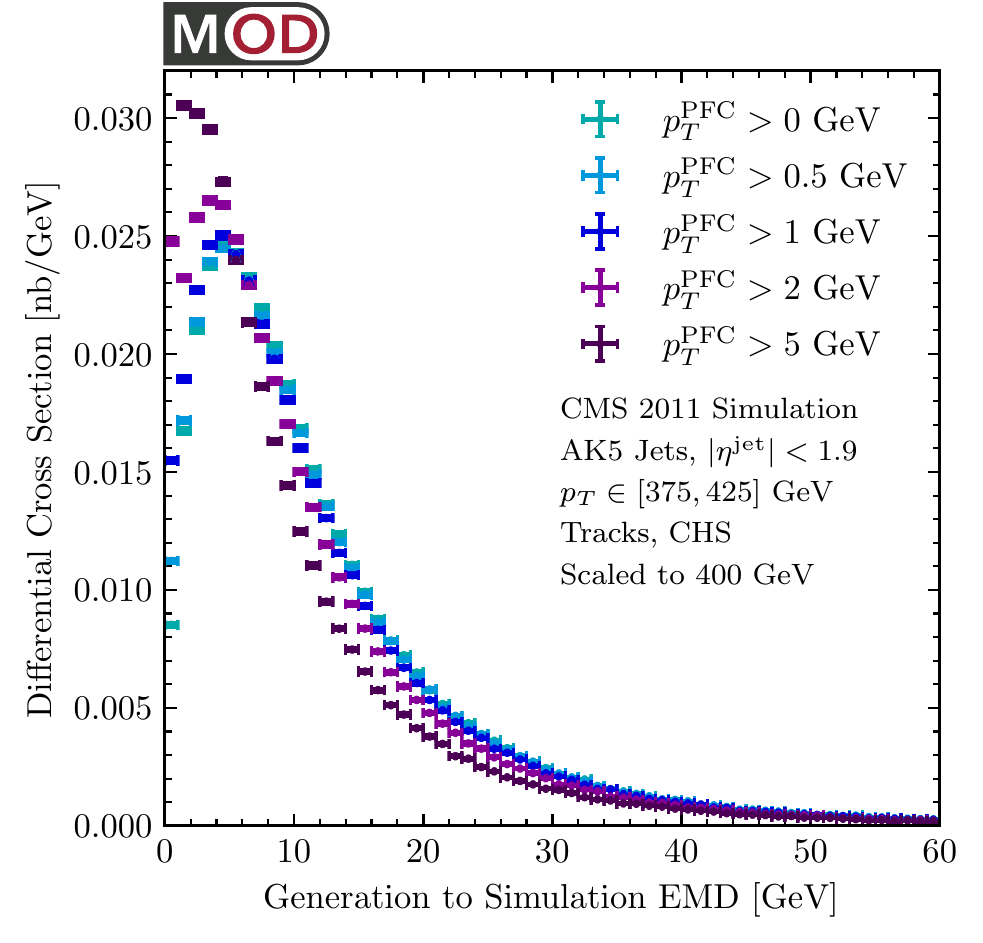}
	\label{fig:emdcmp_gensimEMDpTTrack}
	}
	\\
	  \subfloat[]{
	\includegraphics[height=0.47\textwidth]{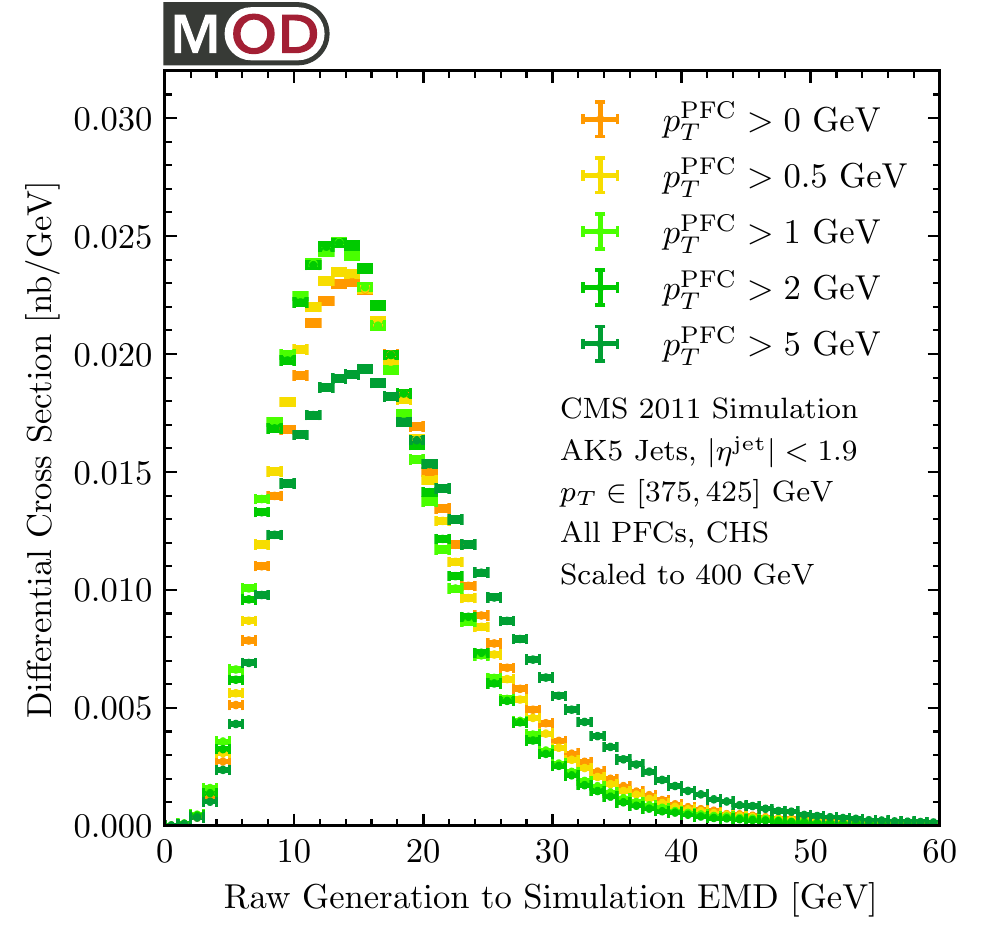}
	\label{fig:emdcmp_genrawsimEMDpTAll}
	}
\subfloat[]{
	\includegraphics[height=0.47\textwidth]{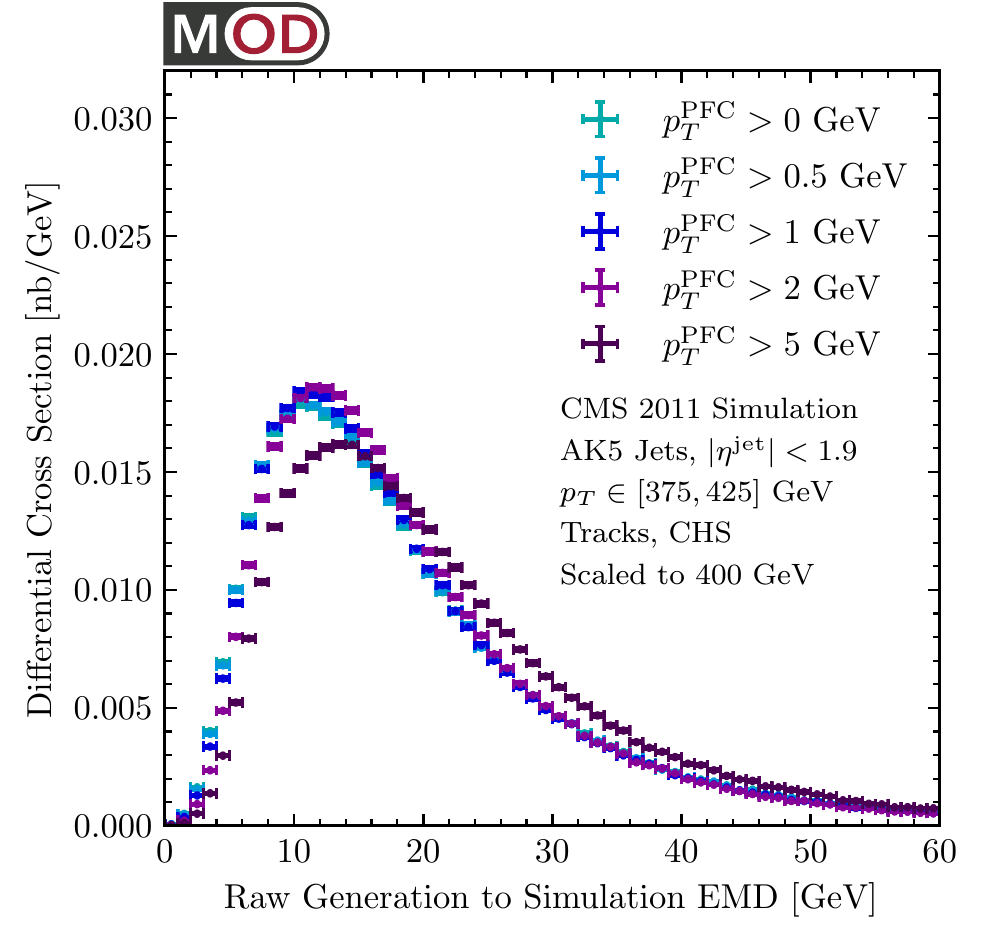}
	\label{fig:emdcmp_genrawsimEMDpTTrack}
	}
  \caption{  \label{fig:emdcmp_pT}
  Generation-to-simulation EMD as different PFC $p_T$ selections are applied to the jet constituents, for (left column) all PFCs and (right column) just tracks.
  (top row)  The baseline generation-level jet has the same $p_T^\text{PFC}$ cut and track selection requirements as the simulation-level jets.
  (bottom row)  The baseline generation-level jet uses all particles (``raw''), with no $p_T$ cuts or track restrictions.
  In all cases, we apply the rescaling factor in \Eq{eq:EMDrescaling}.
}
\end{figure*}

We now use EMD to study the impact of the $p_T^\text{PFC}$ cut in our analysis.
In the top row of \Fig{fig:emdcmp_pT}, we do an apples-to-apples comparison with the same particle selection at generation level and simulation level.
As the $p_T$ cut on the PFCs gets more aggressive, the generation-to-simulation EMD decreases, indicating better agreement.
Of course, this $p_T^\text{PFC}$ cut removes information about jet substructure, so there is a balance between minimizing detector effects and maximizing sensitivity to the underlying radiation pattern.
In the bottom row of \Fig{fig:emdcmp_pT}, the baseline generation-level jet contains all particles (``raw''), regardless of what selections are made at simulation level.
When using all PFCs in \Fig{fig:emdcmp_genrawsimEMDpTAll}, the EMD decreases (i.e.~reconstruction improves) as the $p_T^\text{PFC}$ cut gets more stringent, up until the \SI{2}{GeV} point where we start to see degradation.
When using just charged PFCs in \Fig{fig:emdcmp_genrawsimEMDpTTrack}, the peak of the EMD distribution shifts to lower values but there is a long tail, and the reconstruction always degrades with increasing $p_T^\text{PFC}$ cut.

\begin{figure*}
\centering
\subfloat[]{
\includegraphics[height=0.95\columnwidth]{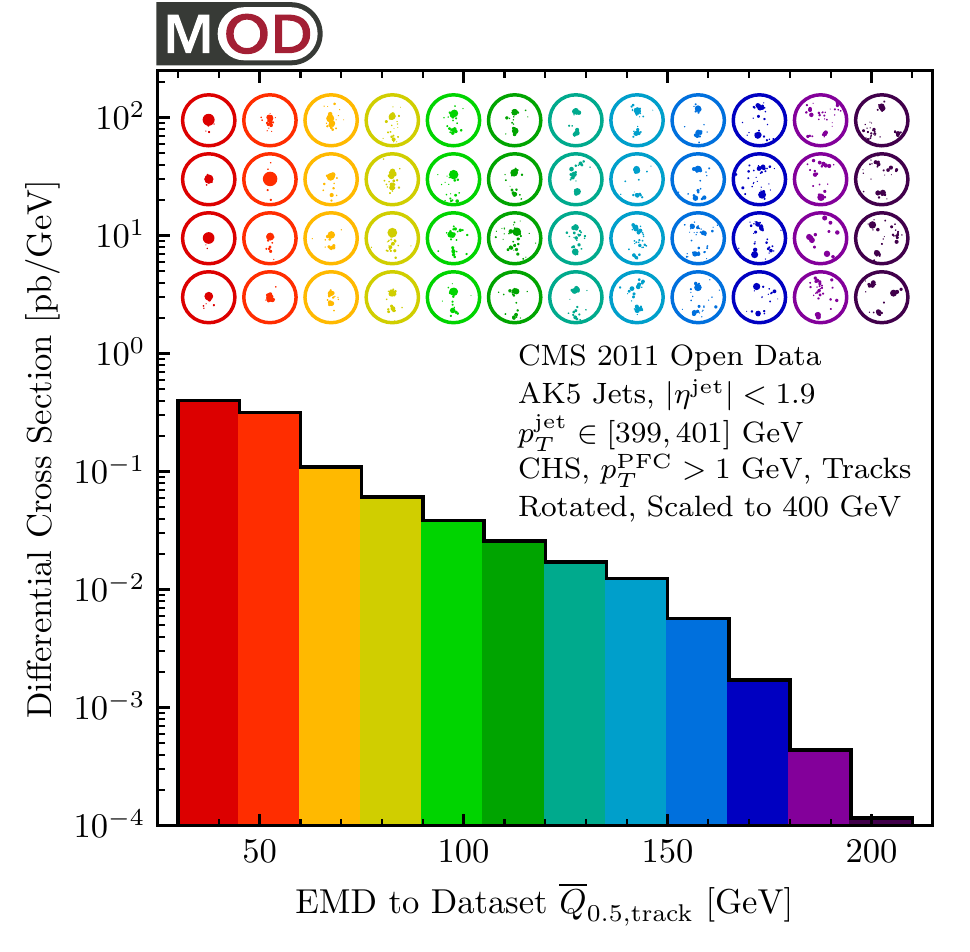}
}
\subfloat[]{
\includegraphics[height=0.95\columnwidth]{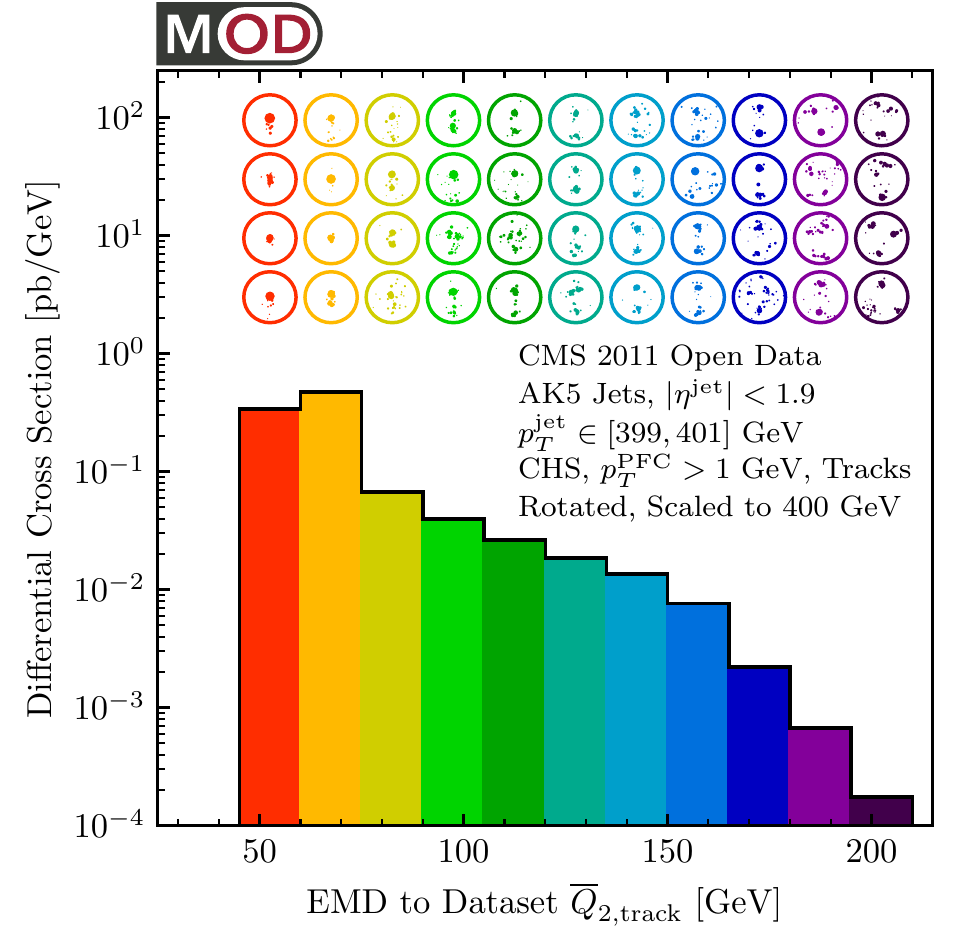}
}
\caption{\label{fig:kmedoidsemdapp} 
Distribution on the CMS Open Data of $\overline{Q}_n$ from \Eq{eq:barQdef} for (a) $n=\frac12$ and (b) $n=2$ along with the 4-medoids in each histogram bin.
See \Fig{fig:kmedoidsemd} for the analogous distribution for $n=1$.
}
\end{figure*}

In \Fig{fig:kmedoidsemdapp}, we study the most anomalous jets according to $\overline{Q}_n$ from \Eq{eq:barQdef} for the additional choices of $n$ of $n=\frac12$ and $n=2$.
The results are comparable to the $n=1$ case shown in \Fig{fig:kmedoidsemd}, with all three choices of $n$ agreeing on the three most anomalous jets.

\bibliography{mod_emd}

\end{document}